\pdfoutput=1
\documentclass[11pt,twoside,a4paper,cmspaper,final,collab]{cms-tdr}
\def\svnVersion{aff0633}\def\svnDate{2026/03/23}\def\cmsCernNoTag{CERN-EP-2025-173}\def\cmsCernDate{\today}\def\cmsMessage{Published in the Journal of High Energy Physics as \href{http://dx.doi.org/10.1007/JHEP03(2026)073}{\doi{10.1007/JHEP03(2026)073}.}}
\begin{document}\cmsNoteHeader{HIG-24-004}

\newcommand{\WW}{\ensuremath{\PW\PW}\xspace}  
\newcommand{\ZZ}{\ensuremath{\PZ\PZ}\xspace}
\newcommand{\Zg}{\ensuremath{\PZ\gamma}\xspace}
\newcommand{\HtoWW}{\ensuremath{\PH\to\PW\PW}\xspace} 
\newcommand{\HtoTT}{\ensuremath{\PH\to\PGt\PGt}\xspace}
\newcommand{\HVV}{\ensuremath{\PH\PV\PV}\xspace}
\newcommand{\glgl}{\ensuremath{\Pg\Pg}\xspace}
\newcommand{\emu}{\ensuremath{\Pe\PGm}\xspace}

\newcommand{\wg}{\ensuremath{\PW{}\PGg}\xspace} 
\newcommand{\mll}{\ensuremath{m_{\Pell\Pell}}\xspace} 
\newcommand{\mth}{\ensuremath{\mT}\xspace} 
\newcommand{\mtltwo}{\ensuremath{\mT^{\Pell_2}}\xspace} 
\newcommand{\mjj}{\ensuremath{m_{\text{jj}}}\xspace}
\newcommand{\ptll}{\ensuremath{\pt^{\Pell\Pell}}\xspace} 

\newcommand{\ptlone}{\ensuremath{\pt^{\Pell_1}}\xspace} 
\newcommand{\ptltwo}{\ensuremath{\pt^{\Pell_2}}\xspace} 
\newcommand{\ptlthree}{\ensuremath{\pt^{\Pell_3}}\xspace} 
\newcommand{\ptjone}{\ensuremath{\pt^{j_ 1}}\xspace} 
\newcommand{\ptjtwo}{\ensuremath{\pt^{j_2}}\xspace} 

\newcommand{\etajone}{\ensuremath{\eta_\mathrm{{j_1}}}\xspace}
\newcommand{\etajtwo}{\ensuremath{\eta_\mathrm{{j_2}}}\xspace}
\newcommand{\etalone}{\ensuremath{\eta_\mathrm{{\Pell_1}}}\xspace}
\newcommand{\etaltwo}{\ensuremath{\eta_\mathrm{{\Pell_2}}}\xspace}
\newcommand{\phijone}{\ensuremath{\phi_\mathrm{{j_1}}}\xspace}
\newcommand{\phijtwo}{\ensuremath{\phi_\mathrm{{j_2}}}\xspace}
\newcommand{\philone}{\ensuremath{\phi_\mathrm{{\Pell_1}}}\xspace}
\newcommand{\philtwo}{\ensuremath{\phi_\mathrm{{\Pell_2}}}\xspace}

\newcommand{\mlj}{\ensuremath{m_{\Pell j}}}
\newcommand{\detajj}{\ensuremath{\Delta\eta_\mathrm{{jj}}}\xspace}
\newcommand{\dphijj}{\ensuremath{\Delta\Phi_\mathrm{{jj}}}\xspace}
\newcommand{\dphill}{\ensuremath{\Delta\phi_\mathrm{{\Pell\Pell}}}\xspace}
\newcommand{\drll}{\ensuremath{\Delta\mathrm{R}_\mathrm{{\Pell\Pell}}}\xspace}

\newcommand{\DVbfGgh}{\ensuremath{\mathcal{D}^\mathrm{(ME)}_\mathrm{VBF\_ggH}}\xspace}
\newcommand{\DVbfVh}{\ensuremath{\mathcal{D}^\mathrm{(ME)}_\mathrm{VBF\_VH}}\xspace}
\newcommand{\DGghVh}{\ensuremath{\mathcal{D}^\mathrm{(ME)}_\mathrm{ggH\_VH}}\xspace}
\newcommand{\DVbfDy}{\ensuremath{\mathcal{D}^\mathrm{(ME)}_\mathrm{VBF\_DY}}\xspace}

\newcommand{\dytt}{DY\ensuremath{\PGt\PGt}\xspace}
\newcommand{\DYtott}{DY\ensuremath{\to\PGt\PGt}\xspace}
\newcommand{\VgS}{V\ensuremath{{\gamma^*}}\xspace}

\newcommand{\Dtwodim}{\ensuremath{\mathcal{D}_\mathrm{VBF,ggH}}\xspace}
\newcommand{\Dvbf}{\ensuremath{\mathcal{D}_\mathrm{VBF}}\xspace}
\newcommand{\Dggf}{\ensuremath{\mathcal{D}_\mathrm{ggH}}\xspace}

\newcommand{\CP}{\ensuremath{CP}\xspace}

\newlength\cmsTabSkip\setlength{\cmsTabSkip}{1ex}
\providecommand{\cmsTable}[1]{\resizebox{\textwidth}{!}{#1}}
\ifthenelse{\boolean{cms@external}}{\providecommand{\CL}{C.L.\xspace}}{\providecommand{\CL}{CL\xspace}}

\cmsNoteHeader{HIG-24-004}
\title{Model-independent measurement of the Higgs boson associated production with two jets and decaying to a pair of W bosons in proton-proton collisions at \texorpdfstring{$\sqrt{s} = 13\TeV$}{sqrt(s) = 13 TeV}}

\author*[cern]{The CMS Collaboration}

\date{\today}

\abstract{A model-independent measurement of the differential production cross section of the Higgs boson decaying into a pair of W bosons, with a final state including two jets produced in association, is presented. In the analysis, events are selected in which the decay products of the two W bosons consist of an electron, a muon, and missing transverse momentum. The model independence of the measurement is maximized by employing a discriminating variable, developed through machine learning, that is agnostic to the signal hypothesis. The analysis is based on proton-proton collision data at $\sqrt{s}=13\TeV$ collected with the CMS detector from 2016--2018, corresponding to an integrated luminosity of $138\fbinv$. The production cross section is measured as a function of the difference in azimuthal angle between the two jets. The differential cross section measurements are used to constrain Higgs boson couplings within the standard model effective field theory framework.}

\hypersetup{
pdfauthor={CMS Collaboration},
pdftitle={Model-independent measurement of the Higgs boson associated production with two jets and decaying to a pair of W bosons in proton-proton collisions at sqrt(s) =13 TeV},
pdfsubject={CMS},
pdfkeywords={CMS, Higgs}}

\maketitle 

\section{Introduction}
\label{sec:introduction}

The observation of a scalar resonance compatible with the long-sought Higgs boson (\PH) was achieved in 2012 by the ATLAS and CMS Collaborations~\cite{Aad:2012tfa,Chatrchyan:2012xdj,Chatrchyan:2013lba} at the CERN LHC. This marked a significant experimental result supporting the standard model (SM) of elementary particles, which initiated a series of precision measurements aimed at determining the properties of the newly discovered particle, including its spin-parity and its couplings to fermions and electroweak (EW) gauge bosons. To date, the Higgs boson has been observed in all accessible production mechanisms through several decay channels, with no significant deviations from the SM expectations~\cite{ATLAS:2022vkf,CMS:2022dwd}. While experimental data indicate that the Higgs boson's couplings to gauge bosons are consistent with the SM predictions, potential modifications of these couplings may still arise.
In beyond-the-SM (BSM) scenarios, the interaction between the Higgs boson and vector bosons, generally denoted as \HVV, may be affected by contributions from anomalous couplings (ACs), leading to new tensor structures in the interaction terms. The combined Charge conjugation and Parity (\CP) properties of these AC contributions can generate distinct characteristics in the Higgs boson production mechanisms, potentially resulting in measurable modifications to the kinematic distributions of the final-state particles. Moreover, the simultaneous presence of a nonzero \CP-odd AC and the SM contribution in the \HVV interaction vertex could lead to \CP violation in Higgs boson interactions.

\begin{figure}[!htb]
    \centering
    \includegraphics[width=0.6\textwidth]{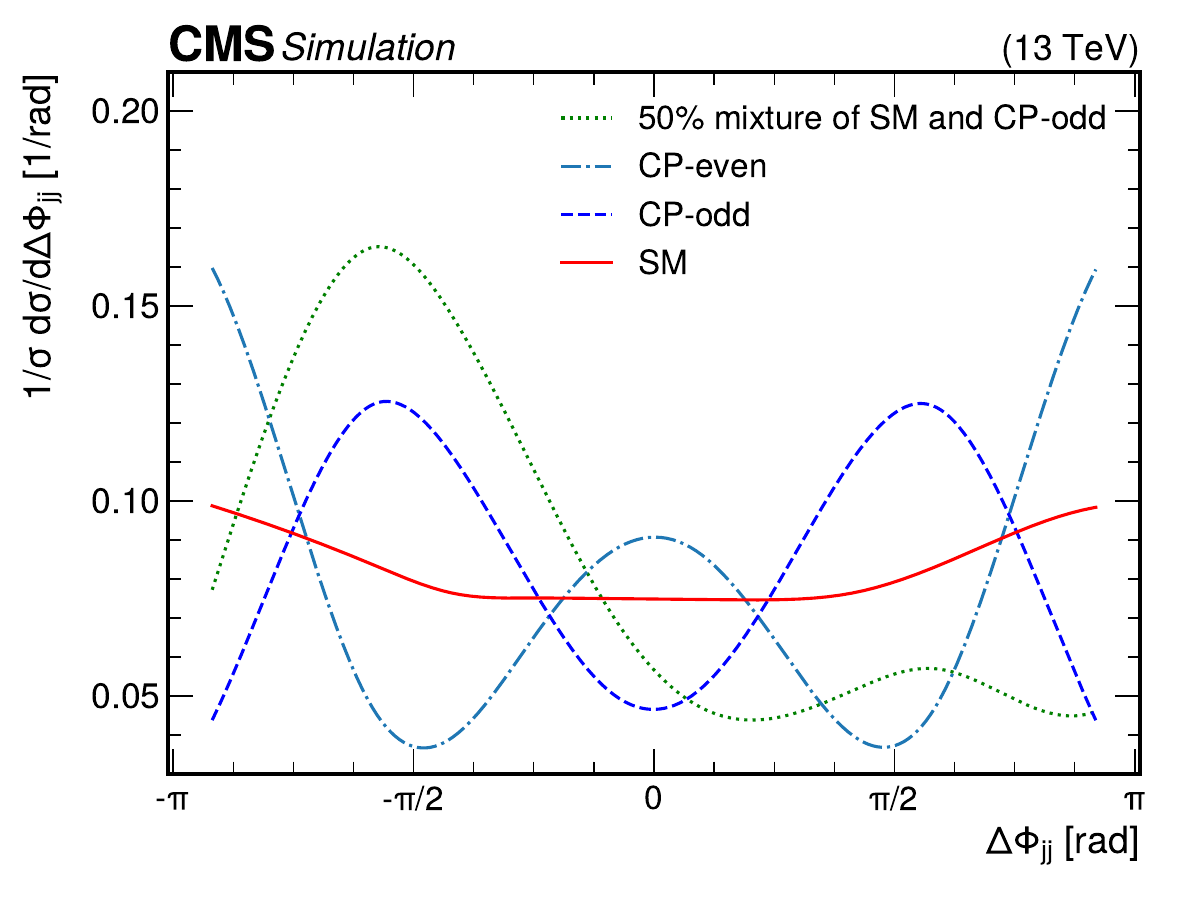}
    \caption{Normalized VBF differential cross section as a function of the signed azimuthal angle difference between the two jets, with the Higgs boson mass assumed to be 125\GeV. Different hypotheses are superimposed corresponding to a mixed \CP\ scenario, a pure \CP-even AC, a pure \CP-odd AC, and an SM coupling in the HVV vertex.}
    \label{fig:sigma_cp_odd_even}
\end{figure}

To probe such effects, this study focuses on a subset of Higgs boson events produced in association with two jets, as typically observed in the vector boson fusion (VBF) mechanism and in the gluon-gluon fusion (ggH) process with associated initial-state radiation (ISR). In this topology, one of the most sensitive observables for detecting BSM effects in the \HVV interaction is the signed azimuthal angle ($\Delta\Phi_\mathrm{{jj}}$) between the two leading jets produced in association with the Higgs boson. Following the definition provided in Ref.~\cite{Hankele:2006ma}, the \dphijj observable is given by the azimuthal angle of the forward jet ($j_\mathrm{f}$) minus that of the backward jet ($j_\mathrm{b}$): 

\begin{equation} \label{eq:def_deltaphi}
    \dphijj = \phi_\mathrm{j_f} - \phi_\mathrm{j_b} 
    \quad \text{with}\quad\eta_\mathrm{j_f} > \eta_\mathrm{j_b}, 
    \quad -\pi < \dphijj < \pi , 
\end{equation}

where $\eta_\mathrm{j_f}$ and $\eta_\mathrm{j_b}$ are the pseudorapidities of the two jets with the highest transverse momenta (\pt), $j_\mathrm{f}$ and $j_\mathrm{b}$, in the event. As illustrated in Fig.~\ref{fig:sigma_cp_odd_even}, in the presence of a purely \CP-odd AC, the Higgs boson production cross section via VBF is suppressed at 0 and $\pm\pi$ radians, while for a \CP-even AC, suppression occurs at $\pm\pi/2$ radians. In contrast, in the SM scenario, the VBF cross section remains nearly flat. When both \CP-odd and \CP-even components are present, the distribution becomes asymmetric around zero.

The definition in Eq.~\eqref{eq:def_deltaphi} also addresses a potential ambiguity in the sign of \dphijj in proton-proton collisions. In such collisions, the sign of a particle's azimuthal angle flips when the event is viewed from the opposite beam direction. By defining \dphijj as the difference between the azimuthal angles of $j_\mathrm{f}$ (with larger pseudorapidity) and $j_\mathrm{b}$ (with smaller pseudorapidity), this ambiguity is resolved. Exchanging the beam directions simultaneously swaps the forward and backward jets, leaving the sign of \dphijj unchanged.

This paper presents a differential measurement of the Higgs boson production cross section as a function of \dphijj, using proton-proton (pp) collision data collected with the CMS detector from 2016--2018 at $\sqrt{s}=13\TeV$, corresponding to an integrated luminosity of 138\fbinv.
The analysis targets the different-flavor dilepton ($\Pe\PGm$) final state from \HtoWW decays, and considers the Higgs boson production through VBF and ggH mechanisms.
The measurement is carried out within a detector fiducial volume to avoid extrapolation to the full phase space based on underlying theoretical assumptions. An unfolding procedure is employed to correct for the detector effects, ensuring that the result can be directly compared to any physics model. Similar differential measurements of the Higgs boson production via VBF and ggH mechanisms in the \HtoWW decay channel have been previously performed by the ATLAS Collaboration using data collected at a center-of-mass energy of 13\TeV~\cite{ATLAS:2023hyd,ATLAS:2023pwa,ATLAS:2025hki}.

Despite the model independence of the unfolded results, model assumptions inevitably enter the analysis through the signal extraction procedure. Typically, the analysis strategy relies on simulations of SM signal processes as benchmark models. This can introduce biases if the measurement is later reinterpreted in terms of BSM signal hypotheses, as the shape of the fit variable distribution generally depends on Higgs boson's properties, such as the \HVV couplings.
To mitigate this dependence, a methodology based on the machine-learning concept of domain adaptation is employed, aiming to reduce the impact of model assumptions by defining a fit variable that is agnostic to the signal hypothesis. This approach enhances the robustness of the measurement, allowing for reinterpretation within different physics scenarios. Finally, the model-agnostic cross section measurement results are reinterpreted within the SM effective field theory (SMEFT) framework, placing constraints on the Wilson coefficients in the Warsaw basis~\cite{Grzadkowski_2010,deFlorian:2227475}.   

This paper is organized as follows. The AC phenomenology used to describe alternative models is discussed in Section~\ref{sec:sigphenomenology}. Section~\ref{sec:cms} provides a brief overview of the CMS apparatus and the event reconstruction procedure. Data sets and Monte Carlo (MC) samples are detailed in Section~\ref{sec:dataset}. Event selection and categorization are discussed in Section~\ref{sec:event_selection}. Methods for estimating background contributions are presented in Section~\ref{sec:background}. The methodology of the model-agnostic classification is described in Section~\ref{sec:methodology}. Sources of systematic uncertainties are summarized in Section~\ref{sec:systematics}. Section~\ref{sec:sigextraction} outlines the signal extraction procedure, while the results are presented in Section~\ref{sec:results}. The reinterpretation of the results within the SMEFT framework is discussed in Section~\ref{sec:SMEFT}. Tabulated results are provided in the HEPData record for this analysis~\cite{hepdata}. Finally, a summary is given in Section~\ref{sec:conclusions}.

\section{Signal phenomenology}
\label{sec:sigphenomenology}

Anomalous coupling effects can modify the production cross sections and kinematics of both ggH and VBF processes, as well as alter the kinematic properties of the \HtoWW decay channel. The general scattering amplitude $\mathcal{A}$ describing the interaction between a spin-0 Higgs boson and two spin-1 gauge bosons, $\PV_{1}\PV_{2}=\WW$, $\ZZ$, $\PZ\PGg$, $\PGg\PGg$, or \glgl, has the form~\cite{hzz_ac_2021, PhysRevD.89.035007, Gritsan:2020pib}: 
\begin{equation} 
\label{eq:fullampl} 
        \mathcal{A} (\PH\PV_1\PV_2) \sim \left[a^{\PV\PV}_1 + \frac{\kappa^{\PV\PV}_1 q^2_{\PV_1}+\kappa^{\PV\PV}_2 q^2_{\PV_2}}{(\Lambda^{\PV\PV}_1)^2}\right]
	m^2_{\PV_1}\epsilon^*_{\PV_1}\epsilon^*_{\PV_2}+\frac{1}{v}a^{\PV\PV}_2 f^{*(1)}_{\upmu\upnu}f^{*(2),\upmu\upnu}+\frac{1}{v}a^{\PV\PV}_3 f^{*(1)}_{\upmu\upnu}\tilde{f}^{*(2),\upmu\upnu},
\end{equation}
where $f^{(i),\upmu\upnu} = \epsilon^\upmu_{\PV_i}q^\upnu_{\PV_i} - \epsilon^\upnu_{\PV_i}q^\upmu_{\PV_i}$ and $\tilde{f}^{(i)}_{\upmu\upnu} = \frac{1}{2} \epsilon_{\upmu \upnu \uprho \upsigma} f^{(i), \uprho \upsigma}$ are, respectively, the field strength tensor and the dual field strength tensor of a spin-1 gauge boson with four-momentum $q_{\PV_i}$, polarization vector $\epsilon_{\PV_i}$ and pole mass $m_{\PV_i}$. In the formula, $v$ stands for the Higgs field vacuum expectation value. The parameter $\Lambda\mathrm{^{VV}_1}$ is the energy scale of BSM physics and is a free parameter of the model. Finally, $a^{\PV\PV}_1$, $a^{\PV\PV}_2$, $a^{\PV\PV}_3$, $\kappa^{\PV\PV}_1$ and $\kappa^{\PV\PV}_2$ are real numbers that modify the corresponding amplitude terms.
In the SM, the only nonvanishing coupling at tree level is $a^{\PV\PV}_1 \ne 0$ for $\PV\PV = \PZ\PZ, \PW\PW$, while all other couplings satisfy $a^{\PV\PV}_{i \ne 1} = 0$. Tree-level couplings involving at least one massless gauge boson ($\PV\PV = \PZ\PGg$, $\PGg\PGg$, \glgl) do not contribute, as the corresponding pole mass vanishes. Additional \ZZ and \WW couplings are attributed to anomalous effects. Anomalous effects arising in the SM by loop-induced interactions $\PH\PZ\PGg$, $\PH\PGg\PGg$, $\PH\glgl$ are described by the \CP-even $a^{\PV\PV}_2$ terms, but they are parametrically suppressed by the electromagnetic coupling constant $\alpha$ and the strong coupling constant \alpS and are therefore not experimentally accessible. Similarly, in the SM, $a^{\PV\PV}_3$ couplings arise only at the three-loop level, making their contributions negligible. However, BSM effects could lead to significantly larger values for these coupling parameters.

A minimal set of independent ACs is considered to represent alternative models in this analysis. Following Refs.~\cite{Gritsan:2020pib,Hayrapetyan:2890530}, considerations of symmetry and gauge invariance lead to $\kappa_1^{\PZ\PZ} = \kappa_2^{\PZ\PZ}$, $\kappa_1^{\PGg\PGg} = \kappa_2^{\PGg\PGg} = \kappa_1^{\Pg\Pg} = \kappa_2^{\Pg\Pg} = \kappa_1^{\PZ\PGg} = 0$. Furthermore, the parameters $a_2^{\PGg\PGg}$, $a_3^{\PGg\PGg}$, $a_2^{\PZ\PGg}$, and $a_3^{\PZ\PGg}$ are set to zero, as they are constrained in the direct decays $\PH\to\PGg\PGg$ and $\PH\to\PZ\PGg$. Additionally, the relationship $a_{i} = a_{i}^{\PZ\PZ} = a_{i}^{\PW\PW}$ is adopted, justified by the lack of kinematic differences between $\PZ\PZ$ and $\PW\PW$ fusion processes, which makes these couplings impossible to disentangle. The condition $a_{1}^{\PZ\PZ} = a_{1}^{\PW\PW}$ holds under the requirement of custodial symmetry. As a result, the number of independent ACs is reduced to one \CP-odd coupling, $a_{3}^{\Pg\Pg}$, which contributes to the $\PH\glgl$ vertex and is accessible via the ggH mechanism, and four couplings accessible via the VBF mechanism: the \CP-even $a_2$, $a_{\Lambda_1} \equiv \kappa_1/(\Lambda_1)^2$, and $a_{\Lambda_1}^{\PZ\PGg} \equiv \kappa_2^{\PZ\PGg}/(\Lambda_1^{\PZ\PGg})^2$; and the \CP-odd $a_3$.

In practice, alternative signal event samples are characterized using the effective fractional cross section, defined as:

\begin{equation}
\label{eq:faitoai}
	f_{a_{i}}=\frac{\abs{a_{i}}^{2}\sigma_{i}}{\sum_{j}\abs{a_{j}}^{2}\sigma_{j}} \ \sign\left(\frac{a_{i}}{a_{1}}\right),
\end{equation}

where $\sigma_{i}$ is the cross section for the process corresponding to $a_{i}=1$ and $a_{j \ne i}=0$, and the summation in the denominator runs over all considered couplings, including $a_{1}$. For the anomalous $\PH\glgl$ vertex with no tree-level contribution, the effective fractional cross section is given by: 

\begin{equation}
\label{eq:faitogg}
f_{a_{3}}^{\Pg\Pg\PH} = \frac{\abs{a_3^{\Pg\Pg}}^2 \sigma_{3}^{\Pg\Pg}}{\abs{a_2^{\Pg\Pg}}^2 \sigma_{2}^{\Pg\Pg} + \abs{a_3^{\Pg\Pg}}^2 \sigma_{3}^{\Pg\Pg}} \ \sign\left(\frac{a_{3}^{\Pg\Pg}}{a_{2}^{\Pg\Pg}}\right),
\end{equation}

where $\sigma_{3}^{\Pg\Pg}$ and $\sigma_{2}^{\Pg\Pg}$ correspond to the cross sections for the cases $a_{3}^{\Pg\Pg}=1$, $a_{2}^{\Pg\Pg}=0$ and $a_{2}^{\Pg\Pg}=1$, $a_{3}^{\Pg\Pg}=0$, respectively.

While the AC formalism is used to describe alternative signal event samples, this analysis does not aim to directly constrain the ACs, as was done in Ref.~\cite{Hayrapetyan:2890530}. The AC signal event samples are used to define a model-agnostic fit variable and the measured differential cross sections are reinterpreted in terms of EFT couplings, as further discussed in Section~\ref{sec:SMEFT}.

\section{The CMS detector and event reconstruction}
\label{sec:cms}

The CMS apparatus~\cite{CMS:2008xjf,Hayrapetyan_2024} is a multipurpose, nearly hermetic detector, designed to identify electrons, muons, photons, and hadrons~\cite{CMS:2015xaf,CMS:2018rym,CMS:2015myp,CMS:2014pgm}. A global reconstruction ``particle-flow" (PF) algorithm~\cite{CMS:2017yfk} combines the information provided by the all-silicon inner tracker and by the crystal electromagnetic (ECAL) and brass-scintillator hadron calorimeters, operating inside a 3.8\unit{T} superconducting solenoid, with data from gas-ionization muon detectors interleaved with the solenoid flux return yoke, to build \PGt leptons, jets, missing transverse momentum (\ptmiss), and other physics objects~\cite{CMS:2018jrd,CMS:2016lmd,CMS:2019ctu}.

Events of interest are selected using a two-tiered trigger system. The first level, composed of custom hardware processors, uses information from the calorimeters and muon detectors to select events at a rate of around 100\unit{kHz} within a fixed latency of 4\mus~\cite{CMS:2020cmk}. The second level, known as the high-level trigger (HLT), consists of a farm of processors running a version of the full event reconstruction software optimized for fast processing, and reduces the event rate to a few \unit{kHz} before data storage~\cite{CMS:2016ngn,CMS:2024aqx}.

The primary vertex is taken to be the vertex corresponding to the hardest scattering in the event, evaluated using tracking information alone, as described in Section 9.4.1 of Ref.~\cite{CMS-TDR-15-02}. The primary vertex must be within 24\unit{cm} of the nominal interaction point along the beam axis and within 2\unit{cm} in the transverse plane.

The PF algorithm aims to reconstruct and identify each individual particle in an event (PF candidate), with an optimized combination of information from the various elements of the CMS detector. Electrons are identified and their momenta are measured in the interval $\abs{\eta} < 2.5$ by combining the energy measurement in the ECAL with the momentum measurement in the tracker. The momentum resolution for electrons with $\pt \approx 45\GeV$ from $\PZ \to \Pe \Pe$ decays ranges from 1.6--5.0\%. It is generally better in the barrel region than in the endcaps, and also depends on the bremsstrahlung energy emitted by the electron as it traverses the material in front of the ECAL~\cite{CMS:2020uim,CMS-DP-2020-021}. Muons are measured in the range $\abs{\eta} < 2.4$, with detection planes made using three technologies: drift tubes, cathode strip chambers, and resistive-plate chambers. The efficiency for muon reconstruction and identification exceeds 96\%. Matching muons to tracks measured in the silicon tracker results in a relative \pt resolution of 1\% in the barrel and 3\% in the endcaps for muons with \pt up to 100\GeV, and of better than 7\% in the barrel for muons with \pt up to 1\TeV~\cite{CMS:2018rym}.  

Jets are reconstructed offline from the PF candidates using the anti-\kt algorithm~\cite{Cacciari:2008gp, Cacciari:2011ma} with a distance parameter of 0.4. 
Jet momentum is determined as the vectorial sum of all particle momenta in the jet, and is found from simulation to be, on average, within 5--10\% of the true momentum over the whole \pt spectrum and detector acceptance. Additional proton-proton interactions within the same or nearby bunch crossings (pileup) can contribute additional tracks and calorimetric energy deposits, increasing the apparent jet momentum. To mitigate this effect, tracks identified to be originating from pileup vertices are discarded and an offset correction is applied to correct for remaining contributions~\cite{CMS:2020ebo}. Jet energy corrections are derived from simulation studies so that the average measured energy of jets becomes identical to that of particle level jets. In situ measurements of the momentum balance in dijet, photon+jet, Z+jet, and multijet events are used to determine any residual differences between the jet energy scale in data and in simulation, and appropriate corrections are made~\cite{CMS:2016lmd}. Additional selection criteria are applied to each jet to remove jets potentially dominated by instrumental effects or reconstruction failures~\cite{CMS:2020ebo}. The jet energy resolution amounts typically to 15--20\% at 30\GeV, 10\% at 100\GeV, and 5\% at 1\TeV~\cite{CMS:2016lmd}.

Jets likely arising from the hadronization of b quarks are identified using b tagging algorithms~\cite{Sirunyan:2017ezt}.
For each jet in the event a score is calculated through a multivariate combination of different jet properties. Jets are considered b-tagged if their associated score exceeds a threshold, corresponding to a certain tagging efficiency as measured in simulated \ttbar events. In the analysis, the \textsc{DeepJet} algorithm~\cite{Bols_2020} is used with its loose working point, which corresponds to a probability of approximately 10\% for mistagging a jet originating from a light quark or a gluon as the b jet.

The vector \ptvecmiss is computed as the negative vector \pt sum of all the PF candidates in an event, and its magnitude is denoted as \ptmiss~\cite{CMS:2019ctu}. The \ptvecmiss is modified to account for corrections to the energy scale of the reconstructed jets in the event.

The pileup-per-particle-identification algorithm~\cite{Bertolini:2014bba} is applied to reduce the pileup dependence of the \ptvecmiss observable. The \ptvecmiss is computed from the PF candidates weighted by their probability to originate from the primary interaction vertex~\cite{CMS:2019ctu}.

\section{Data sets and simulations}
\label{sec:dataset}

The analysis is performed using data sets recorded with the CMS detector in 2016, 2017, and 2018 and corresponding to integrated luminosities of 36.3, 41.5, and 59.8\fbinv, respectively~\cite{CMS-LUM-17-003,CMS-PAS-LUM-17-004,CMS-PAS-LUM-18-002}.

The events are selected through HLT algorithms that require the presence of either a single lepton or both an electron and a muon, satisfying isolation and identification requirements. In the 2016 data set, the single-electron trigger \pt is set at 25\GeV for electrons with ${\abs{\eta} < 2.1}$, and at 27\GeV for ${2.1 < |\eta| < 2.5}$. The \pt threshold for the single-muon trigger is 24\GeV for ${\abs{\eta} < 2.4}$. For the dilepton \emu trigger, the \pt thresholds are 23\GeV for the leading and 12\GeV for the subleading lepton. During the first part of data taking in 2016, a lower \pt threshold of 8\GeV was applied to the subleading muon. In the 2017 data set, the \pt thresholds for the single-electron and single-muon triggers are increased to 35 and 27\GeV, respectively. In the 2018 data set, these thresholds are set to 32 and 24\GeV, respectively. The \pt thresholds for the dilepton triggers for both the 2017 and 2018 remain consistent with those used in the latter part of the 2016 data set.

Signal modeling and background estimation rely on MC simulated events. In order to account for changes in the CMS detector and pileup conditions, four independent sets of simulated events are produced, corresponding to the 2017 and the 2018 data sets, as well as the first and second parts of the 2016 data set.

Various processes are generated using different event generators, but the same set of parton distribution functions (PDFs), underlying event (UE) tune and parton shower (PS) configuration is used for all events. The NNPDF31 PDF set~\cite{Ball_2017} at next-to-next-to-leading order accuracy in perturbative quantum chromodynamics (pQCD) is used in all samples. All matrix element generators are interfaced to \PYTHIA v8.240~\cite{Sjostrand:2014zea} for the simulation of the parton shower, the hadronization, and the UE interactions. The description of the UE and multiple interactions is based on the CP5 tune~\cite{Sirunyan:2019dfx}.

The SM Higgs boson samples are simulated with \POWHEG v2~\cite{Nason:2004rx,Frixione:2007vw,Alioli:2010xd} at next-to-leading order (NLO) precision in pQCD. In particular, the ggH signal sample is generated using the \textsc{powheg}+\textsc{MiNLO} approach~\cite{Kardos_2014} to simulate gluon fusion Higgs boson production with two additional jets at NLO precision. For the VBF sample, the dipole recoil approach~\cite{sjostrand_2018} is used to model ISR, as it provides a more accurate description of additional QCD emissions in VBF processes compared to the default recoil scheme in the \PYTHIA parton shower~\cite{Jager:2020hkz}. The \textsc{JHUGen} generator~\cite{PhysRevD.86.095031} is used to simulate the decay of the Higgs boson into two \PW bosons and subsequently into leptons. The decay into two \PGt leptons is simulated with \PYTHIA. Other Higgs boson decay channels are not considered in this analysis, as they have a negligible contribution.

One of the most important backgrounds is the production of $\PWp\PWm$ pairs without involving a Higgs boson (nonresonant \WW production), with the \PW bosons decaying leptonically and two jets coming from ISR. This background arises predominantly from quark-initiated production, with a minor contribution from gluons. The EW vector boson scattering (VBS) process ($\PWp\PWm\mathrm{jj}$) is also considered, as it has a nonnegligible contribution within the 2-jet phase space of this analysis. The gluon-induced \WW background is simulated using the \textsc{mcfm} v7.0~\cite{Campbell:1999ah,Campbell:2011bn,Campbell:2015qma} event generator at leading order (LO) precision and the inclusive cross section is reweighted to match NLO precision~\cite{Caola:2016trd}. Quark-initiated \WW production is generated with \POWHEG at NLO and reweighted to match Next-to-NLO precision with next-to-next-to-leading logarithmic (NNLL) resummation. Finally, the non-resonant EW VBS production of \WW pairs with two additional jets is simulated at LO precision with \MGvATNLO v2.4.2~\cite{Alwall2014} using the MLM jet matching and merging scheme~\cite{Alwall:2007fs}.
The dominant contribution of Drell--Yan (DY) production populates the region where ${m_{\PGt\PGt}>50\GeV}$. In this phase space, dedicated samples of \DYtott are simulated using the \MGvATNLO generator at NLO precision with up to two additional partons in the matrix element (ME) calculations, using the FxFx jet matching and merging scheme~\cite{Frederix:2012ps}. In the region ${10 < m_{\PGt\PGt} < 50\GeV}$, a \DYtott sample is simulated at LO precision. The triboson production events are simulated using \MGvATNLO at NLO precision. The \wg events are simulated with \MGvATNLO at NLO precision with up to one additional parton in the ME calculations and FxFx jet merging scheme. Additional diboson processes (such as $\PW\PZ$ and $\PZ\PZ$ production) resulting in leptonic final states are generated using \POWHEG at NLO precision, while lepton+jets final states are simulated with \MGvATNLO at NLO precision with the FxFx jet merging scheme. Top quark events are produced using \POWHEG at NLO, with the exception of the single top quark production in the \textit{s} channel that is generated using \MGvATNLO at the same order of precision. 

Samples for the VBF and ggH production processes with AC scenarios are also simulated. For the VBF samples, the effective fractional cross sections, as defined in Eq.~\eqref{eq:faitoai}, are set to 0.5 and 1 for all the ACs, except for the $a_{\Lambda_1}^{\Zg}$ coupling which is simulated only with fractional cross section equal to 0.5. The Higgs boson decay into W bosons, followed by their subsequent decay into leptons, is simulated with the coupling values at the decay vertex matching those used at the production vertex. The ggH samples are simulated with ACs affecting only the production vertex, assuming the effective fractional cross section defined in Eq.~\eqref{eq:faitogg} equal to 0.5 and 1. The \textsc{JHUGen} MC generator is used to simulate the VBF processes at LO and \textsc{powheg}+\textsc{JHUGen} is used to simulate ggH events at NLO.

The detector response is simulated using a detailed description of the CMS detector, based on the \GEANTfour package~\cite{Agostinelli:2002hh}. The distribution of the number of pileup interactions in the simulation is reweighted to match the one observed in data. 

The efficiency of the trigger system is evaluated in data on a per-lepton basis by selecting di-lepton events compatible with originating from a \PZ boson.
The per-lepton efficiencies are then combined probabilistically to obtain the overall efficiencies of the trigger selections used in the analysis.
The procedure is repeated in data and simulation to extract data-to-simulation correction scale factors. Trigger selections are applied also to simulated events, and residual differences between efficiencies in data and simulation are corrected by applying the aforementioned scale factors.

\section{Event selection}
\label{sec:event_selection}

The event selection targets the Higgs boson production in the \HtoWW channel in association with two jets. Events are required to contain the two highest-\pt leptons (leading and subleading candidates) with opposite charge and different flavor ($\Pe\PGm$), in order to suppress DY background. The transverse momentum of the leading lepton \ptlone is required to be greater than 25\GeV, whereas the threshold on \pt of the subleading lepton is set to 13\GeV. In the 2016 data sets, this threshold is lowered to 10\GeV for muons, due to different HLT requirements with respect to 2017 and 2018. The \pt of the third lepton, if present, is required to be below 10\GeV in order to suppress minor backgrounds, such as $\PW\PZ$ and triboson production. The invariant mass (\mll)  and transverse momentum (\ptll) of the dilepton system are required to be greater than 12 and 20\GeV, respectively. Due to the presence of neutrinos in the final state, a \ptmiss threshold of 20\GeV is set.
The final state is then selected by requiring, in addition to the $\Pe\PGm$ pair, two jets with ${\pt > 30\GeV}$ each and an invariant mass (\mjj) higher than 120\GeV.

The kinematic phase space is then divided into a \HtoWW signal region (SR) and two control regions (CRs), namely the top quark CR and the DY CR. The two CRs are designed to be populated by $\ttbar$+$\PQt\PW$ and \DYtott events, respectively, and are used to check the agreement between data and simulation and to constrain the normalization of the corresponding background. The SR is defined by adding further requirements on the transverse mass, defined as

\begin{equation}
    \mth = \sqrt{{2\ptll\ptmiss[1-\cos{\Delta\Phi(\vec{p}_{\mathrm{T}}^{\ \Pell\Pell},\ptvecmiss)}]}},
    \label{eq:mth}
\end{equation}

and on the transverse mass built with \ptmiss and the subleading lepton momentum:

\begin{equation}
    \mtltwo = \sqrt{{2\ptltwo\ptmiss[1-\cos{\Delta\Phi(\vec{p}_{\mathrm{T}}^{\ \Pell_2},\ptvecmiss)}]}}.
    \label{eq:mtl2}
\end{equation}

Events with $\mth > 60\GeV$ and $\mtltwo > 30\GeV$ are selected. Finally, no b-tagged jet must be present in the SR in order to reduce events from \ttbar and $\PQt\PW$ production. The top quark CR is made orthogonal to the SR by requiring at least one b-tagged jet with $\pt > 20\GeV$. The \mtltwo requirement is common to the SR, whereas the \mll threshold is increased to 50\GeV. In the DY CR the b jet veto requirement is restored, but the \mth requirement is inverted with respect to the SR and the invariant mass of the dilepton system \mll must be between 40 and 80\GeV, without any constraints on \mtltwo. The purities of the top quark and DY CRs are approximately 97 and 72\%, respectively. The complete list of selection criteria defining the analysis phase space is provided in Table~\ref{tab:analysis_regions}. Events satisfying these criteria are categorized into reconstructed-level bins of the observable \dphijj, following the binning scheme outlined in Table~\ref{tab:Dphijj_bins}.

\begin{table}[!htb]
    \centering
    \topcaption{{Definition of the analysis phase spaces.}}
    \begin{tabular}{l l}
    Region & Requirements \\ \hline
    \multirow{9}{*}{Global selection}         &   oppositely charged $\Pe\PGm$ final state  \\                                                         
                                              &   $\ptlone > 25\GeV$  \\                                                     
                                              &   $\ptltwo > 13\GeV$ (10\GeV for 2016 data) \\                                             
                                              &   $\ptlthree < 10\GeV$  \\    
                                              &   $\mll > 12\GeV$ \\ 
                                              &   $\ptll > 30\GeV$ \\
                                              &   $\ptmiss > 20\GeV$ \\       
                                              &   at least two jets with $\pt>30\GeV$ and $\abs{\eta} < 4.7$ \\                                     
                                              &   $\mjj>120\GeV$  \\ [\cmsTabSkip] 
    \multirow{3}{*}{SR}                       &   $\mth > 60\GeV$ \\                                                                             
                                              &   $\mtltwo > 30\GeV$ \\                                                                         
                                              &   no b-tagged jets with $\pt > 20\GeV$  \\   [\cmsTabSkip]                                                        
    \multirow{3}{*}{Top quark CR}                   &   $\mll > 50\GeV$  \\                                                                       
                                              &   $\mtltwo > 30\GeV$  \\                                                                      
                                              &   at least one b-tagged jet  \\   [\cmsTabSkip]                                                                 
    
    \multirow{3}{*}{DY CR}                    &   $40 < \mll < 80\GeV$  \\                                                            
                                              &   $\mth < 60\GeV$  \\                                                                        
                                              &   no b-tagged jets with $\pt > 20\GeV$  \\                                                     
    \end{tabular}
    \label{tab:analysis_regions}
\end{table}

\begin{table}[!htb]
    \centering
    \caption{{Definition of the \dphijj bins.}}
    \begin{tabular}{llll}
    
    Bin 0 & Bin 1 & Bin 2 & Bin 3\\ \hline
    \multirow{2}{*}{$-\pi<\dphijj \leq - \frac{\pi}{2}$} & \multirow{2}{*}{$- \frac{\pi}{2}<\dphijj \leq0$} & \multirow{2}{*}{$0<\dphijj\leq\frac{\pi}{2}$} & \multirow{2}{*}{$\frac{\pi}{2}<\dphijj\leq\pi$} \\
    & & &\\
    \end{tabular}
\label{tab:Dphijj_bins}
\end{table}

The Higgs boson differential cross section is measured in a generator-level fiducial phase space designed to minimize the dependence of the measurement on the underlying model of the Higgs boson production. The set of requirements to define the fiducial phase space has been determined by considering two correlated quantities: the reconstruction efficiency for signal events originating within the fiducial phase space and the purity of the selected sample, \ie, the fraction of reconstructed signal events that originate from within the fiducial phase space over the total number of reconstructed signal events. The full set of requirements used to define the fiducial phase space is reported in Table~\ref{tab:fiducial_volume}. For the VBF signal, the selection yields a purity of approximately 70\% and a reconstruction efficiency of about 30\%, while for the ggH signal the corresponding values are approximately 55 and 20\%. Leptons are defined as ``dressed'', meaning that the momenta of radiated photons within a cone of $\Delta R = \sqrt{\smash[b]{(\Delta\eta)^2+(\Delta\phi)^2}} < 0.1$ around the lepton direction are summed to the lepton momentum. This approach replicates the definition of reconstructed electrons, where radiated photons contribute to the energy cluster in the ECAL, and their energy is thus included in the electron measurement. Furthermore, muons and electrons originating from leptonic \PGt decays are excluded. Generator-level jets are defined using the same algorithm as at the reconstructed level, as described in Section~\ref{sec:cms}. Finally, at generator level, the same bin boundaries as those at reconstructed level, listed in Table~\ref{tab:Dphijj_bins}, are applied to the \dphijj distributions.

\begin{table}[!htb]
    \centering
    \topcaption{Definition of the fiducial phase space. Observables are defined using generator-level quantities.}
    \begin{tabular}{c}
    Fiducial requirements\\ \hline
     oppositely charged $\Pe\PGm$ (not from $\PGt$ lepton decay) final state\\
     at least two jets with $\pt>30\GeV$ and $\abs{\eta} < 4.7$ \\
     $\ptlone > 25\GeV$, $\ptltwo > 13\GeV$\\ 
     $\abs{\etalone} < 2.5$, $\abs{\etaltwo} < 2.5$\\
     $\ptll > 30\GeV$, $\mll > 12\GeV$ \\
     $\mjj > 120\GeV$, $\mth > 60\GeV$\\
    \end{tabular}
\label{tab:fiducial_volume}
\end{table}

It is important to clarify that signal events outside the fiducial phase space (out-of-fiducial events) are categorized into two types, based on whether they have a well-defined \dphijj observable or not. In the first case, two jets with $\pt > 30\GeV$ and $\abs{\eta}<4.7$ are present at generator level, but the other fiducial requirements are not satisfied. In the second case, events do not contain two jets with $\pt > 30\GeV$ and $\abs{\eta} < 4.7$ at generator level, making the computation of the \dphijj observable impossible, and the event is labeled as out-of-fiducial with no \dphijj.

With the selection outlined in Table~\ref{tab:fiducial_volume}, the out-of-fiducial signal events with a well-defined \dphijj that enter the analysis selection constitute 10 and 13\% of the total number of reconstructed events for the VBF and ggH signal, respectively. These fractions are significant relative to the total number of reconstructed events. Therefore, out-of-fiducial events with a well-defined \dphijj value are measured alongside the fiducial ones. Specifically, the out-of-fiducial signal component within a given \dphijj bin is measured together with the corresponding fiducial component in the same bin. On the contrary, the out-of-fiducial signal events without a well-defined \dphijj that enter the analysis selection constitute 20 and 32\% of the total number of reconstructed events for the VBF and ggH signal, respectively. These kinds of events are treated as an overall background in this analysis.

\section{Background estimation}
\label{sec:background}

The backgrounds from top quark production and DY processes are estimated using a combination of MC simulations and dedicated CRs. Their normalizations are treated as free parameters in the fit, with an independent parameter assigned to each \dphijj bin of the corresponding CR.

In contrast, the normalization of the nonresonant \WW contribution is not determined in a dedicated CR. Instead, theoretical uncertainties associated with the modeling of this process are incorporated into the measurement. 

Background contributions also arise from events with a single \PW boson produced with jets and from lepton+jets \ttbar decays, where a \PW boson decays into a \qqbar pair. These processes become relevant when a jet constituent is misidentified as a lepton, mimicking the second prompt lepton in the final state. Such nonprompt leptons are typically accompanied by nearby particles. Since prompt leptons from hard scattering processes are expected to be isolated, this background is reduced by applying specific isolation criteria to signal electrons and muons. Nonprompt background is directly estimated from data as described in detail in Ref.~\cite{Sirunyan:2018egh}. This method involves measuring the probability that a nonprompt lepton passing a loose selection also satisfies a tight selection (misidentification rate) and the corresponding probability for a prompt lepton (prompt rate). The misidentification rate is determined from a dijet-enriched data sample, while the prompt rate is obtained using a ``tag-and-probe'' method~\cite{Khachatryan2011} in a DY-enriched sample. These measured rates are then applied as event weights to account for the probability of selecting a non-prompt lepton alongside a lepton candidate that meets the standard selection criteria.

In addition to the main backgrounds, minor contributions include Higgs boson production with \HtoTT, associated Higgs production with vector bosons or top quarks in the \HtoWW channel, \PW/\PZ boson production with an on- or off-shell photon, and multiboson production. These backgrounds are modeled using MC simulations.

\section{Model-agnostic classification} \label{sec:methodology}

To maximize the sensitivity of the measurement, a classification algorithm is employed to separate signal processes from background contributions. To maintain the model independence of the measurement, the classification algorithm must accurately distinguish signal-like events from the background processes, while reducing its reliance on the specific physics models used for signal during training. For this purpose, an adversarial deep neural network (ADNN)~\cite{Camaiani_2022} is employed. This machine learning approach introduces an adversarial component to suppress the network ability to learn specific features of a given signal hypothesis. In this analysis, the Higgs boson signal is extracted from a fit to the ADNN score distribution. Consequently, the ADNN is designed to be agnostic to the couplings of the Higgs boson to vector bosons, ensuring that the shape of the ADNN score remains stable across different model assumptions for the \HVV coupling.

The ADNN consists of two neural networks, a classifier (C) and an adversary (A), trained in a competitive way to perform different tasks. The classifier is designed to distinguish between signal and background events and is trained on a data set that includes background events and signal events generated under different signal hypotheses. The adversary, on the other hand, aims to infer the underlying physics model of a signal event from the second-to-last layer of the classifier. The two networks are trained using a competitive learning approach, where the classifier is optimized to distinguish signal from background events while being simultaneously penalized if the data representation is sensitive to the physics hypothesis of the signal events. The penalty is imposed by the adversary, which compels the classifier to base its main classification task on features that are independent of the assumed signal model. Such training approach is implemented using a two-step procedure on a labeled data sample. In each epoch, the classifier is first trained with a combined loss function $\mathcal{L} = \mathcal{L}_\mathrm{C}-\alpha\ \mathcal{L}_\mathrm{A}$, where $\mathcal{L}_\mathrm{C}$ represent the classifier cross-entropy loss, and $\mathcal{L}_\mathrm{A}$ corresponds to the adversary loss, scaled by a tunable hyperparameter $\alpha$ that controls the relative importance of the two networks. During this step, the adversary weights remain fixed, and the minimization is performed solely with respect to the classifier parameters. In the subsequent step within the same epoch, the adversary is trained with the $\mathcal{L}_\mathrm{A}$ loss function. More details on the specific setup used in this work can be found in Ref.~\cite{Camaiani_2022}.

In order to discriminate both the VBF and ggH signals with a model-independent approach, two distinct ADNNs have been implemented. The first network, referred to as the VBF-ADNN, is designed to target the VBF production process. It has been trained using samples corresponding to varying $a_{\Lambda_1}$, $a_2$, and $a_3$ couplings, both in pure ($f_{a_i} = 1$) and mixed ($f_{a_i} = 0.5$) scenarios, as well as the $a_{\Lambda_1}^{\Zg}$ coupling in a mixed scenario ($f_{a_{\Lambda_1}^{\Zg}} = 0.5$). For the ggH signal, a separate ADNN has been trained, including alternative samples corresponding to the pure BSM scenario ($f_{a_{3}}^{\Pg\Pg} = 1$), and the mixed SM-BSM hypothesis ($f_{a_{3}}^{\Pg\Pg} = 0.5$). This network is referred to as the GGH-ADNN in the following. 

The ADNNs have been implemented using the \textsc{Keras}~\cite{chollet2015keras} and \textsc{TensorFlow}~\cite{abadi2016tensorflow} libraries, employing the Adam gradient-descent optimizer~\cite{kingma2017adammethodstochasticoptimization} during the learning process of the algorithm. The training is performed in the inclusive SR detailed in Table~\ref{tab:analysis_regions}, using MC simulations of the signal process generated under both the SM and the AC assumptions, alongside the main background contributions. The SM ggH process is treated as background in the training of the VBF-ADNN, and vice versa, the SM VBF process is treated as background in the training of the GGH-ADNN. Each background contribution is incorporated into the data set according to the proportions dictated by the SM, ensuring that the total number of events in the background class matches that of the signal class. Additionally, signal events are selected to maintain an equal number for each of the considered physics hypotheses. For the GGH-ADNN, the composition of the background class has been adjusted to enhance the discrimination power between the ggH and VBF processes. Maintaining the proportions of the SM cross sections, only a few events of the VBF process would end up to populate the background class, making their discrimination from the ggH signal unfeasible. Therefore, the background class of the GGH-ADNN consists of half VBF events and half events from other backgrounds, selected according to the SM proportions. Finally, the training sample has been randomly divided in two parts, one designated to the actual training and the other reserved for the validation of the algorithm. The validation set constitutes approximately 20\% of the entire training sample.

The input features of the ADNN are listed in Table~\ref{tab:input_features}, and have been chosen among the kinematic observables that can highlight the Higgs boson signal characteristics with respect to background processes within the 2-jets phase space. Input features include a quark-gluon likelihood ($qgl$) discriminant~\cite{CMS-DP-2016-070}, which is evaluated separately for the leading ($qgl_\mathrm{j_1}$) and subleading jets ($qgl_\mathrm{j_2}$) in the event. Kinematic discriminants based on ME calculations are also included, to enhance the separation of signal and background processes. These discriminants are constructed from the probability density $\mathcal{P}(\vec{\Omega})$ of an event originating from a given process, utilizing the likelihood ratio approach to optimize discrimination between hypotheses. Four ME-based discriminants, computed using the ME likelihood approach~\cite{Gritsan:2020pib,PhysRevD.86.095031,PhysRevD.81.075022,PhysRevD.89.035007,PhysRevD.94.055023,Martini_2021,Davis_2022} and \textsc{MoMEMta}~\cite{Brochet:2018pqf} frameworks, are incorporated into the ADNN input set: \DVbfGgh, \DVbfVh, \DGghVh, and \DVbfDy. While the first three focus on distinguishing between VBF, ggH, and VH processes, \DVbfDy targets the separation of VBF from DY events, leveraging the distinct kinematic properties of jets originating from ISR and from VBF production. In addition to the kinematics observables listed in Table~\ref{tab:input_features}, boolean indicators are used to specify the data-taking year (2016, 2017, or 2018), ensuring that possible year-dependent effects are accounted for in the training.

\begin{table}[!htb]
    \centering
    \topcaption{{Set of ADNN input features.}}
    \renewcommand{\arraystretch}{1.3}
    \begin{tabular}{ll}
    Variable                                                                                    & Description\\ 
    \hline
    \ptjone, \ptjtwo                                                                            & Magnitudes of the transverse momentum of the leading jets\\
     \etajone, \etajtwo                                                                         & Pseudorapidities of the two leading jets\\
     \mjj                                                                                       & Invariant mass of the dijet system\\
    \detajj                                                                                     & Pseudorapidity gap between the leading jets\\
     \phijone, \phijtwo                                                                         & Azimuthal angles of the two leading jets\\
     \ptlone, \ptltwo                                                                           & Magnitudes of the transverse momentum of the leading leptons\\
     \ptll                                                                                      & Magnitude of the transverse momentum of the dilepton system\\
     \etalone, \etaltwo                                                                         & Pseudorapidities of the two leading leptons\\
     \philone, \philtwo                                                                         & Azimuthal angles of the two leading leptons\\
     \mll                                                                                       & Invariant mass of the dilepton system\\
     \dphill, \drll                                                                             & Angular and radial separations between the leading leptons\\
     \mlj                                                                                       & Invariant mass of the lepton-jet system ($\Pell=\{\Pell_1, \ \Pell_2\}$, $j=\{j_1, \ j_2\}$)\\
     $C_{\text{tot}}$                                                                         & Centrality, defined as $C_{\text{tot}}=\log\Bigl(\sum\limits_{\Pell_1, \Pell_2}\abs{(2\eta_{\Pell}-\sum\limits_{\mathrm{j_1, j_2}}\eta_\mathrm{j})}/\abs{\Delta\eta_{\mathrm{jj}}}\Bigr)$ \\
    \ptmiss                                                                                        & Missing transverse momentum\\
     $qgl_\mathrm{j_1}$, $qgl_\mathrm{j_2}$                                                     & Quark-gluon likelihood discriminant for the two leading jets\\
     \mth                                                                                       & Transverse mass built with \ptmiss and \ptll\\
     \mtltwo                                                                                    & Transverse mass built with \ptmiss and \ptltwo \\
     $\Delta\phi(\vec{p}^{\ \Pell\Pell}_\mathrm{T}, \ptvecmiss)$        & Azimuthal opening angle between $\vec{p}^{\ \Pell\Pell}_\mathrm{T}$ and \ptvecmiss \\
     \HT & Hadronic activity, defined as the scalar sum of $\pt^j$ over all jets in the event \\
     \DVbfGgh                                                                                   & ME-based discriminant separating the VBF and ggH productions\\ 
     \DVbfVh                                                                                    & ME-based discriminant separating the VBF and VH productions\\ 
     \DGghVh                                                                                    & ME-based discriminant separating the ggH and VH productions \\ 
     \DVbfDy                                                                                    & ME-based discriminant separating the VBF and DY productions
    \end{tabular}
 \label{tab:input_features}
\end{table}

The training and validation sets of the VBF- and GGH-ADNN are preprocessed by standardizing the input features, subtracting the mean and dividing by the standard deviation (s.d.). To prevent overlap between training and deployment data sets, the data is further subdivided based on the parity of the sequential number associated to each MC generated event used for the training. One network is trained on odd-numbered events, while the other is trained on even-numbered events. At runtime, each network is applied to the opposite data set, maintaining separation between training and deployment sets.

The VBF-ADNN and GGH-ADNN have been optimized separately for both signal vs. background discrimination and the decorrelation of the classifier score from the signal hypothesis. This optimization was achieved by maximizing the binary accuracy of the classifier while simultaneously minimizing the two-sample Kolmogorov--Smirnov test statistic~\cite{Hodges1958TheSP} between the distributions of the classifier output evaluated on signal events simulated under different physics assumptions. The optimization was carried out using the \textsc{Optuna}~\cite{optuna_2019} package to determine the optimal values for the $\alpha$ parameter, the number of nodes in the hidden layers, and the number of hidden layers in both the classifier and the adversary.

\begin{figure}[!htb]
    \centering
    {\includegraphics[width=0.453\textwidth]{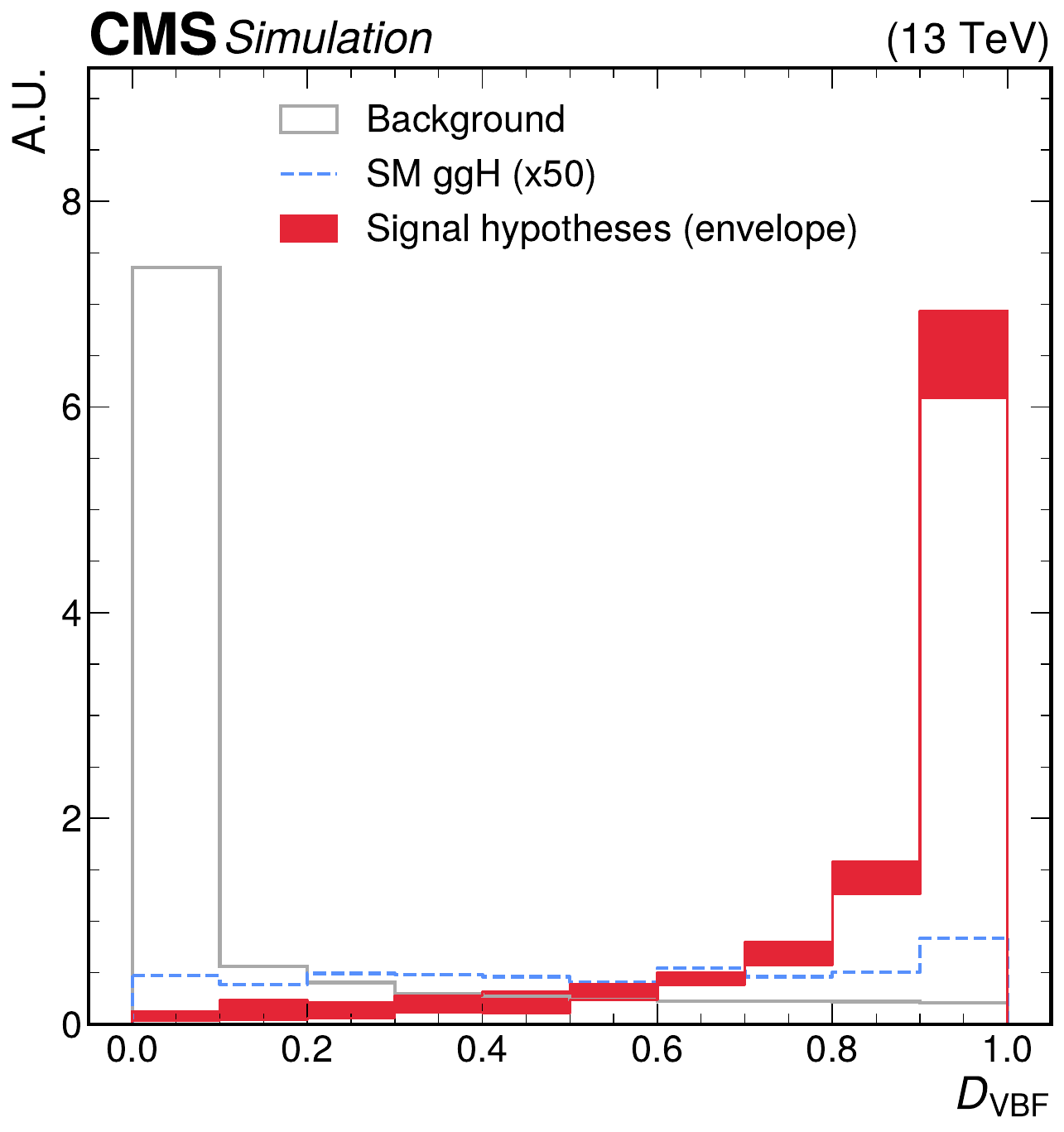}} \quad
    {\includegraphics[width=0.45\textwidth]{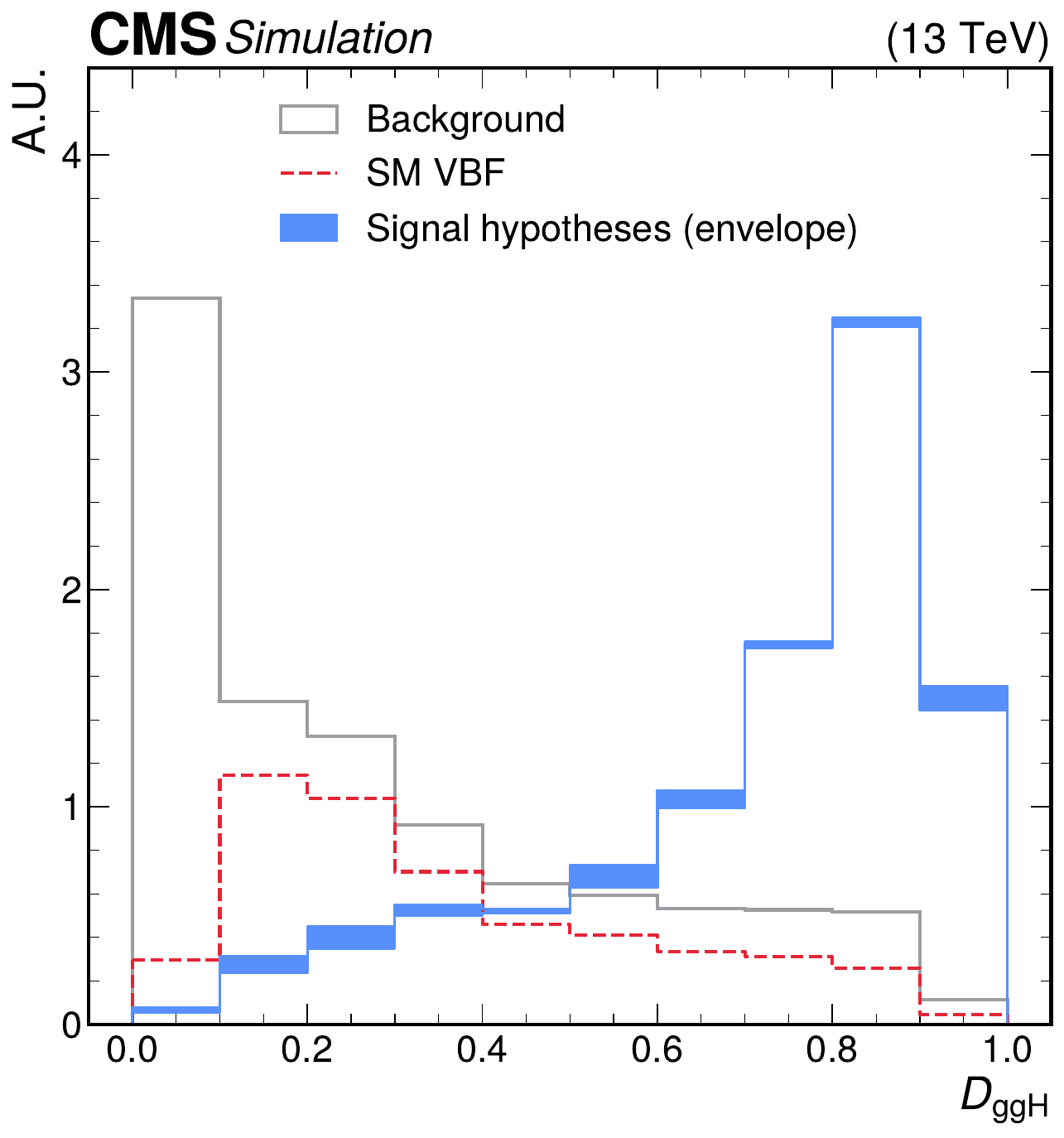}}
    \caption{Normalized distributions of \Dvbf (left) and \Dggf (right), evaluated on signal and background events using the ADNNs trained on even-numbered MC events. The signal predictions are displayed as an envelope representing the range of algorithm outputs across all signal hypotheses included in the training.
    The background class contributions from SM ggH (left) and SM VBF (right) events are highlighted using dashed lines. For \Dvbf, the background class contains ggH events according to the SM proportion, and the corresponding contribution is rescaled by a factor of 50 to enhance visibility in the plot; for \Dggf, the background class contains 50\% VBF events.
    }
  \label{fig:VBF_ADNN_even_all_sig}
\end{figure}

Figure~\ref{fig:VBF_ADNN_even_all_sig} shows the distributions of the VBF-ADNN score (\Dvbf) and the GGH-ADNN score (\Dggf) for signal and background events. 
The performance of the algorithm on the different signal samples is assessed by displaying the envelope of the algorithm predictions, computed as the minimum and maximum values of the predicted scores across all signal hypotheses included in the training.  The narrow envelope suggests that the classifier is unable to distinguish between the different underlying physics hypotheses of the signal process. Nonetheless, both networks exhibit strong discriminating power between signal and background events, achieving a binary accuracy of 91\% for the VBF-ADNN and 79\% for the GGH-ADNN.

\section{Systematic uncertainties}
\label{sec:systematics}

The systematic uncertainties arise from experimental and theoretical sources. The impact of the different sources of systematic uncertainty on the final measurement result is summarized in Table~\ref{tab:syst_contributions}. The results shown in this table correspond to fit configuration 3, described in Section~\ref{sec:sigextraction}. The larger relative MC statistical uncertainty observed in the highest \dphijj bin reflects both a small difference in the absolute MC statistical uncertainty and the variation of the total uncertainty across the bins.

\begin{table}[!htb]
    \centering
    \topcaption{Contributions of different sources of uncertainty in the differential cross section measurement, expressed as a percentage of the total uncertainty ($\Delta \sigma_i / \Delta \sigma_{\text{tot}} \times 100$). For asymmetric errors, the largest of the up and down uncertainties is reported. The systematic component includes all sources except for background normalization, which is part of the statistical component. Results are shown for each of the four \dphijj bins and correspond to fit configuration 3, described in Section~\ref{sec:sigextraction}.}
    \cmsTable{\begin{tabular}{lcccc}
    Uncertainty source & {$-\pi< \dphijj \leq - \frac{\pi}{2}$} & {$- \frac{\pi}{2}< \dphijj\leq 0$} & {$0 < \dphijj\leq \frac{\pi}{2}$} & {$\frac{\pi}{2}< \dphijj\leq \pi$} \\ \hline
    Theory (signal) & 18 & 16 &  16 & 17 \\
    Theory (background) & 30 & 17 & 16 & 27 \\
    Integrated luminosity & 16 & 17 & 15 & 14 \\
    b tagging & 16 & 14 & 15 & 13\\
    Jets & 15 & 15 & 13 & 14 \\
    Nonprompt leptons & 14 & 15 & 15 & 17 \\
    Pileup & 15 & 15 & 13 & 16 \\
    Leptons & 17 & 15 & 13 & 15 \\
    \ptmiss & 14 & 15 & 13 & 14 \\
    Backg. norm. & 14 & 15 & 13 & 14 \\
    MC stat. & 31 & 31 &  32 & 40 \\ [\cmsTabSkip]
    Statistical & 91 & 93 & 93 & 86 \\
    Systematic & 46 & 38 & 39 & 52 \\
    \end{tabular}}
\label{tab:syst_contributions}
\end{table}

The integrated luminosities for the 2016, 2017, and 2018 data-taking years have 1.2--2.5\% individual uncertainties, while the overall uncertainty for the 2016--2018 period is 1.6\%~\cite{CMS-LUM-17-003,CMS-PAS-LUM-17-004,CMS-PAS-LUM-18-002}.

The uncertainties in the trigger efficiency and lepton reconstruction and identification efficiencies are evaluated as functions of the lepton \pt and $\eta$, independently for electrons and muons. These uncertainties cause both a normalization and a shape change of the signal and background templates and are kept uncorrelated  among the four data sets. Their effect is of approximately 2\% for electrons and approximately 1\% for muons.

The resolution of the lepton momentum scale, jet energy scale, and unclustered energy scale cause the migration of the simulated events inside or outside the analysis acceptance, as well as migrations across the bins of the signal and background templates. The impact of these sources in the template normalizations is 0.6--1.0\% for the electron momentum scale, 0.2\% for the muon momentum scale, and 1--10\% for \ptmiss.

The uncertainty in the jet energy resolution smearing applied to simulated samples to match the \pt resolution measured in data causes both a normalization and a shape uncertainty of the templates. This uncertainty has a minor impact on all the bins of the \dphijj observable (effect below 1\%) and is uncorrelated among the four data sets. 

The pileup jet identification is affected by a systematic uncertainty that can modify both the shape and normalization of the templates. This uncertainty is uncorrelated across the four data sets and its effect is estimated to be less than 1\%.

The uncertainty related to the pileup interactions is estimated by varying the total inelastic $\Pp\Pp$ cross section by 4.6\% around the nominal value of 69.2\unit{mb}, which has been measured from data~\cite{PUATLAS2016,Sirunyan:2018nqx}.

The uncertainty associated with the b tagging efficiency is modeled by 10 different sources, divided according to the reconstructed jet flavor (five for \PQb and \PQc jets and five for light-flavor quark and gluon jets). The theoretical component of the uncertainty is correlated among different data sets, whereas statistical uncertainties are left uncorrelated.

The estimation of the nonprompt-lepton background is affected by the systematic uncertainty due to the limited size of the control samples used to determine the misidentification rate. An additional uncertainty is applied to account for differences in the jet flavor composition between the multijet CR, where the misidentification rate is measured, and the phase space used in this analysis. Both sources may vary the shape of the nonprompt-lepton distributions, with an effect from a few percent to 10\% depending on the \dphijj bin of the SR. These uncertainties are treated as uncorrelated between electrons and muons, as well as across different data sets. Finally, a conservative 30\% normalization uncertainty is assigned to nonprompt-leptons background to cover any residual discrepancies between MC prediction and data in a same-sign validation region, which is defined by applying the same selection as in the inclusive signal region while requiring the two leptons to have the same electric charge. The statistical uncertainty due to the limited number of simulated events is associated with each bin of the simulated signal and background templates~\cite{BARLOW1993219}.

The theoretical uncertainties relevant to the simulated MC samples have different sources: the choice of the PDF set and the coupling constant \alpS and missing higher-order corrections in the perturbative expansion of the MEs in simulation. Template variations, both in shape and normalization, associated with the aforementioned sources are parametrized with nuisance parameters correlated among the four data sets. The uncertainties in the PDF set and \alpS choice are found to have in general a negligible effect on the simulated templates, as the effect of the shape variation on the expected uncertainties was found to be below 1\%. Therefore only the normalization change is considered, taking into account the effect due to the cross section and acceptance variation. These uncertainties are not considered for backgrounds with a normalization constrained through data in dedicated CRs. For the Higgs boson signal processes, these theoretical uncertainties are computed by the LHC Higgs Cross Section Working Group~\cite{deFlorian:2227475} for each production mechanism, except for the ggH and VBF signal given that the corresponding cross sections are measured in the final fit. The effect of missing higher-order corrections for the background processes is estimated by reweighting the simulated events with alternative event weights, where the renormalization ($\mu_\mathrm{R}$) and factorization ($\mu_\mathrm{F}$) scales are varied by a factor of 0.5 or 2, and the envelopes of the varied templates are taken as the variation corresponding to one standard deviation. The extreme variations where both scales are shifted in opposite direction are excluded. Uncertainties in the Higgs boson production processes are split into those in the overall cross section and in the event acceptance. The former is provided by the LHC Higgs Cross Section Working Group, whereas the latter is calculated by means of event reweighting. For the VBF and ggH signals, the uncertainty in the acceptance is applied, together with two distinct shape uncertainties related to the variation of $\mu_{\mathrm{R}}$ and $\mu_{\mathrm{F}}$. Each of these uncertainties is calculated by varying one scale by a factor 2 and 0.5, while keeping the other scale at its nominal value and vice versa. Finally, they are renormalized in order to affect only the shape of the distributions, leaving the normalization of the distributions unchanged in each \dphijj fiducial bin. Additional shape uncertainties considered for the signal processes include electroweak corrections for the VBF process, as well as uncertainties related to the resummation scale and the top quark mass for the ggH production process. The theoretical uncertainty due to the modeling of the PS and UE is taken into account for all the simulated samples. The uncertainty in PS modeling primarily affects the jet multiplicity and is computed using \PYTHIA. In particular, the $\mu_{\mathrm{R}}$, that regulates the dynamics of the parton splitting evolution, is varied by a factor of two for both the ISR and final-state radiation. For signal processes, only the shape effect is included. The uncertainty in the UE is estimated with \PYTHIA and is found to have a minimal effect on the template shapes. However, it affects the normalization by approximately 1.5\%. 

Additional theoretical uncertainties in specific background processes are also taken into account. An uncertainty of 8\% is assigned to the relative fraction of single top quark and \ttbar processes. The top quark background normalization is constrained from data in the dedicated CR independently for each \dphijj bin. Uncertainties on the $\mu_{\mathrm{R}}$ and $\mu_{\mathrm{F}}$, can modify the ratio of the expected yields between the SR and the CR, as an acceptance factor. This uncertainty is estimated to be 1\% and is applied to the CR. A similar uncertainty is applied to the \dytt sample, and is estimated to be 2\% and considered correlated across all data sets in the DY CR. Variation of the NNLL weights are used to assess the theoretical uncertainty of the quark-induced WW production cross section. The weights are varied by shifting the $\mu_{\mathrm{R}}$, $\mu_{\mathrm{F}}$ and resummation scale. The latter determines the scale below which QCD radiations are resummed. The gluon-induced WW process is generated at LO and normalized to the NLO cross section. A theoretical uncertainty of 15\% in the scale factor is considered. Finally, the normalization corrections applied to the low- and high-mass \VgS process have associated uncertainties of 25\% and 16\%, respectively, which are treated as correlated among the different data sets.

\section{Signal extraction}
\label{sec:sigextraction}

The differential production cross sections of the Higgs boson are inferred from the signal strength modifiers $\boldsymbol{\mu}$, which are extracted through a simultaneous maximum likelihood fit to the signal and background distributions across the four \dphijj bins of the SR and the two CRs. The migration matrix is nearly diagonal and hence the unfolding procedure is directly performed inside the likelihood fit.
 
The fit provides the complete set of $\hat{\boldsymbol{\mu}}$, corresponging to unfolded ratios between observed and expected cross sections per generator-level bin, as well as the associated correlation matrix. The results have been determined using the CMS statistical analysis tool \textsc{Combine}~\cite{CMS:2024onh}, which is based on the \textsc{RooFit}~\cite{Verkerke:2003ir} and \textsc{RooStats}~\cite{Moneta:2010pm} frameworks.

The signal is extracted with three different fit configurations. The first fit extracts the overall cross section of VBF and ggH processes by introducing four independent signal strength parameters, one per \dphijj bin, that scale the combined contribution of the two signal processes. In the second configuration, the VBF and ggH cross sections are measured simultaneously by introducing a set of four independent signal strength parameters for each production process, allowing a differential measurement of both processes across the \dphijj bins. This has the advantage to extract additional information from the data, trading off with a reduced sensitivity to the individual Higgs production mechanisms. A third fit measures the VBF cross section, with four signal strength parameters, while fixing the ggH process to the SM prediction. This allows to place more stringent constraints on anomalous HVV couplings, thereby enhancing the precision of the VBF-specific parameters. On the other hand, in this phase space, a significantly better measurement of the ggH contribution alone cannot be achieved, and therefore this case is not considered.

The fit variable chosen for the signal extraction procedure depends on the specific cross section being measured. In fit configuration 3, used to measure the VBF differential cross section, the fit variable is the \text{\Dvbf} observable. This observable has been chosen due to its strong discriminating power between the VBF signal and background processes, including the ggH contribution. Moreover, it is designed to be model agnostic with respect to potential anomalous couplings that may affect the HVV vertex. The binning scheme of the \Dvbf distribution has been optimized to maximize the sensitivity of the analysis while ensuring a sufficient number of MC events per bin. This optimization process includes imposing a maximum threshold on the relative statistical uncertainty for the MC background predictions and a minimum number of signal events required per bin of the template, and it has been performed in each \dphijj bin of the SR and for each data set individually. The threshold values are 0.15 for the relative statistical uncertainty on the MC background events in each bin and 1.5 for the minimum number of signal events in each bin. The overall production cross section of the VBF and ggH mechanisms is measured by performing a maximum likelihood fit to a two-dimensional template for signal and background in the (\Dvbf, \Dggf) plane. This two-dimensional variable is referred to as \Dtwodim, and is employed in fit configurations 1 and 2. The suitability of \Dtwodim for these measurements arises from the construction of its components: \Dvbf is designed to be model independent with respect to anomalous couplings affecting VBF production, while \Dggf is constructed to be agnostic with respect to anomalous couplings impacting the ggH process. Additionally, both components exhibit strong discriminating power against background processes. Consequently, the combined variable mitigates model dependence simultaneously for both production mechanisms, while ensuring a good separation between signals and backgrounds. The binning scheme of the \Dtwodim variable is optimized using the same strategy as that employed for \Dvbf, performed on the unrolled one-dimensional distribution. Each 2D bin is flattened into a 1D histogram, where the bin labels on the $x$ axis of the 1D histogram correspond to the bin numbering of the original 2D map.

\begin{figure}[!htb]
    {\includegraphics[width=0.45\textwidth]{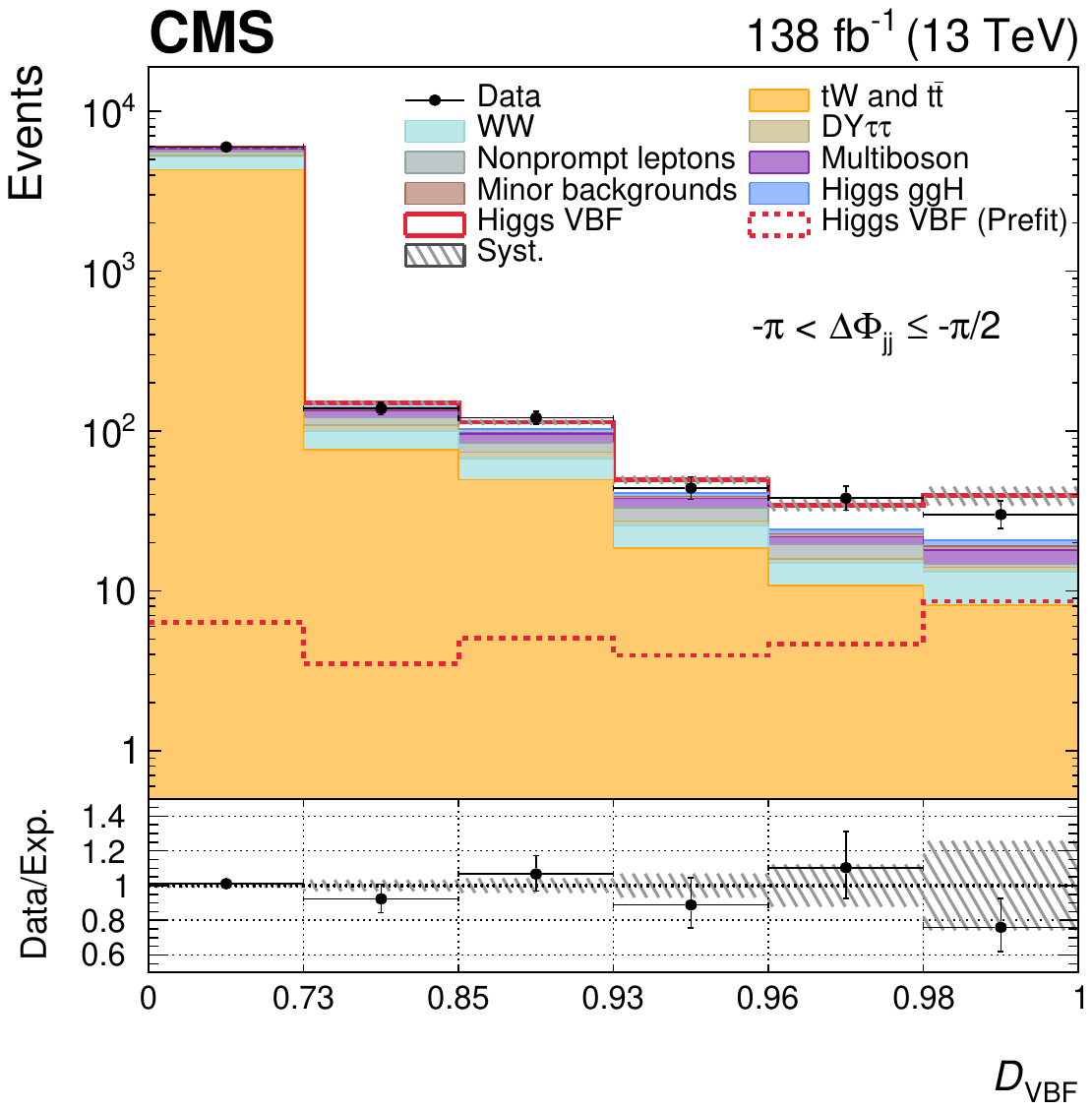}}
    {\includegraphics[width=0.45\textwidth]{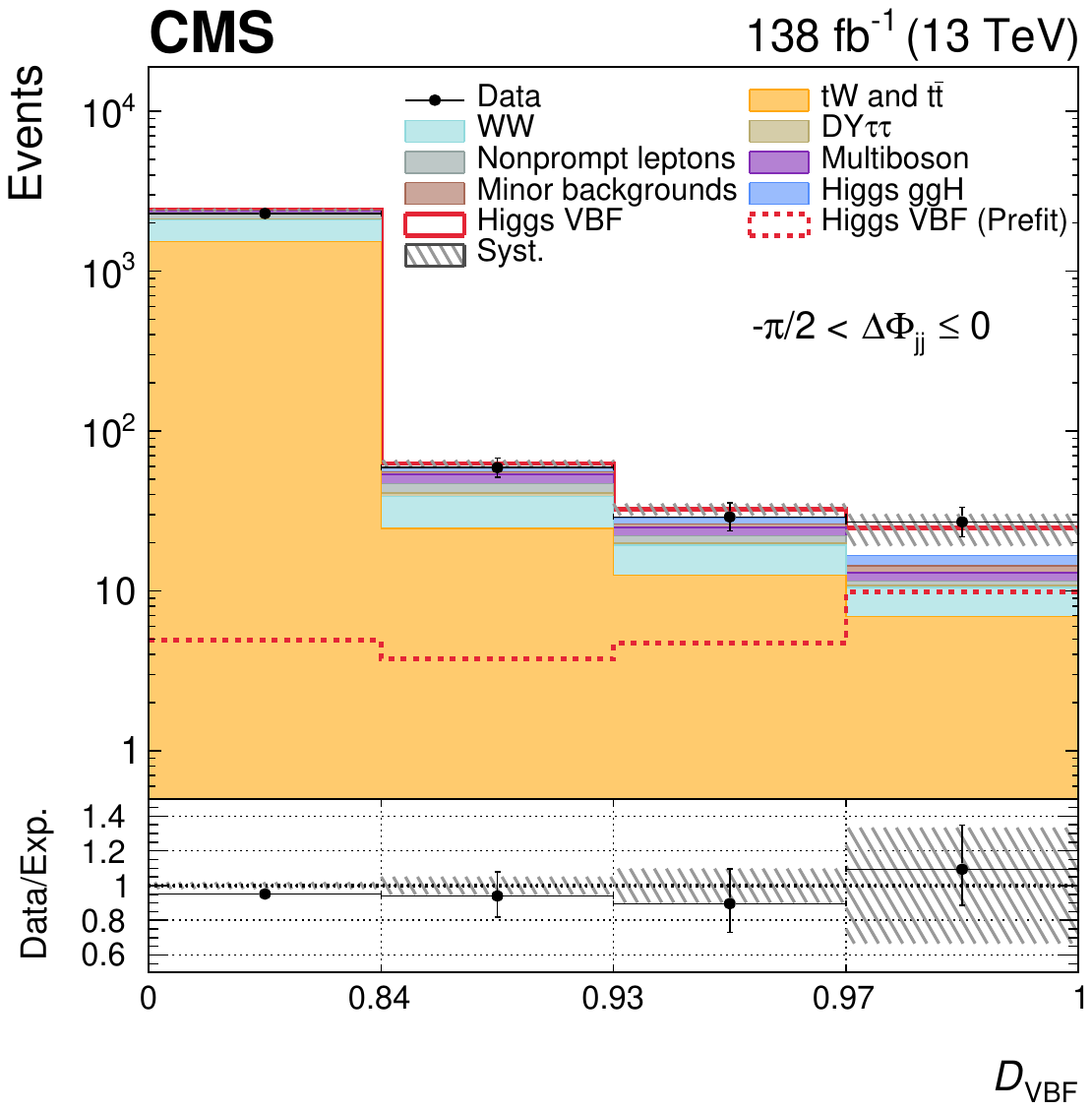}}\\
    {\includegraphics[width=0.45\textwidth]{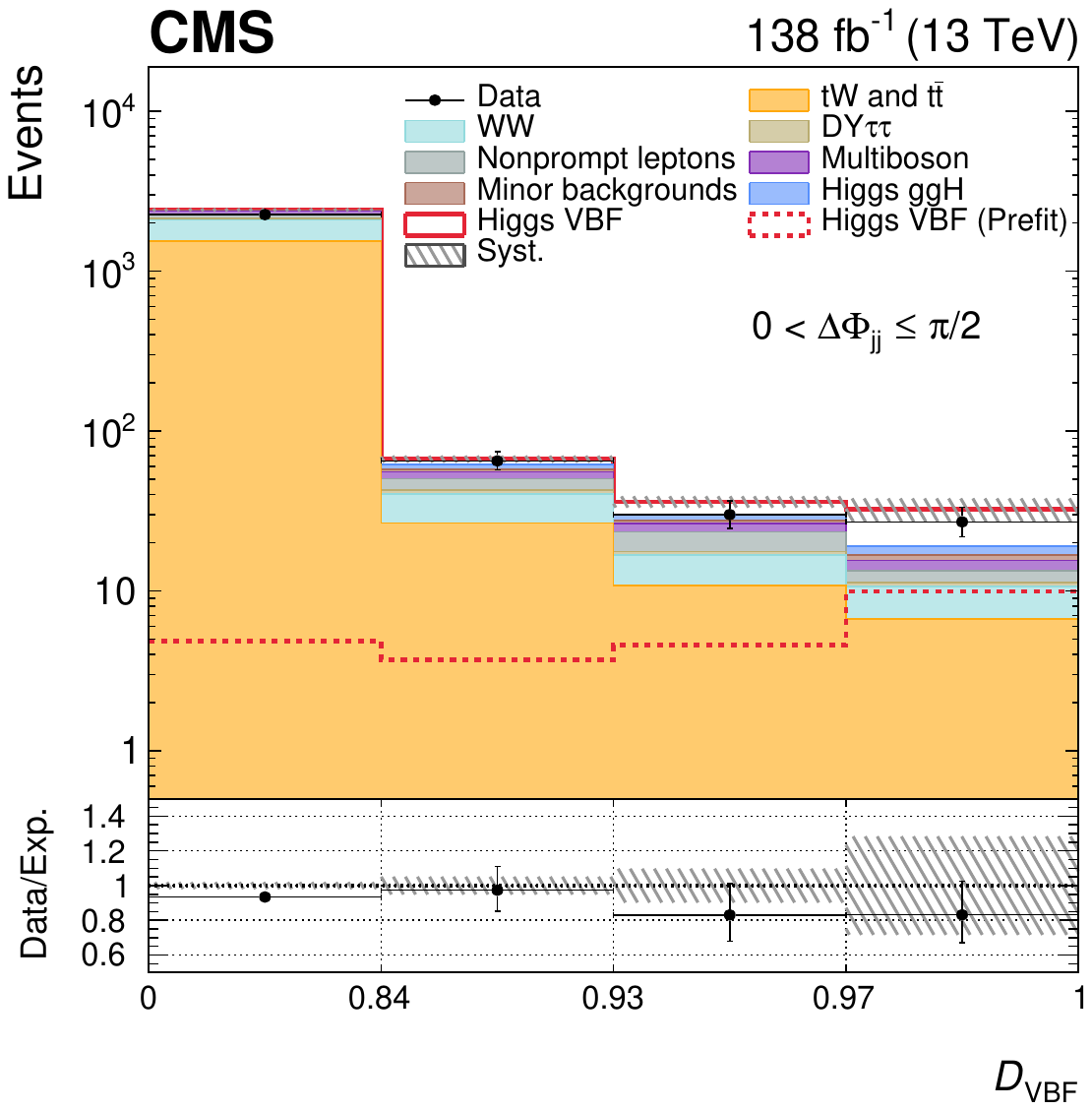}}
    {\includegraphics[width=0.45\textwidth]{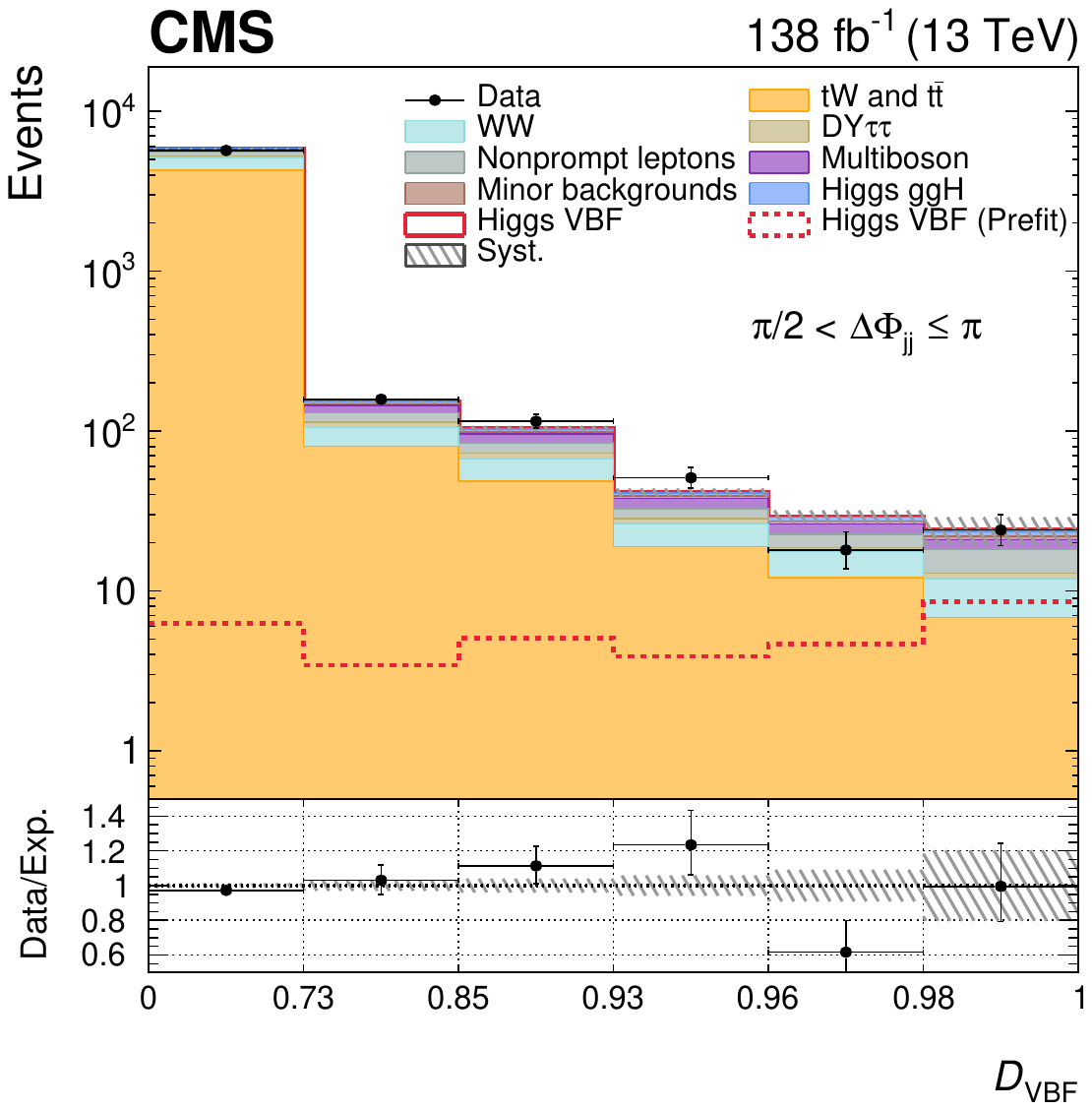}}
    \caption{
      Post-fit \Dvbf distributions in the \dphijj bins of the SR for the 2016--2018 data set, corresponding to fit configuration 3. Systematic uncertainties are shown as dashed gray bands. The pre-fit signal is shown superimposed as a dotted line, while the post-fit signal is included in the stacked histograms on top of the background templates. A uniform binning is applied for visualization, with the true binning range indicated on the $x$ axis. The binning scheme optimized for the 2018 data set is used. The lower panel shows the ratio of data to the total post-fit expected yield (Data/Exp.), where the signal contribution is included in the expectation.
    }
    \label{fig:SR_adnn_run2}
\end{figure}

\begin{figure}[!htb]
    {\includegraphics[width=0.45\textwidth]{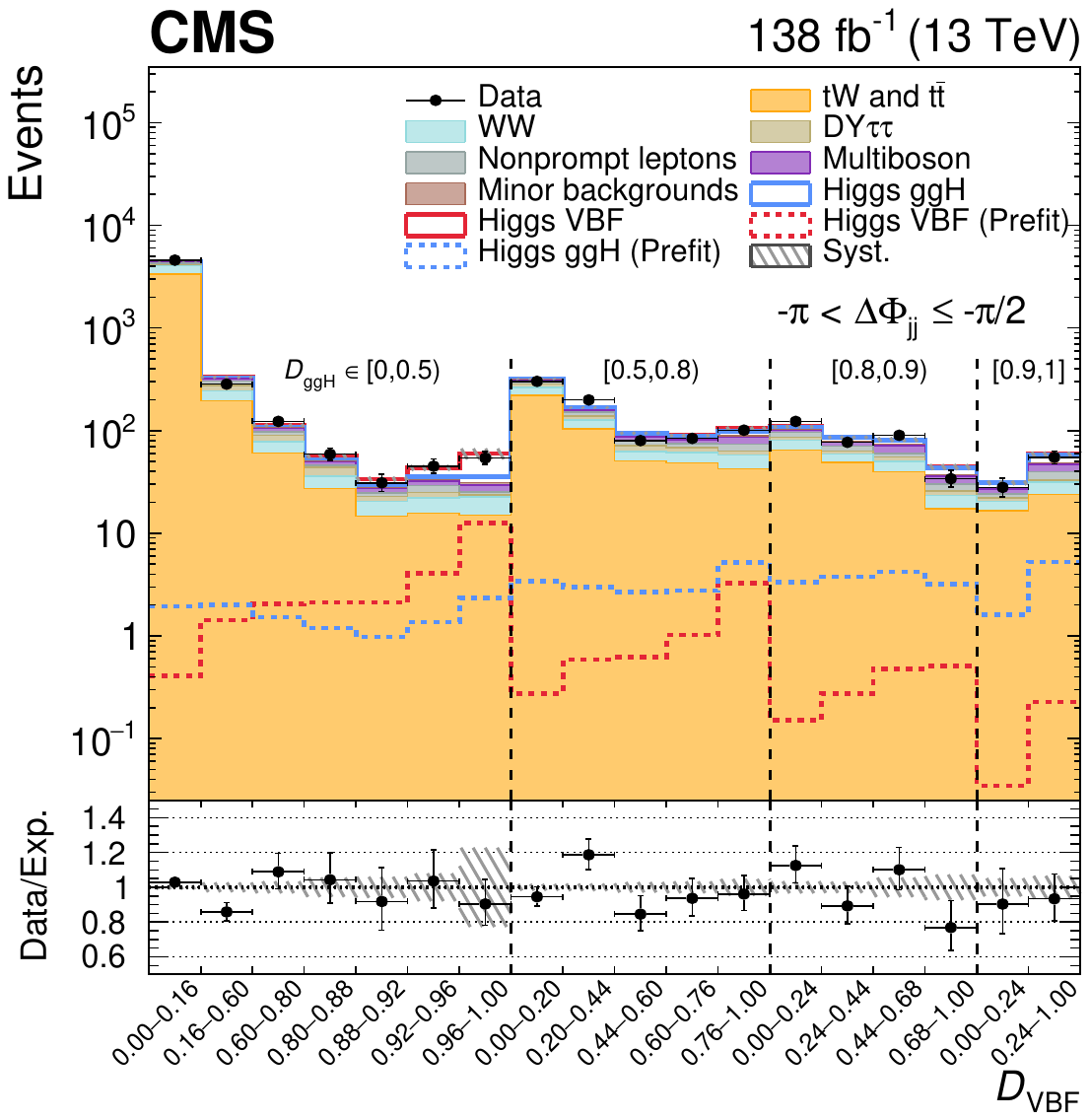}}
    {\includegraphics[width=0.45\textwidth]{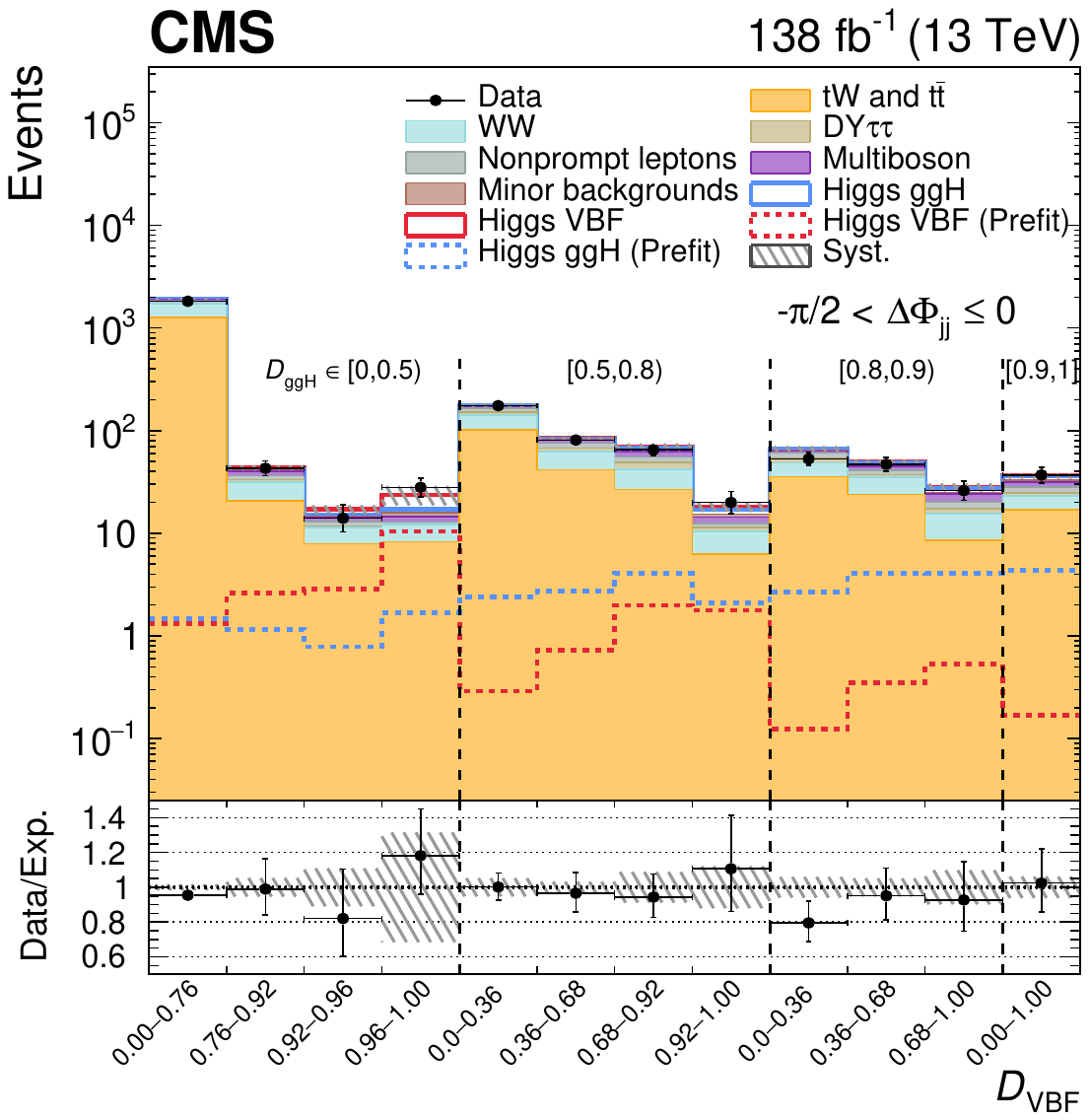}}\\
    {\includegraphics[width=0.45\textwidth]{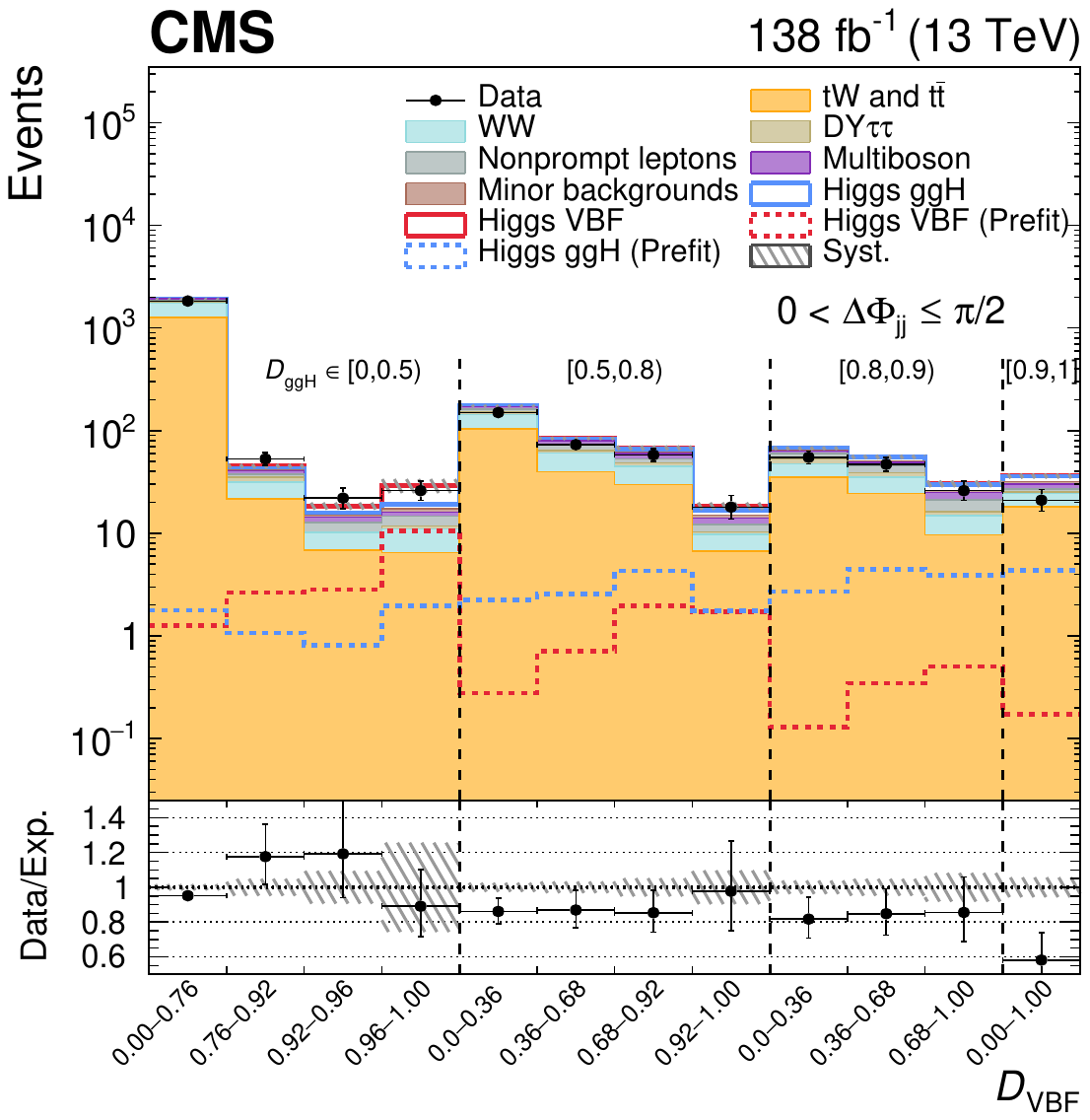}}
    {\includegraphics[width=0.45\textwidth]{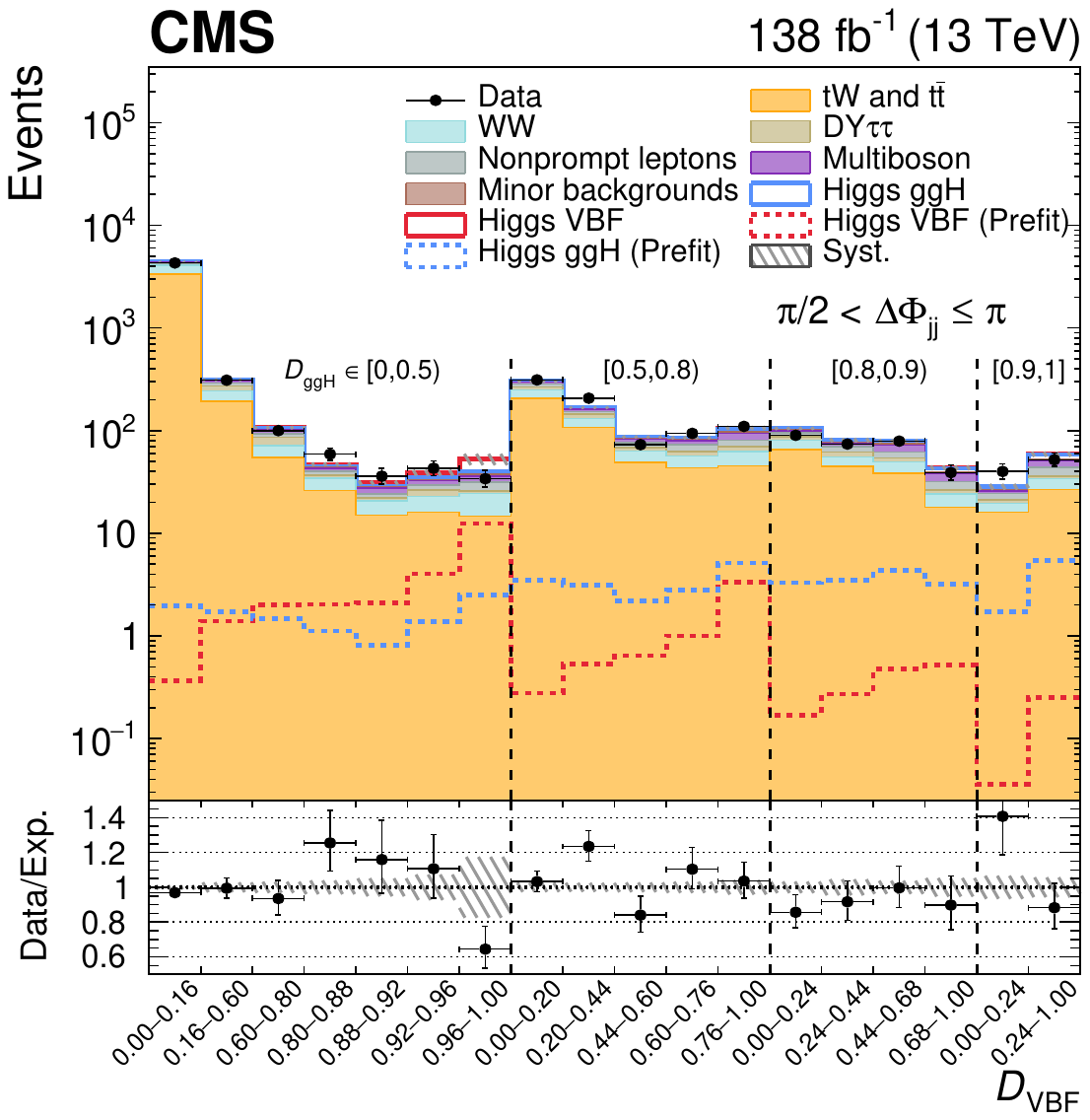}}
    \caption{
      Post-fit \Dtwodim distributions in the \dphijj bins of the SR for the 2016--2018 data set, corresponding to fit configuration 1. The 2D distribution is unrolled into a 1D histogram, where the $x$ axis represents the \Dvbf variable. A uniform binning is adopted for visualization. The boundaries of the \Dggf intervals are marked by dashed black vertical lines, with the corresponding ranges explicitly labeled. Within each \Dggf interval, the binning reflects the \Dvbf subdivisions, and the true \Dvbf ranges are indicated on the $x$ axis. Systematic uncertainties are shown as dashed gray bands. The pre-fit signal is shown superimposed as a dotted line, while the post-fit signal is included in the stacked histograms on top of the background templates. The binning scheme optimized for the 2018 data set is used. The lower panel shows the ratio of data to the total post-fit expected yield (Data/Exp.), where the signal contribution is included in the expectation.
    }
    \label{fig:SR_adnn2D_run2}
\end{figure}

The observed post-fit distributions of the \Dvbf and \Dtwodim are shown in Figs.~\ref{fig:SR_adnn_run2} and~\ref{fig:SR_adnn2D_run2}, respectively, for each \dphijj bin of the SR. Figure~\ref{fig:SR_adnn_run2} corresponds to fit configuration 3, while Fig.~\ref{fig:SR_adnn2D_run2} refers to fit configuration 1. In both figures, the signal prediction corresponding to the SM expectation is shown superimposed on the background templates as a dotted line, while the post-fit signal is stacked on top of the background templates. Systematic uncertainties are indicated by dashed gray bands. The binning scheme optimized for the 2018 data set is used. The ratio panels illustrate the level of agreement between the observed data and the total expected yields, including signal. The pre-fit signal purity in the most sensitive bins ranges from 25 to 45\% for the \Dvbf distribution, and from 20 to 40\% for the \Dtwodim distribution.

\section{Results}
\label{sec:results}

\begin{figure}[!htb]
    \centering
    {\includegraphics[width=0.5\textwidth]{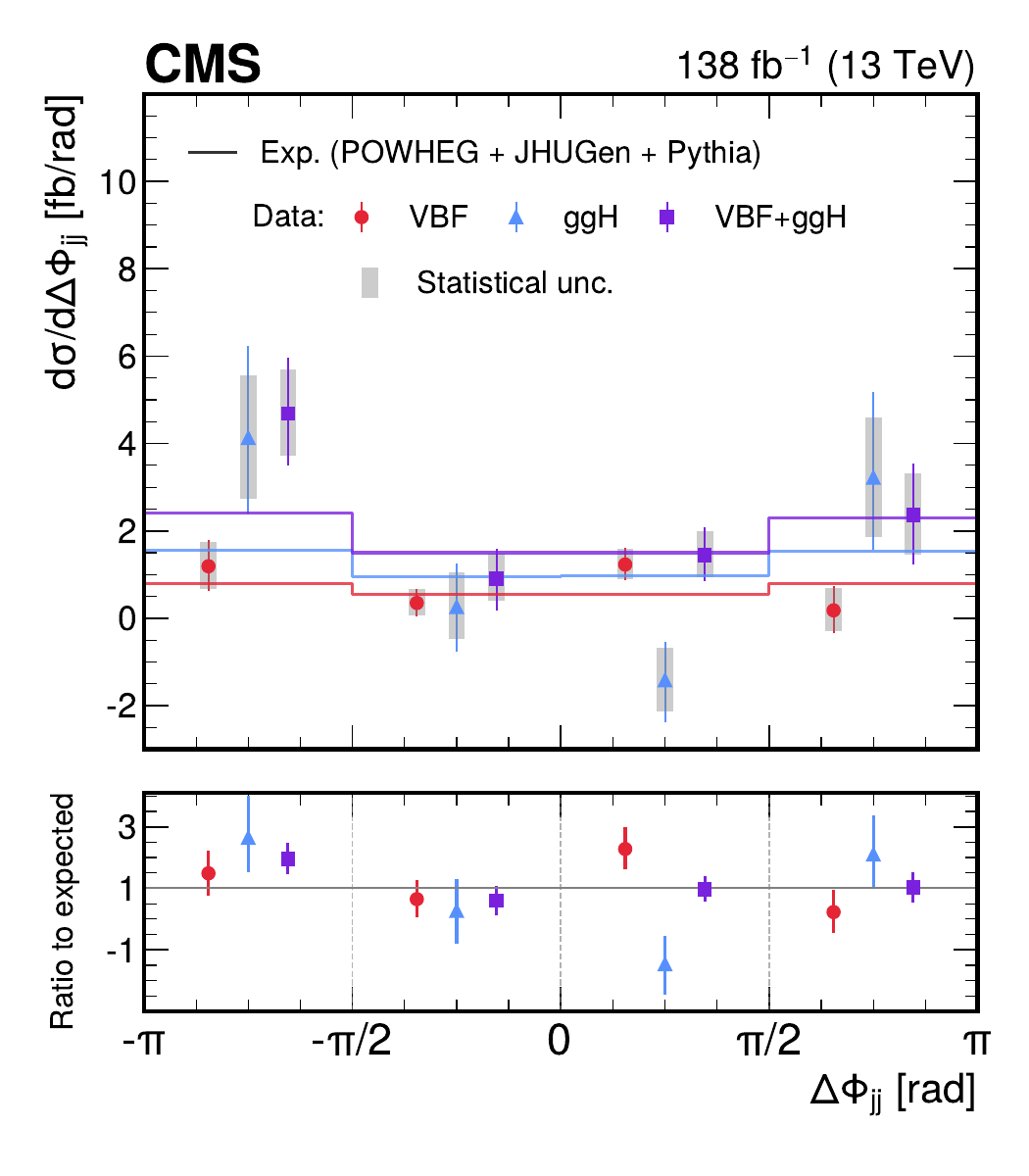}} 
    \caption{Measured fiducial cross section of the VBF and ggH production processes. Colored markers represent the extracted cross section values from data, with error bars showing the combined statistical and systematic uncertainties: red for VBF, light blue for ggH, and violet for the sum of VBF + ggH. The gray bands indicate the statistical uncertainties. The colored histogram corresponds to the expected SM prediction, simulated with \POWHEG + \textsc{JHUGen} + \PYTHIA generators. The lower panel displays the ratio of the measured values to the SM expectation.}
    \label{fig:measured_xsec_2_plots}
\end{figure}

The differential cross sections measured from fit configurations 1 and 2 are presented in Tables~\ref{tab:results_sigma_vbf_ggh_together_unblinded} and~\ref{tab:results_sigma_vbf_ggh_separately_unblinded}, respectively. Total and statistical uncertainties are quoted at the 68\% confidence level (\CL), along with the observed significance relative to the background-only hypothesis. The statistical uncertainty, incorporating both the Poisson component and background normalization, remains the dominant source of uncertainty in both measurements. These results are also illustrated in Fig.~\ref{fig:measured_xsec_2_plots}. In fit configuration 2, the best fit value of the ggH signal strength in the \dphijj bin 2 is negative, resulting in a negative cross section. This unphysical result can occur when the signal is small compared to backgrounds and statistical fluctuations dominate. More generally, in some \dphijj bins the lower bound of the total uncertainty extends into the negative region. These negative values or uncertainty intervals crossing zero do not correspond to physical cross sections but are artifacts of the statistical treatment, especially in regions with limited sensitivity. The differential cross section measured from fit configuration 3 is presented in Table~\ref{tab:results_sigma_vbf_unblinded}. 

\begin{table}[!htb]
    \centering
    \topcaption{{Measured fiducial cross section summing VBF and ggH production processes, corresponding to fit configuration 1. The total (statistical and systematic) and statistical uncertainties corresponding to the 68\% \CL are shown. The observed significance with respect to the background-only hypothesis is computed accounting for the total uncertainty.}}
    \begin{tabular}{cccccc} 
        Bin & ${\sigma}^\text{fid}$ [fb/rad] &  Total unc. & Statistical unc.  & Significance (s.d.)\\ \hline
        $-\pi< \dphijj \leq - \frac{\pi}{2}$           &            4.69        &     $-1.19 / +1.28 $    &  $-0.96 / +1.00 $  & 4.1\\
        $- \frac{\pi}{2}< \dphijj\leq 0$           &            0.91        &     $-0.74 / +0.69 $    &  $-0.52 / +0.56 $  & 1.4 \\
        $0 < \dphijj\leq \frac{\pi}{2}$           &            1.44        &     $-0.58 / +0.64 $    &  $-0.51 / +0.55 $  & 2.6 \\
        $\frac{\pi}{2}< \dphijj\leq \pi$           &            2.36        &     $-1.12 / +1.18 $    &  $-0.91 / +0.95 $  & 2.1\\
        \end{tabular}
\label{tab:results_sigma_vbf_ggh_together_unblinded}
\end{table}

\begin{table}[!htb]
    \centering
    \topcaption{Measured fiducial cross section of VBF and ggH production processes, corresponding to fit configuration 2. The measurement is performed through a simultaneous fit, where the contributions from VBF and ggH production are determined independently in each bin. The observed significance with respect to the background-only hypothesis is computed accounting for the total uncertainty. Negative values or lower uncertainty bounds extending below zero are artifacts of the fit and reflect statistical fluctuations in regions with limited sensitivity.}
    \begin{tabular}{cccccc}
    Signal              & Bin   &   ${\sigma}^\text{fid}$ [fb/rad]    &     Total unc. &  Statistical unc. & Significance (s.d.) \\ \hline
    \multirow{4}{*}{VBF} &     $-\pi< \dphijj \leq - \frac{\pi}{2}$        &                   1.19                &     $-0.57 / +0.60 $   & $-0.52 / +0.55 $   &  2.1\\
                        &      $- \frac{\pi}{2}< \dphijj\leq 0$            &                   0.35                &     $-0.31 / +0.33 $   & $-0.28 / +0.31 $   &  1.2\\
                        &      $0 < \dphijj\leq \frac{\pi}{2}$             &                   1.23                &     $-0.35 / +0.38 $   & $-0.33 / +0.35 $   &  3.9\\
                        &      $\frac{\pi}{2}< \dphijj\leq \pi$            &                   0.18                &     $-0.53 / +0.56 $   & $-0.48 / +0.51 $   &  0.3\\ [\cmsTabSkip]
    \multirow{4}{*}{ggH} &     $-\pi< \dphijj \leq - \frac{\pi}{2}$        &                   4.13                &     $-1.74 / +2.10 $   & $-1.40 / +1.44 $   & 2.5\\
                        &      $- \frac{\pi}{2}< \dphijj\leq 0$            &                   0.26                &     $-1.03 / +0.99 $   & $-0.74 / +0.79 $   & 0.3\\
                        &      $0 < \dphijj\leq \frac{\pi}{2}$             &                   $-1.42$             &     $-0.96 / +0.88$    & $-0.71 / +0.75$    &    0.0\\
                        &      $\frac{\pi}{2}< \dphijj\leq \pi$            &                   3.22                &     $-1.66 / +1.95 $   & $-1.35 / +1.38 $   & 2.0\\
    \end{tabular}
\label{tab:results_sigma_vbf_ggh_separately_unblinded}
\end{table}

\begin{table}[!htb]
    \centering
    \topcaption{{Measured fiducial cross section of VBF production process while fixing the ggH process to the SM prediction, corresponding to fit configuration 3. The total (statistical and systematic) and statistical uncertainties corresponding to the 68\% \CL are shown. The observed significance with respect to the background only hypothesis is computed accounting for the total uncertainty. Negative values or lower uncertainty bounds extending below zero are artifacts of the fit and reflect statistical fluctuations in regions with limited sensitivity.}}
    \begin{tabular}{cccccc}
    Bin & ${\sigma}^\text{fid}$ [fb/rad] &Total unc.  & Statistical unc.    & Significance (s.d.) \\ \hline
    $-\pi< \dphijj \leq - \frac{\pi}{2}$ & 1.79 & $-0.51 / +0.56 $ &  $-0.46 / +0.49 $ & 3.8       \\
    $- \frac{\pi}{2}< \dphijj\leq 0$     & 0.43 & $-0.30 / +0.33 $ &  $-0.28 / +0.31 $ &  1.5      \\
    $0 < \dphijj\leq \frac{\pi}{2}$      & 0.75 & $-0.30 / +0.33 $ &  $-0.28 / +0.30 $ & 2.6       \\
    $\frac{\pi}{2}< \dphijj\leq \pi$     & 0.00 & $-0.46 / +0.49 $ &  $-0.39 / +0.43 $  &  0.0         \\
    \end{tabular}
\label{tab:results_sigma_vbf_unblinded}
\end{table}

The model dependence of the measured cross section is assessed by evaluating potential biases that arise when interpreting the data under different signal hypotheses. This is tested by generating pseudodata under different BSM signal hypotheses combined with the expected SM background contributions. Each pseudodata set is fitted using SM signal and background templates to extract the signal strength in each \dphijj bin. The extracted signal strength is then compared to the theoretical ratio of fiducial cross sections predicted for the BSM and SM scenarios. Any deviation between these two defines the total bias, which encompasses all sources of model dependence in the analysis, including shape effects in the fit variable, systematic shifts from the unfolding procedure, and variations in acceptance and selection efficiencies. This bias is then used to quantify the model dependence of the measured cross section, by evaluating its impact on the expected cross section. For each \dphijj bin, the largest contribution among those derived from the different BSM hypotheses in the mixed SM-BSM scenario is taken as the model dependence of the result. In the case of the overall Higgs boson production cross section in the two-jet phase space, the model dependence ranges from 20--40\% of the expected cross section, depending on the \dphijj bin. For the VBF cross section, when measured simultaneously with the ggH one, it varies between 20--70\%, while for the ggH cross section, it falls within 10--20\%. Finally, when the VBF cross section is measured alone, the model dependence varies between 10 and 40\% of the expected cross section. The ADNN approach reduces the model dependence by 30--70\%, depending on the \dphijj bin, compared to a deep neural network without an adversarial component. The remaining model dependence is typically smaller than the total uncertainty and it primarily arises from acceptance effects that are not corrected by the ADNN approach. 

Following the measurement of the differential cross sections, additional insight into the \dphijj observable can be gained by studying its asymmetry. While the measured cross sections provide a detailed picture of Higgs boson production as a function of \dphijj, the asymmetry offers a complementary probe of potential \CP-violating effects. Indeed, a nonzero asymmetry value in the \dphijj distribution, in either VBF or ggH events, would indicate \CP\ violation in the Higgs sector. The \dphijj asymmetry is defined as the difference between the number of events with ${0 < \dphijj \leq \pi}$ and those with ${-\pi  < \dphijj \leq 0}$, normalized to the total number of events:

\begin{equation}\label{eq:asymmetry}
    A = \frac{N(\dphijj>0)-N(\dphijj\leq0)}{N(\dphijj>0)+N(\dphijj\leq0)}.
\end{equation} 

Being a ratio of observed quantities, the main advantage of the asymmetry measurement is its independence on most systematic uncertainties, in particular those that have the same effect for all the signal strengths. The observed value of the \dphijj asymmetry is:

\begin{equation}
    A = -0.43^{+0.27}_{-0.32}\,\text{(tot)}= -0.43^{+0.25}_{-0.28}\stat^{+0.10}_{-0.16}\syst,
\end{equation}

with a compatibility with the SM prediction ($A=0$) corresponding to a $p$-value of 11\%.

\section{The SMEFT interpretation}
\label{sec:SMEFT}

The differential cross section measurements can be used to constrain anomalous contribution to the Higgs boson interaction using an EFT approach. The main idea of the EFT is that the BSM physics can be studied by extending the SM in a systematic and possibly model-independent way. Assuming the existence of an unknown phenomenon at a given energy scale $\Lambda$ above the energy that is currently experimentally accessible ($\Lambda \gg E_{\text{LHC}}$), effects of the BSM physics may manifest themselves via effective interactions between SM fields. The effective Lagrangian can be written as:

\begin{equation}
    \mathcal{L}_{\text{SMEFT}}=\mathcal{L}_{\text{SM}}+\mathcal{L}_{\text{EFT}}=\mathcal{L}_{\text{SM}}+\sum_{i=5}^{\infty}\sum_{j=0}^{N_{i}} \frac{c_{j}^{(i)}}{\Lambda^{i-4}}\mathcal{O}_{j}^{(i)},
\label{eq:Lsmeft}
\end{equation}

where $i$ runs over the number of dimensions, $j$ runs over the number of operators $N_i$ of dimension $i$, $\mathcal{L}_{\mathrm{SM}}$ has (mass) dimension 4 and the operator $\mathcal{O}_{j}^{(i)}$ has dimension $i \ge 5$. The parameters $c_j^{(i)}$, known as the Wilson coefficients, specify the strength of the BSM interaction induced by the corresponding operators, and are suppressed by powers of the energy scale $\Lambda$ at which the new physics is expected to occur.

The new operators in Eq.~\eqref{eq:Lsmeft} are constructed using the same fields and symmetries as the SM, ensuring theoretical consistency at low energies. Consequently, the effective Lagrangian is often denoted as the SMEFT Lagrangian. Among all possible operators, only those of dimension $i=6$ are considered, as dimension $i=5$ and 7 operators violate lepton and baryon number conservation. The impact of higher-dimension operators is suppressed by higher powers of the $\Lambda$ scale. Furthermore, the analysis focuses on a subset of dimension-6 operators that both belong to the classes ${\mathcal{L}_6^{(4)}-X^{2}H^{2}}$ and ${\mathcal{L}_6^{(3)}-H^{4}D^{2}}$, as classified in the Warsaw basis~\cite{Grzadkowski_2010,deFlorian:2227475}, and are relevant to the study of HVV vertex for the VBF and ggH production and \HtoWW decay. The operators describing the Higgs boson interaction with quarks that belong to the ${\mathcal{L}_6^{(7)}-\psi^{2}H^{2}D}$ class were also considered to assess possible sensitivity via VBF production. A summary of the operators and their corresponding Wilson coefficients used in this analysis is presented in Table~\ref{tab:wilson_coeff}. 
The coefficients $c_{\PH\PW},\ c_{\PH\PW\mathrm{B}},\ c_{\PH\mathrm{B}},\ c_{\PH\mathrm{G}}$ are associated with \CP-even operators describing the Higgs boson interactions with vector bosons, while $c_{\PH\tilde{\PW}},\ c_{\PH\tilde{\PW}\mathrm{B}},\ c_{\PH\tilde{\mathrm{B}}},\ c_{\PH\tilde{\mathrm{G}}}$ correspond to \CP-odd operators. The presence of \CP-odd operators together with any of the other \CP-even operators, will violate \CP symmetry.
The coefficients $c_{\PH\mathrm{G}}$ and $c_{\PH\tilde{\mathrm{G}}}$ primarily affect ggH production, whereas the remaining operators influence VBF production and Higgs boson decay. Operators corresponding to the coefficients $c_{\PH\Box}$ and $c_{\PH\mathrm{D}}$ are included to probe sensitivity to anomalous corrections in the kinetic terms of the scalar fields. The former coefficient marked with the d’Alembert symbol $\Box$ is related to the operator quadratic in partial derivatives.     

\begin{table}[h!]
    \centering
	\topcaption{List of $X^{2}H^{2}$, $H^{4}D^{2}$, and $\psi^{2}H^{2}D$ operators and their corresponding Wilson coefficients.}
    \begin{tabular}{ l l c }
        Class & Operator & Wilson coefficient \\
        \hline
	$\mathcal{L}_6^{(4)}-X^{2}H^{2}$ & $H^{\dagger}H W^{i}_{\mu\nu}W^{i\mu\nu}$ & $c_{\PH\PW}$ \\
	(\CP-even)                       & $H^{\dagger}H W^{i}_{\mu\nu}B^{i\mu\nu}$ & $c_{\PH\PW\mathrm{B}}$ \\
                                     & $H^{\dagger}H B_{\mu\nu}B^{\mu\nu}$ & $c_{\PH\mathrm{B}}$ \\
                                     & $H^{\dagger}H G^{a}_{\mu\nu}G^{a\mu\nu}$ & $c_{\PH\mathrm{G}}$ \\
        [\cmsTabSkip]
	$\mathcal{L}_6^{(4)}-X^{2}H^{2}$ & $H^{\dagger}H \tilde{W}^{i}_{\mu\nu}W^{i\mu\nu}$ & $c_{\PH\tilde{\PW}}$ \\
	    (\CP-odd)    & $H^{\dagger}H \tilde{W}^{i}_{\mu\nu}B^{i\mu\nu}$ & $c_{\PH\tilde{\PW}\mathrm{B}}$ \\
					 & $H^{\dagger}H \tilde{B}_{\mu\nu}B^{\mu\nu}$ & $c_{\PH\tilde{\mathrm{B}}}$ \\
					 & $H^{\dagger}H \tilde{G}^{a}_{\mu\nu}G^{a\mu\nu}$ & $c_{\PH\tilde{\mathrm{G}}}$ \\
        [\cmsTabSkip]    
	    $\mathcal{L}_6^{(3)}-H^{4}D^{2}$ & $(H^{\dagger}H)\Box(H^{\dagger}H)$ & $c_{\PH\Box}$ \\
                                         & $(D^{\mu}H^{\dagger}H)(H^{\dagger}D_{\mu}H)$ & $c_{\PH\mathrm{D}}$ \\
	[\cmsTabSkip]
	    $\mathcal{L}_6^{(7)}-\psi^{2}H^{2}D$ & $(H^{\dagger}i\overleftrightarrow{D}_{\mu}H)(\bar{u}_{p}\gamma^{\mu}u_{r})$ & $c_{\PH\mathrm{u}}$ \\
	                     & $(H^{\dagger}i\overleftrightarrow{D}_{\mu}H)(\bar{d}_{p}\gamma^{\mu}d_{r})$ & $c_{\PH\mathrm{d}}$ \\
						 & $(H^{\dagger}i\overleftrightarrow{D}_{\mu}H)(\bar{q}_{p}\gamma^{\mu}q_{r})$ & $c_{\text{Hj1}}$ \\
						 & $(H^{\dagger}i\overleftrightarrow{D}^{i}_{\mu}H)(\bar{q}_{p}\sigma^{i}\gamma^{\mu}q_{r})$ & $c_{\text{Hj3}}$ \\
    \end{tabular}
    \label{tab:wilson_coeff}
\end{table}

Using the narrow-width approximation, the signal strength parameter for the $j$th \dphijj bin can be expressed as:

\begin{equation}
\label{eq:signalParamBSM}        
	\mu_j(\vec{c}\,) = \frac{\sigma_j^{\text{SMEFT}}}{\sigma_j^{\text{SM}}} \frac{\mathcal{B}_{j,\text{SMEFT}}^{\HtoWW}}{\mathcal{B}_{j,\text{SM}}^{\HtoWW}} = \mu_{j,\text{prod}}(\vec{c}\,) \mu_{j,\text{decay}}(\vec{c}\,),
\end{equation}

where the vector $\vec{c}$ represents the set of Wilson coefficients associated with the SMEFT operators and $\mu_{j,\text{prod}}(\vec{c}\,)$ and $\mu_{j,\text{decay}}(\vec{c}\,)$ are production and decay scaling functions, respectively. Assuming only diagrams with a single insertion of a BSM vertex, all BSM matrix elements entering the formula for the production cross section, $\sigma_j^{\text{SMEFT}}$, are linear in Wilson coefficients $c_{k}$, therefore it is possible to write:

\begin{equation}
\label{eq:signalParamXS}
	\mu_{j,\text{prod}}(\vec{c}\,) = 1+\frac{\sigma_j^{\text{int}}}{\sigma_j^{\text{SM}}}+\frac{\sigma_j^{\text{BSM}}}{\sigma_j^{\text{SM}}},
\end{equation}

where an impact of the Wilson coefficients on the cross section contribution from the SM-BSM interference and pure BSM effects can be derived by assuming the following parametrization:
\begin{equation}
    \frac{\sigma_j^{\text{int}}}{\sigma_j^{\text{SM}}} = \sum_k A_k^{j}c_k, \quad 
    \frac{\sigma_j^{\text{BSM}}}{\sigma_j^{\text{SM}}} = \sum_{kl} B_{kl}^{j}c_{k}c_{l},
\end{equation}	

where $A_k^j$ and $B_{kl}^j$ are denoted as linear and quadratic constants. A similar assumption holds for the \HtoWW branching fraction, $\mathcal{B}_{j,\text{SMEFT}}^{\HtoWW}$, with the parametrization derived for both total and partial decay widths, meaning the decay scaling function can be written as:  

\begin{equation}
\label{eq:signalParamDecay}
	\mu_{j,\text{decay}}(\vec{c}\,) = \frac{\Gamma_{\text{SMEFT}}^{\HtoWW}/\Gamma_{\text{SM}}^{\HtoWW}}{\Gamma_{\text{SMEFT}}^{\text{H}}/\Gamma_{\text{SM}}^{\text{H}}} = \frac{1+\sum_k A_k^{j,\HtoWW}c_k+\sum_{kl} B_{kl}^{j,\HtoWW}c_{k}c_{l}}{1+\sum_k A_k^{\text{H}}c_k+\sum_{kl} B_{kl}^{\text{H}}c_{k}c_{l}}.
\end{equation}

All linear $A_k$ and quadratic $B_{kl}$ constants are extracted using the \textsc{EFT2Obs} tool~\cite{brooijmans2020leshouches2019physics}, which also accounts for the normalization effect that is not sensitive to \CP violation. This tool employs widely used software packages to derive the EFT parametrization of a given process as a function of selected kinematic observables. In particular, it relies on \MGvATNLO v2.6.7~\cite{Alwall2014} for event generation and \PYTHIA v8.240 for parton showering and hadronization. The \textsc{Rivet} v3.0.1 framework~\cite{10.21468/SciPostPhys.8.2.026} is then used to define the fiducial phase space and extract the distribution of the chosen observable while accounting for acceptance effects. 

The overall scaling function from Eq.~(\ref{eq:signalParamBSM}), $\mu_j(\vec{c}\,|A_{k},B_{kl})$, is approximated by a Taylor series expansion up to quadratic terms in the Wilson coefficients and the likelihood function is re\-pa\-ra\-me\-trized accordingly. In case of the operators governing the Higgs boson interaction with vector bosons, a simultaneous profile likelihood fit of the \CP-even and \CP-odd Wilson coefficients was performed to determine the corresponding confidence intervals for the following pairs:

\begin{itemize} 
\item $c_\text{HW}$ and $c_{\text{H}\tilde{\text{W}}}$, 
\item $c_\text{HWB}$ and $c_{\text{H}\tilde{\text{W}}\text{B}}$,
\item $c_\text{HB}$ and $c_{\text{H}\tilde{\text{B}}}$, 
\item $c_\text{HG}$ and $c_{\text{H}\tilde{\text{G}}}$. 
\end{itemize}
The simultaneous fit was also performed for the pairs of Wilson coefficients corresponding to the \CP-even Higgs boson interaction with quarks:
\begin{itemize}
	\item $c_\text{Hu}$ and $c_{\text{Hd}}$,
    \item $c_\text{Hj1}$ and $c_{\text{Hj3}}$.
\end{itemize}

\begin{figure}[!htb]
\centering
\includegraphics[width=0.45\textwidth]{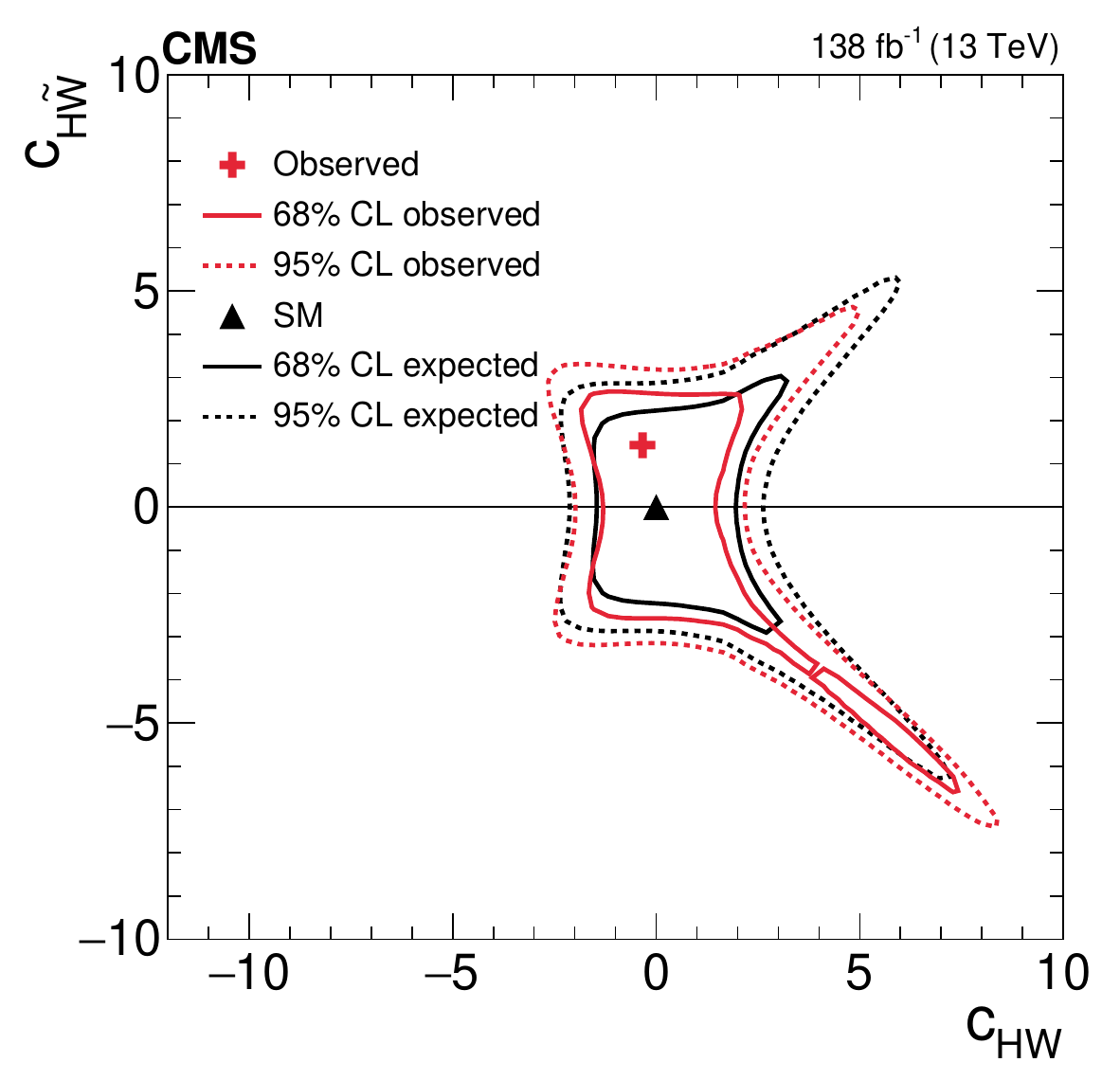}
\includegraphics[width=0.45\textwidth]{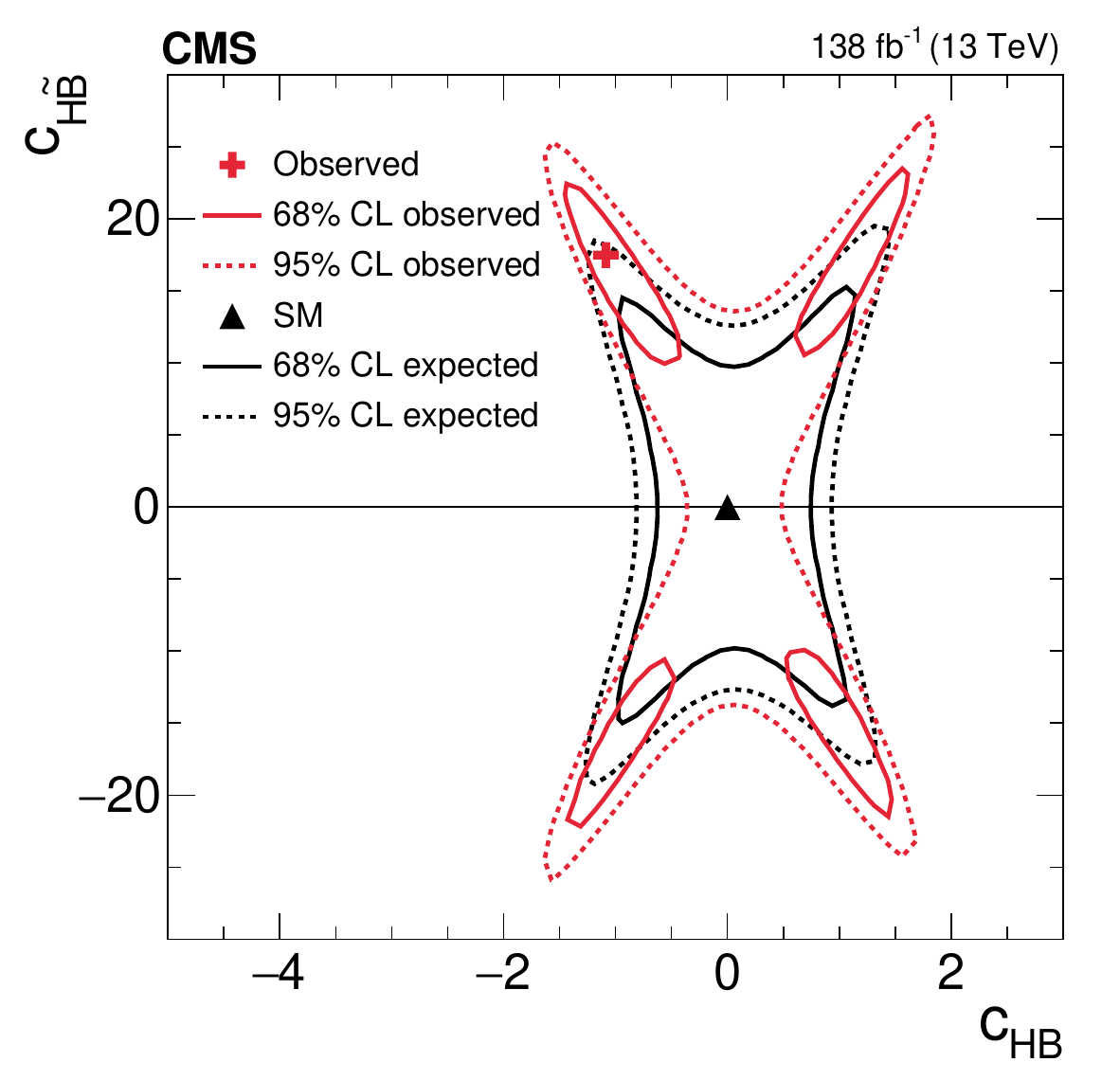} \\
\includegraphics[width=0.45\textwidth]{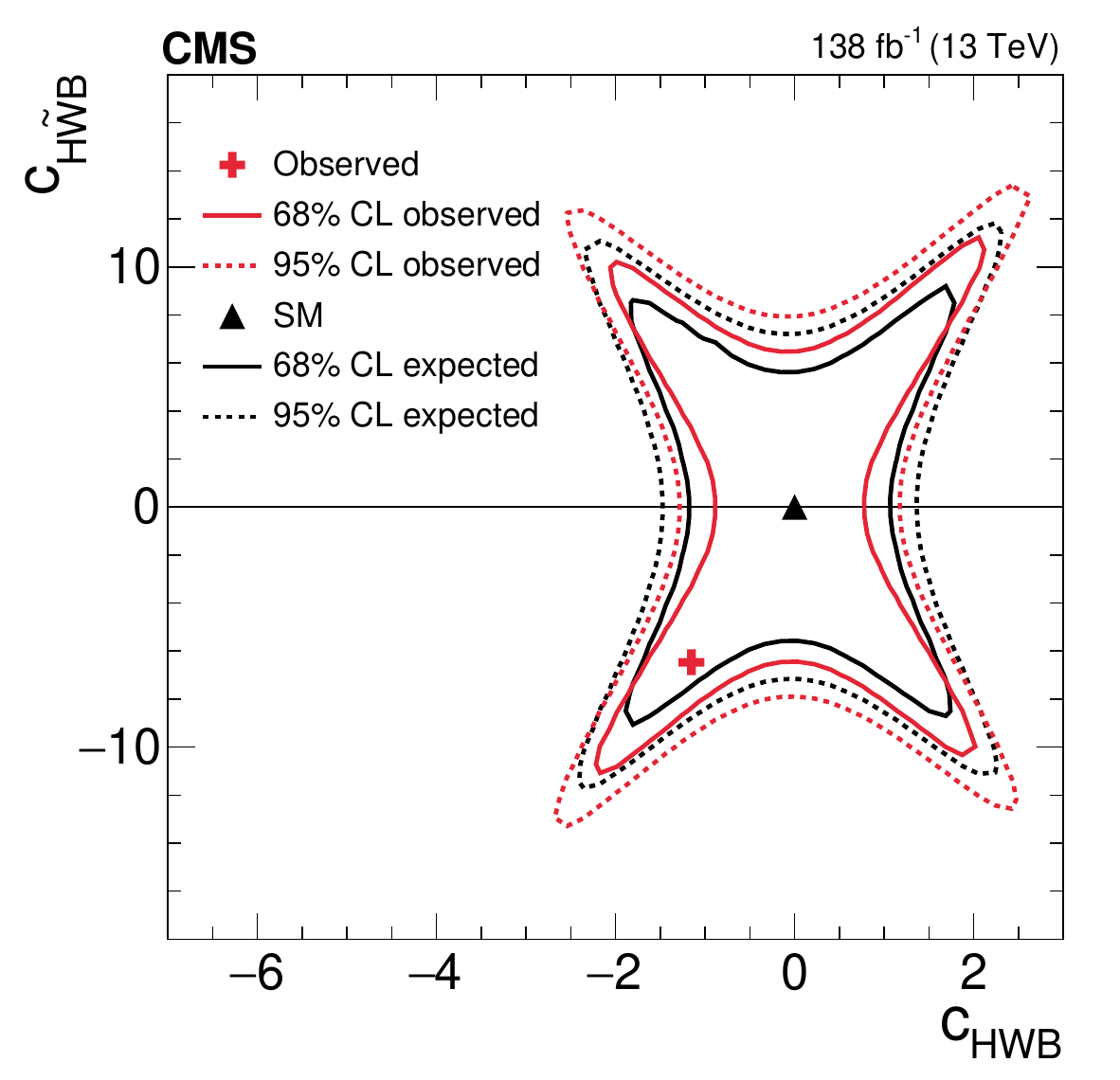}
\includegraphics[width=0.45\textwidth]{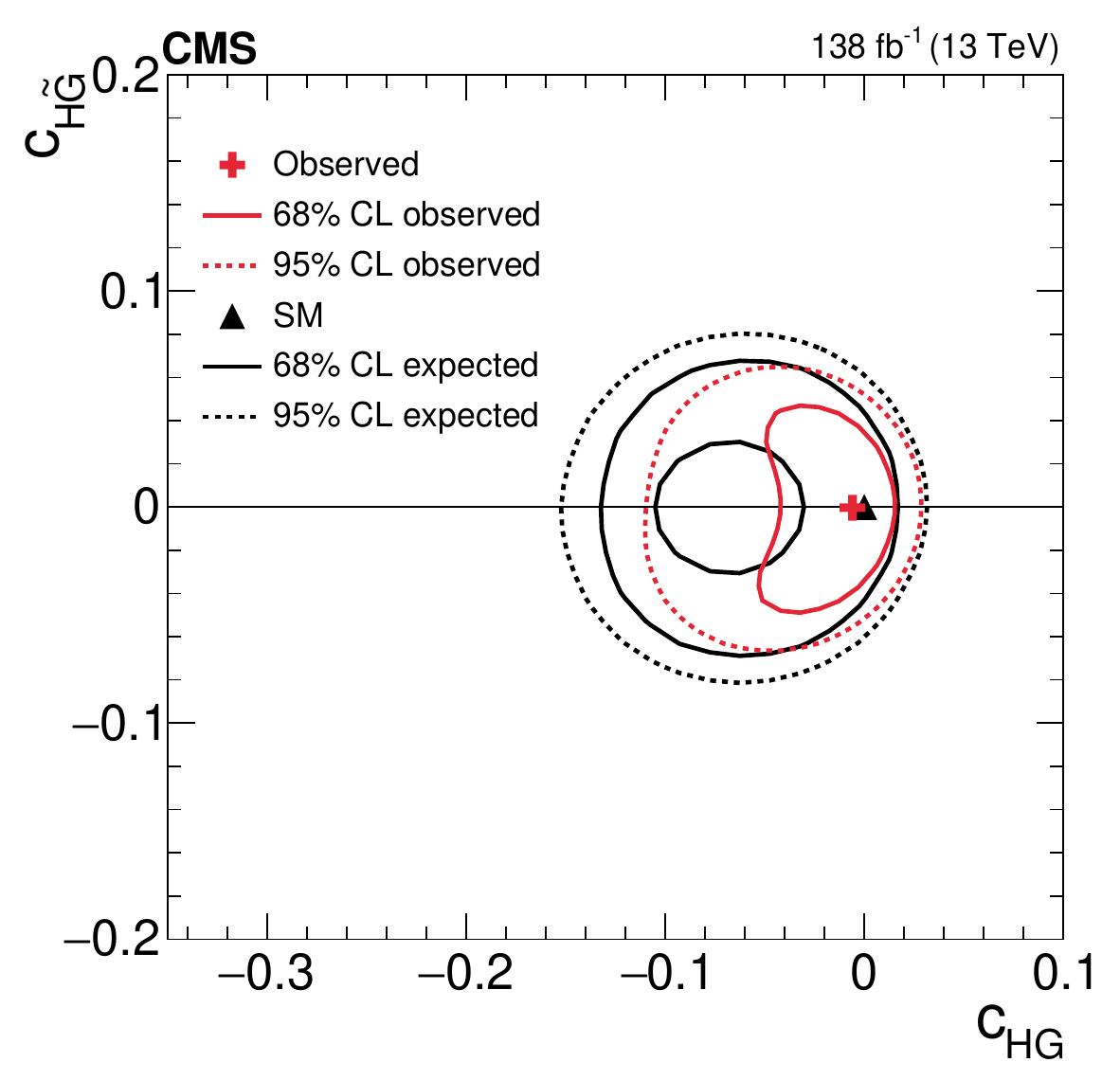} \\
\includegraphics[width=0.45\textwidth]{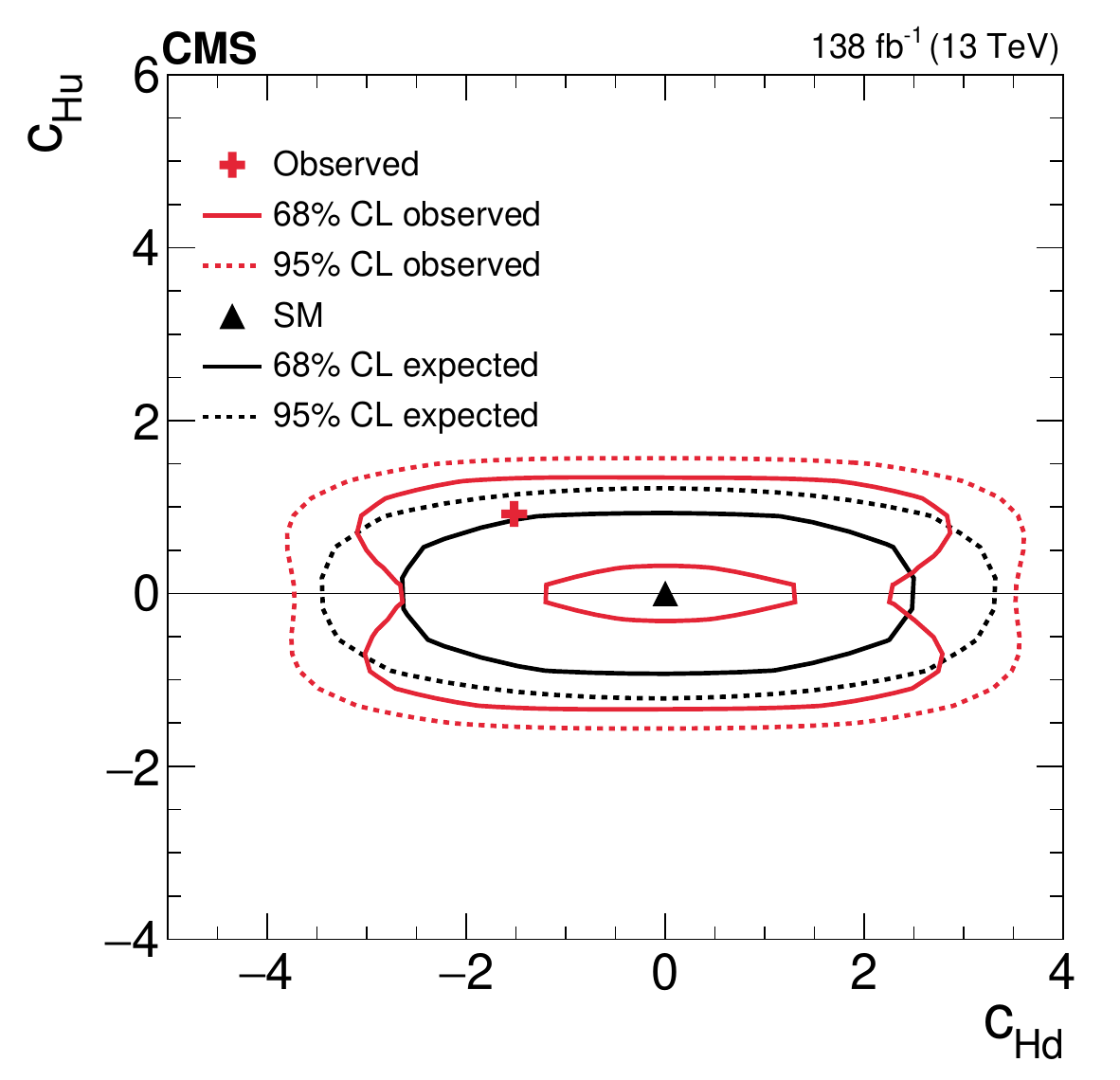}
\includegraphics[width=0.45\textwidth]{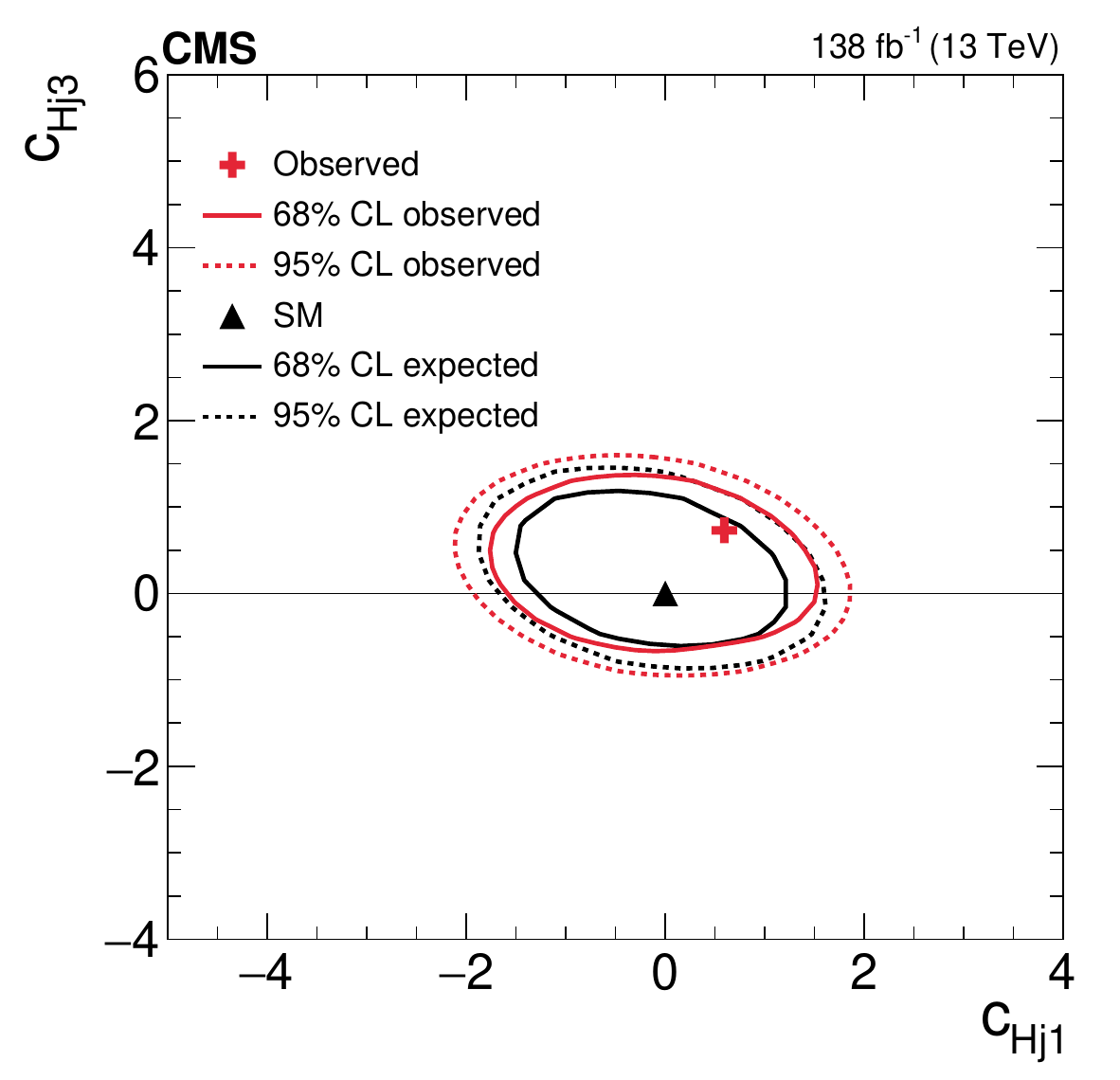}
	\caption{Two-dimensional expected and observed scans for the pairs of Wilson coefficients $c_\text{HW}$, $c_{\text{H}\tilde{\text{W}}}$ (upper left), $c_\text{HB}$, $c_{\text{H}\tilde{\text{B}}}$ (upper right), $c_\text{HWB}$, $c_{\text{H}\tilde{\text{W}}\text{B}}$ (middle left), $c_\text{HG}$, $c_{\text{H}\tilde{\text{G}}}$ (middle right), $c_{\text{Hu}}$, $c_{\text{Hd}}$ (lower left) and $c_{\text{Hj1}}$, $c_{\text{Hj3}}$ (lower right). Solid (dotted) lines correspond to the 68\% (95\%) \CL contours.}
\label{fig:scans2DExpAndObs}
\end{figure}

Sensitivity of the VBF production mechanism to the first three coefficient pairs and to the quark-related operators was assessed using \Dvbf as a fit variable, while keeping the ggH cross section fixed to its SM expectation in each \dphijj bin. On the contrary, in the fits for $c_\text{HG}$ and $c_{\text{H}\tilde{\text{G}}}$, the two-dimensional \text{\Dtwodim} distribution was used while keeping the VBF cross section at its SM expectation. When a single pair was analyzed, all other Wilson coefficients were set to their SM values, \ie, to zero. The corresponding two-dimensional constraints shown in Fig.~\ref{fig:scans2DExpAndObs} are obtained from profile likelihood contours constructed under the Wilks’ theorem approximation, and define the allowed regions for the assumed Wilson coefficients. The results are consistent with the SM at the 95\% \CL for the $(c_\text{HB},c_{\text{H}\tilde{\text{B}}})$ pair and at the 68\% \CL for all other pairs. 

\begin{figure}[!htb]
\centering
\includegraphics[width=0.475\textwidth]{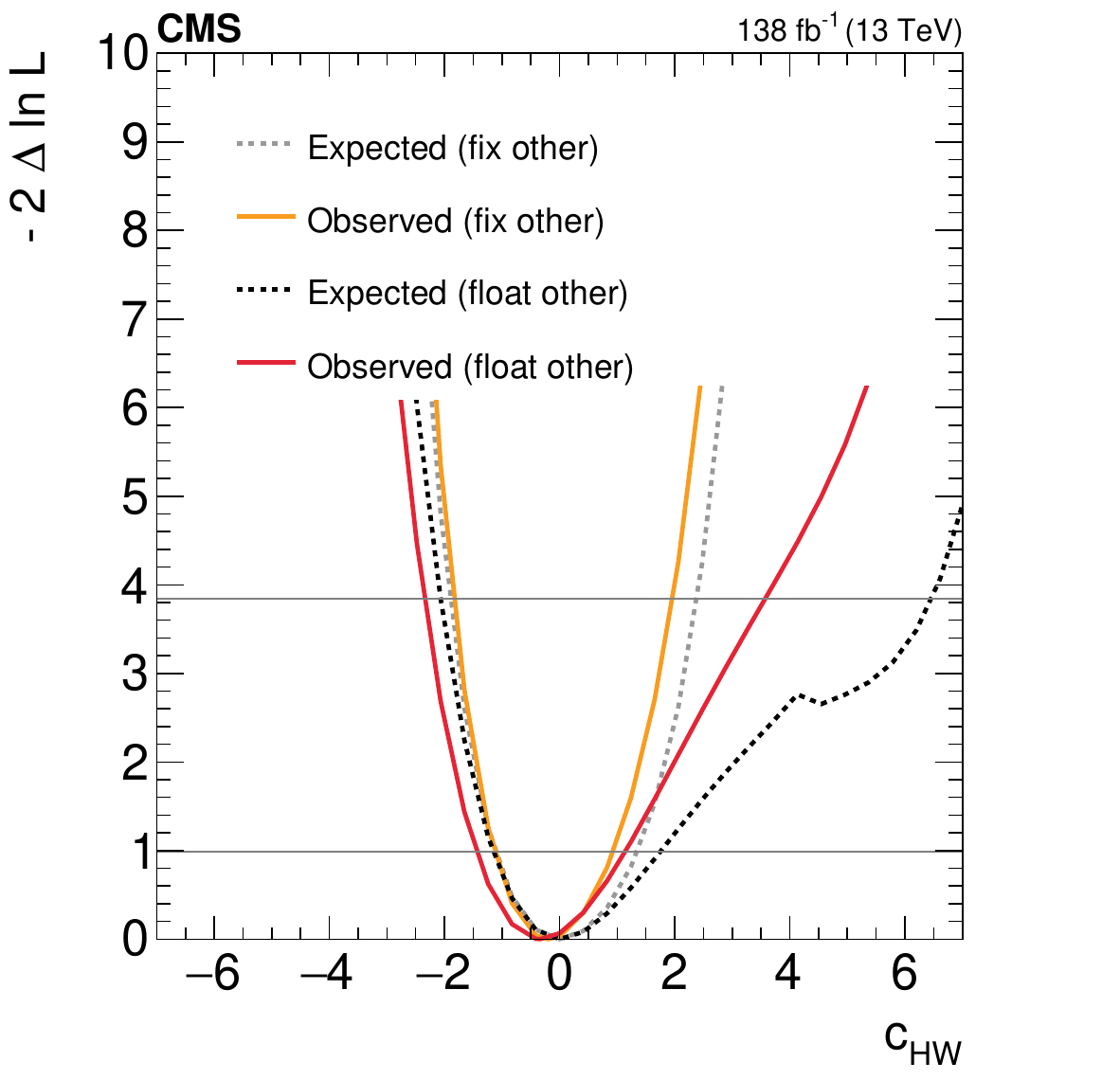}
\includegraphics[width=0.475\textwidth]{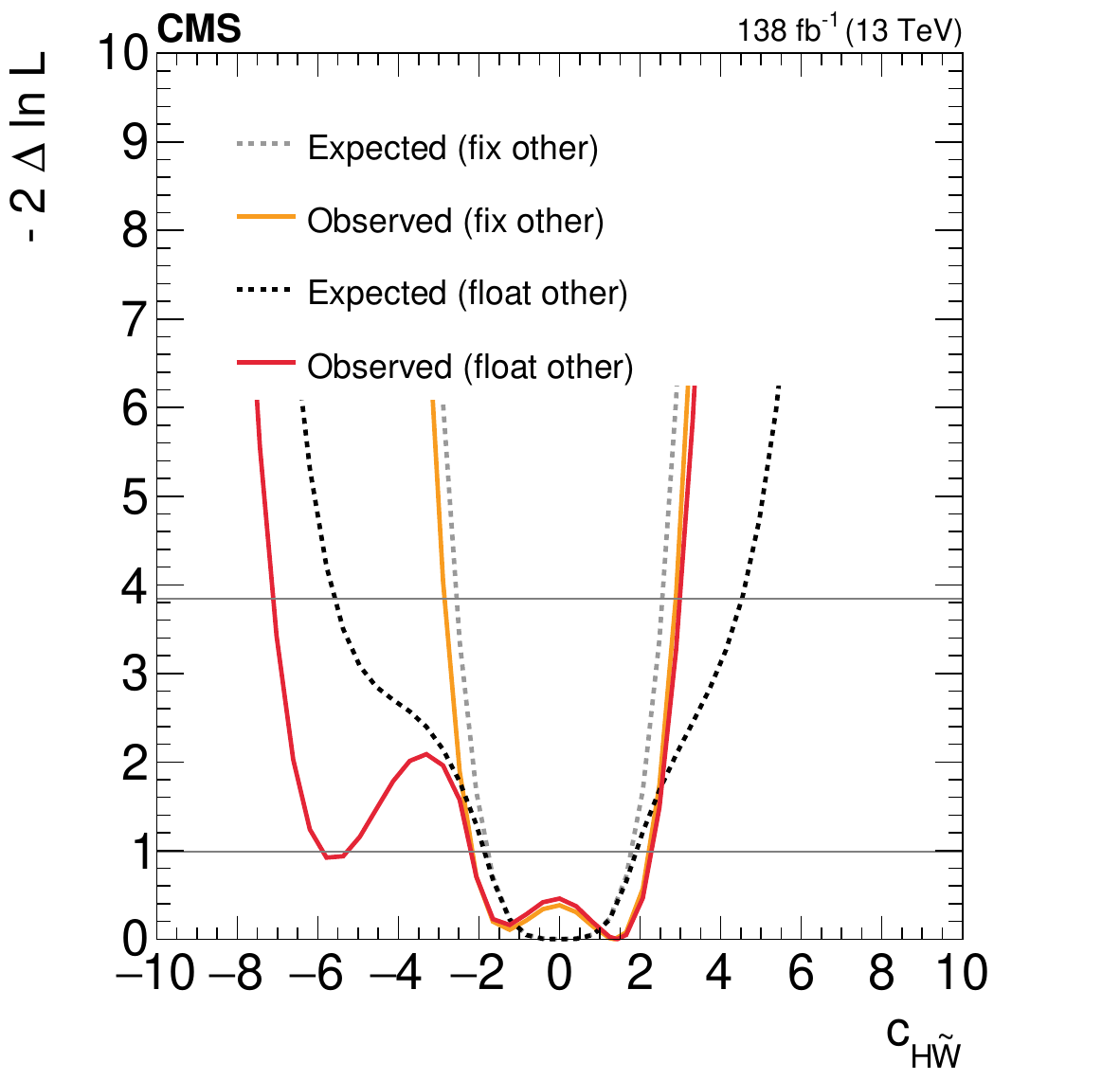} \\
\includegraphics[width=0.475\textwidth]{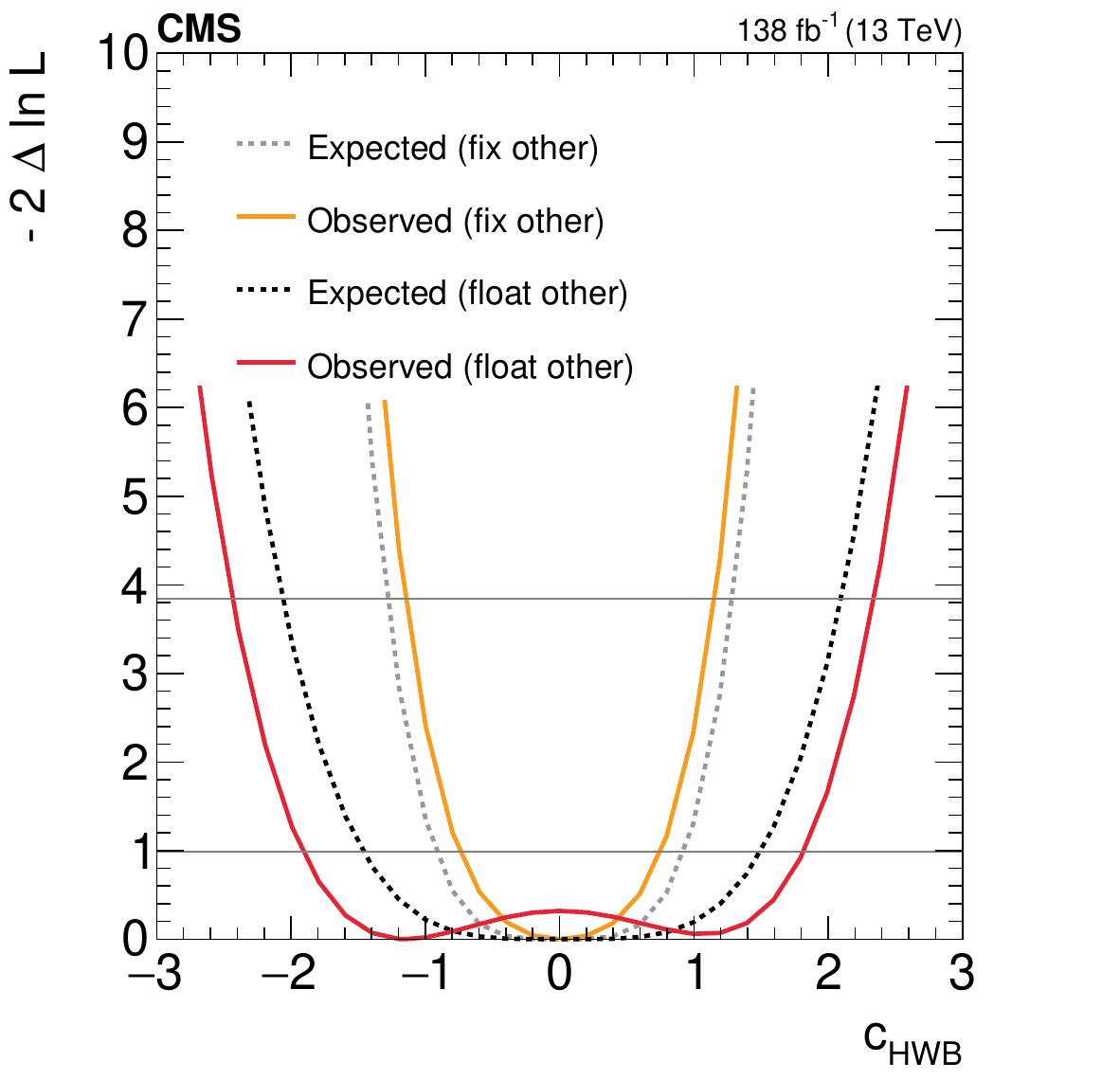}
\includegraphics[width=0.475\textwidth]{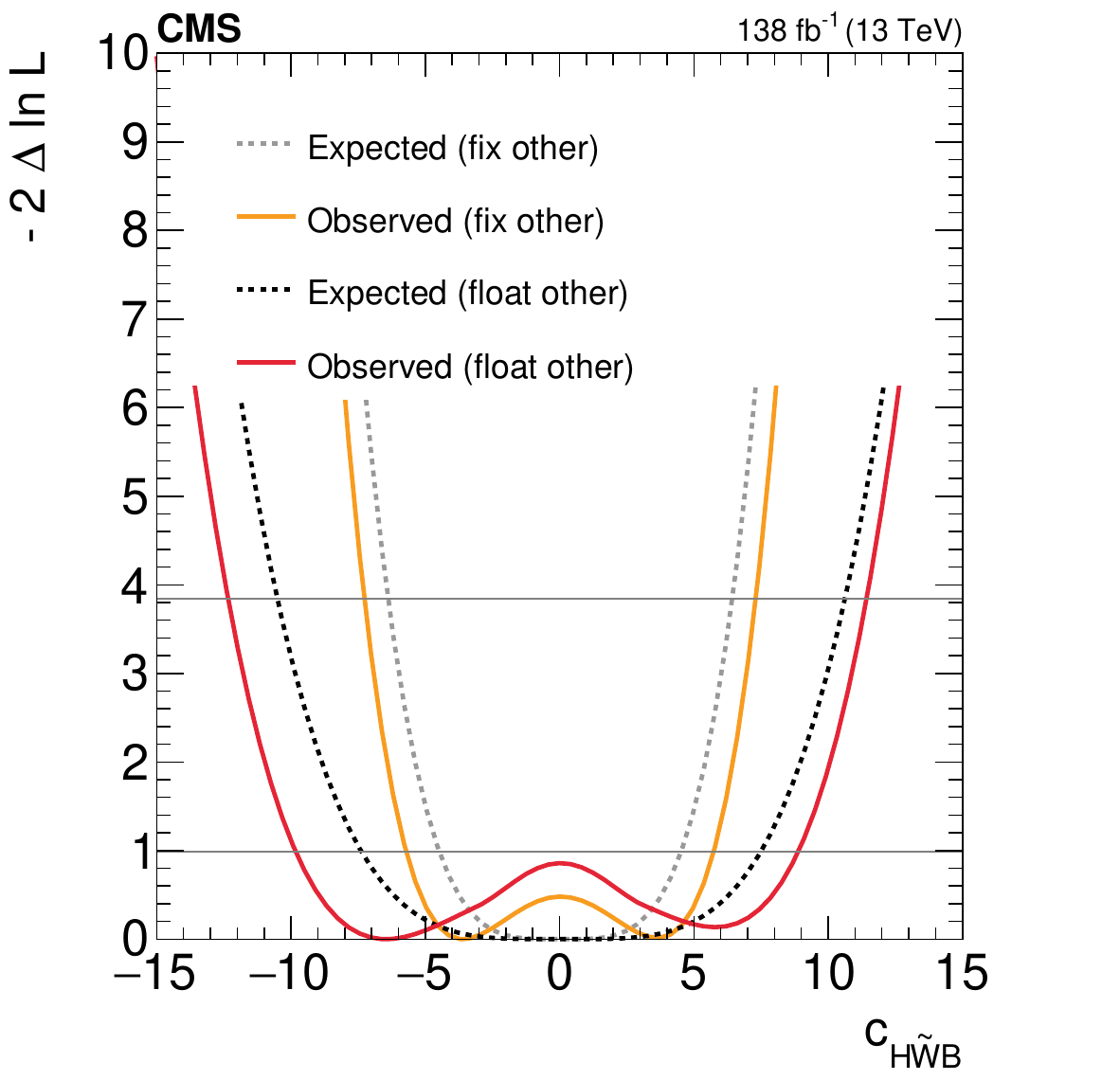} \\
\includegraphics[width=0.475\textwidth]{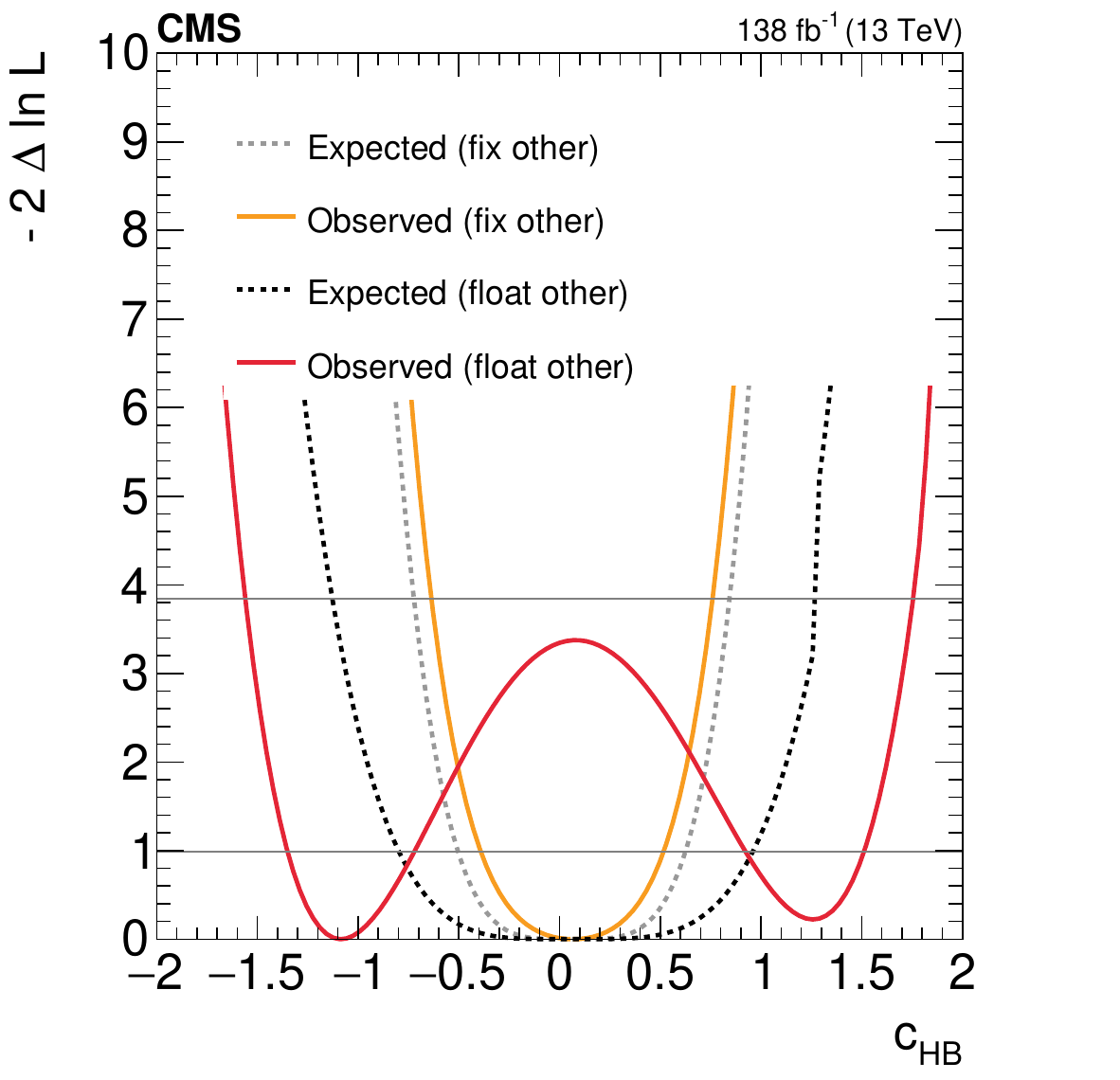}
\includegraphics[width=0.475\textwidth]{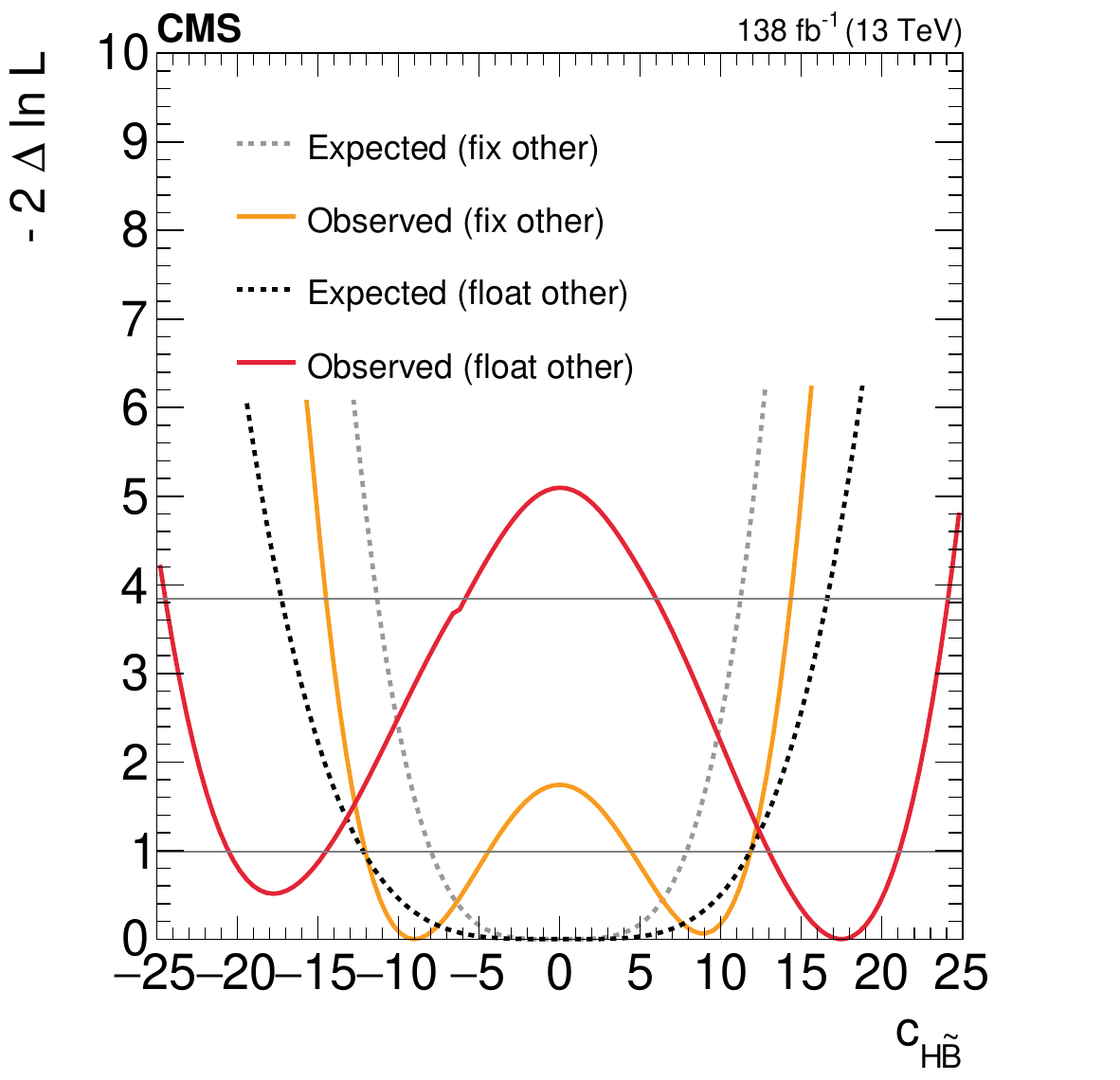}
	\caption{Expected and observed scans for the Wilson coefficients $c_\text{HW}$ (upper left), $c_\text{HWB}$ (middle left), $c_\text{HB}$ (lower left), $c_{\text{H}\tilde{\text{W}}}$ (upper right), $c_{\text{H}\tilde{\text{W}}\text{B}}$ (middle right) and $c_{\text{H}\tilde{\text{B}}}$ (lower right). The results are presented for two scenarios: one where all other coefficients are fixed to their SM values (grey) and another where the coefficient with opposite \CP-parity is allowed to float in the fit (black). Horizontal lines indicate the one-dimensional 68\% and 95\% \CL values.}
\label{fig:scans1DVBF}
\end{figure}

\begin{figure}[!htb]
\centering
\includegraphics[width=0.475\textwidth]{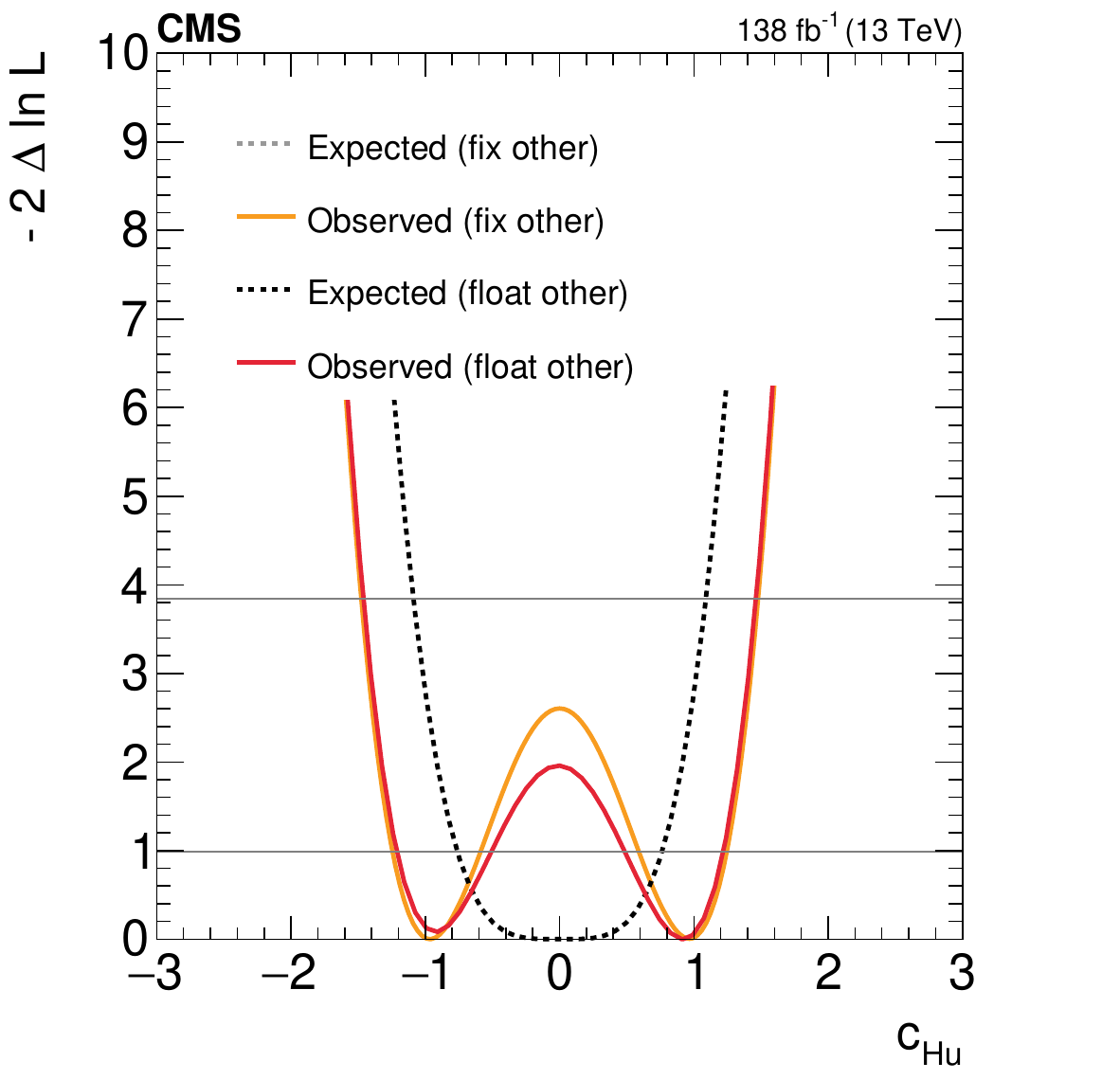}
\includegraphics[width=0.475\textwidth]{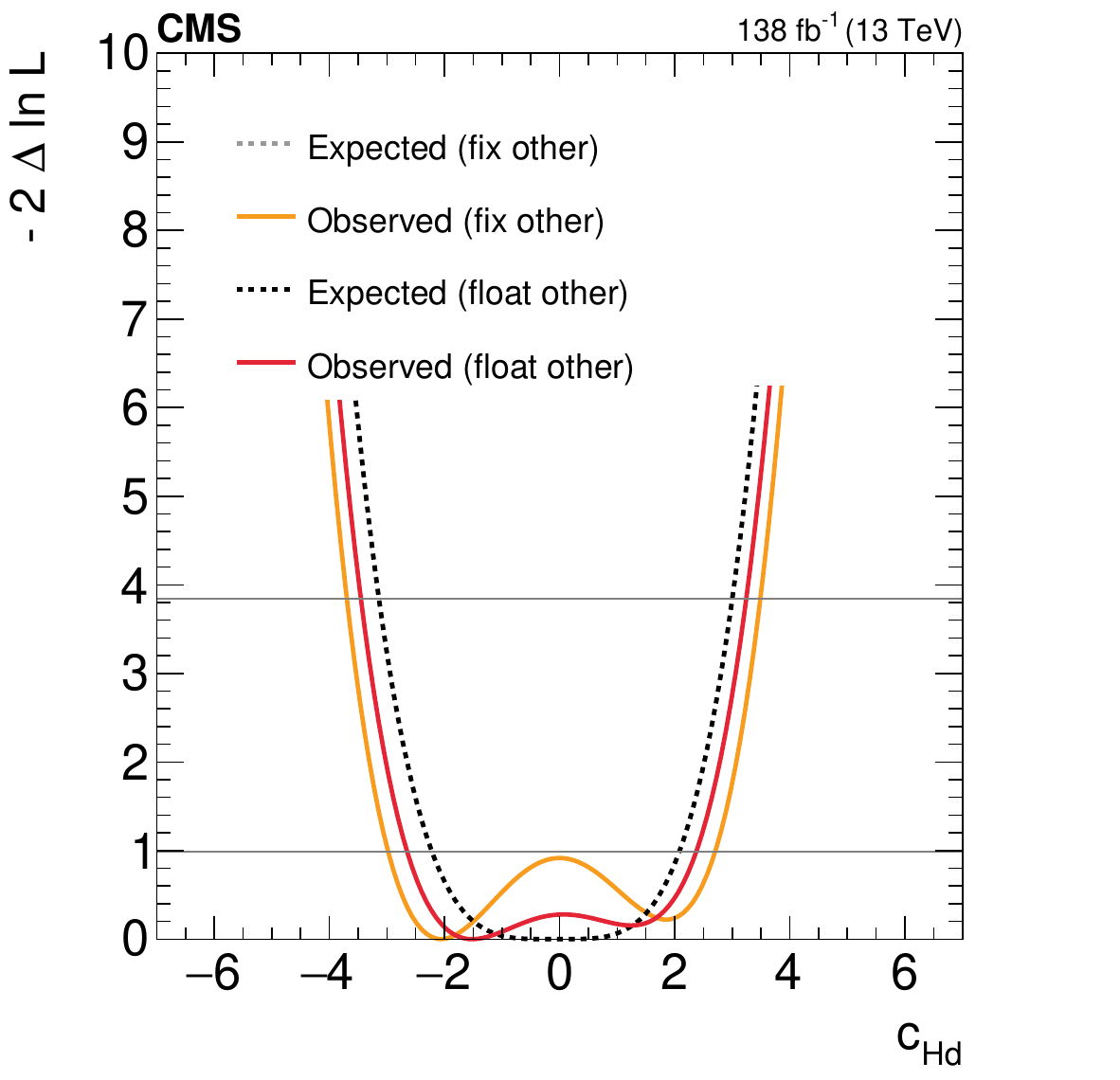} \\
\includegraphics[width=0.475\textwidth]{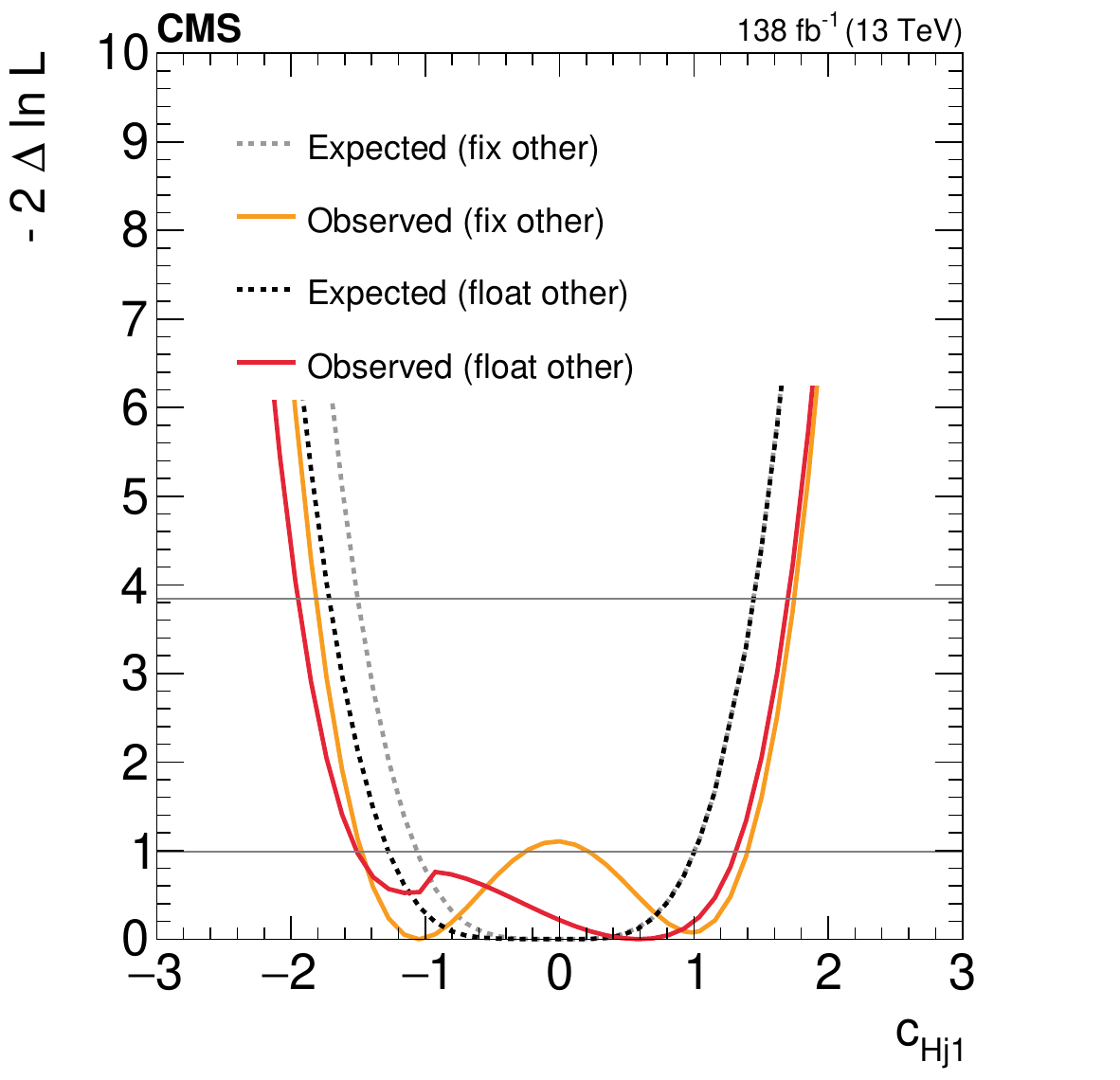}
\includegraphics[width=0.475\textwidth]{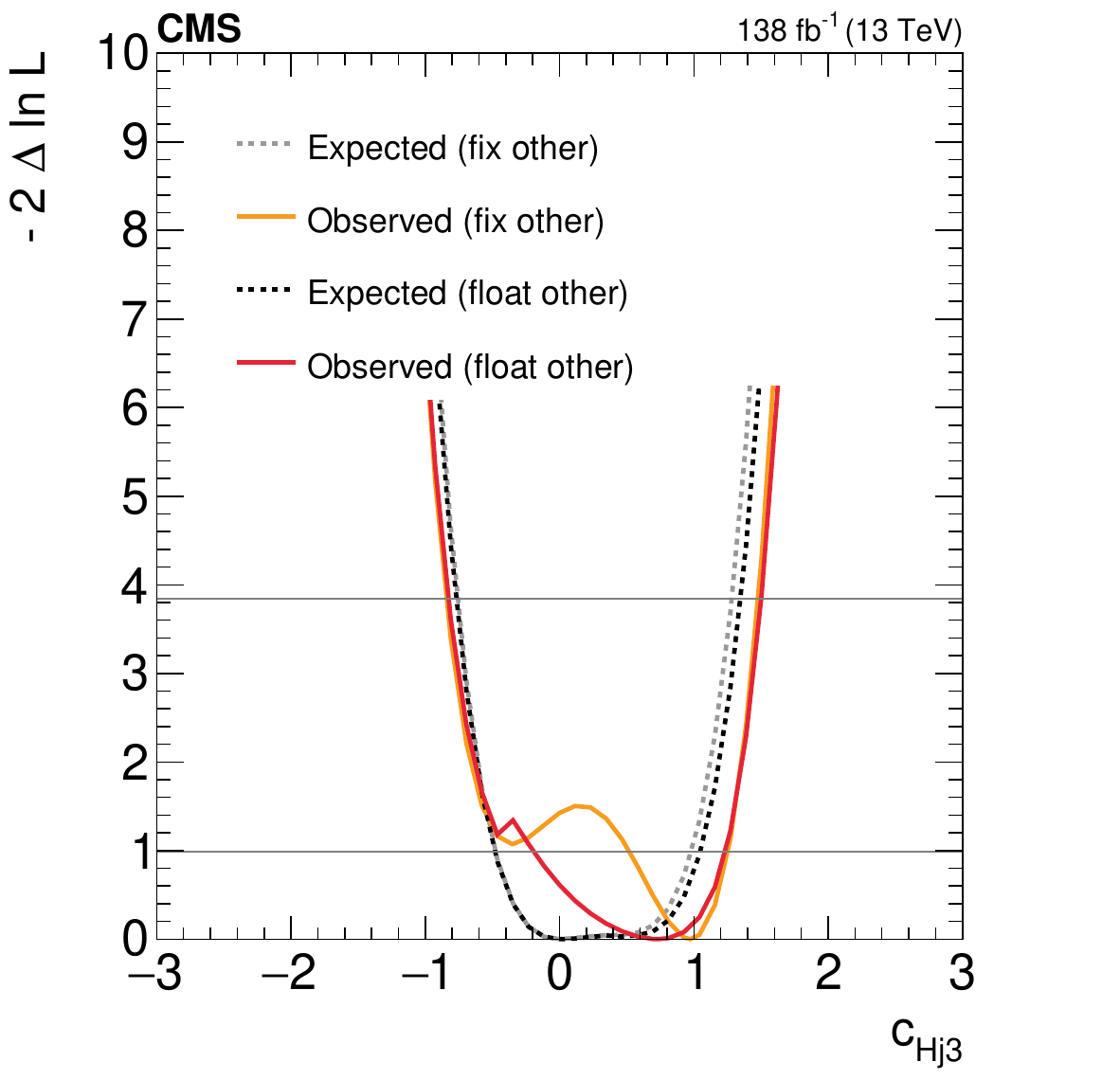}
	\caption{Expected and observed scans for the Wilson coefficients $c_\text{Hu}$ (upper left), $c_{\text{Hd}}$ (upper right), $c_{\text{Hj1}}$ (lower left) and $c_{\text{Hj3}}$ (lower right). The results are presented for two scenarios: one where all other coefficients are fixed to their SM values (grey) and another where the other coefficient from the $(c_\text{Hu},c_{\text{Hd}})$ and $(c_\text{Hj1},c_{\text{Hj3}})$ pairs, respectively, is allowed to float in the fit (black). Horizontal lines indicate the one-dimensional 68\% and 95\% \CL values.}
\label{fig:scans1DVBFHqq}
\end{figure}

\begin{figure}[!htb]
\centering
\includegraphics[width=0.475\textwidth]{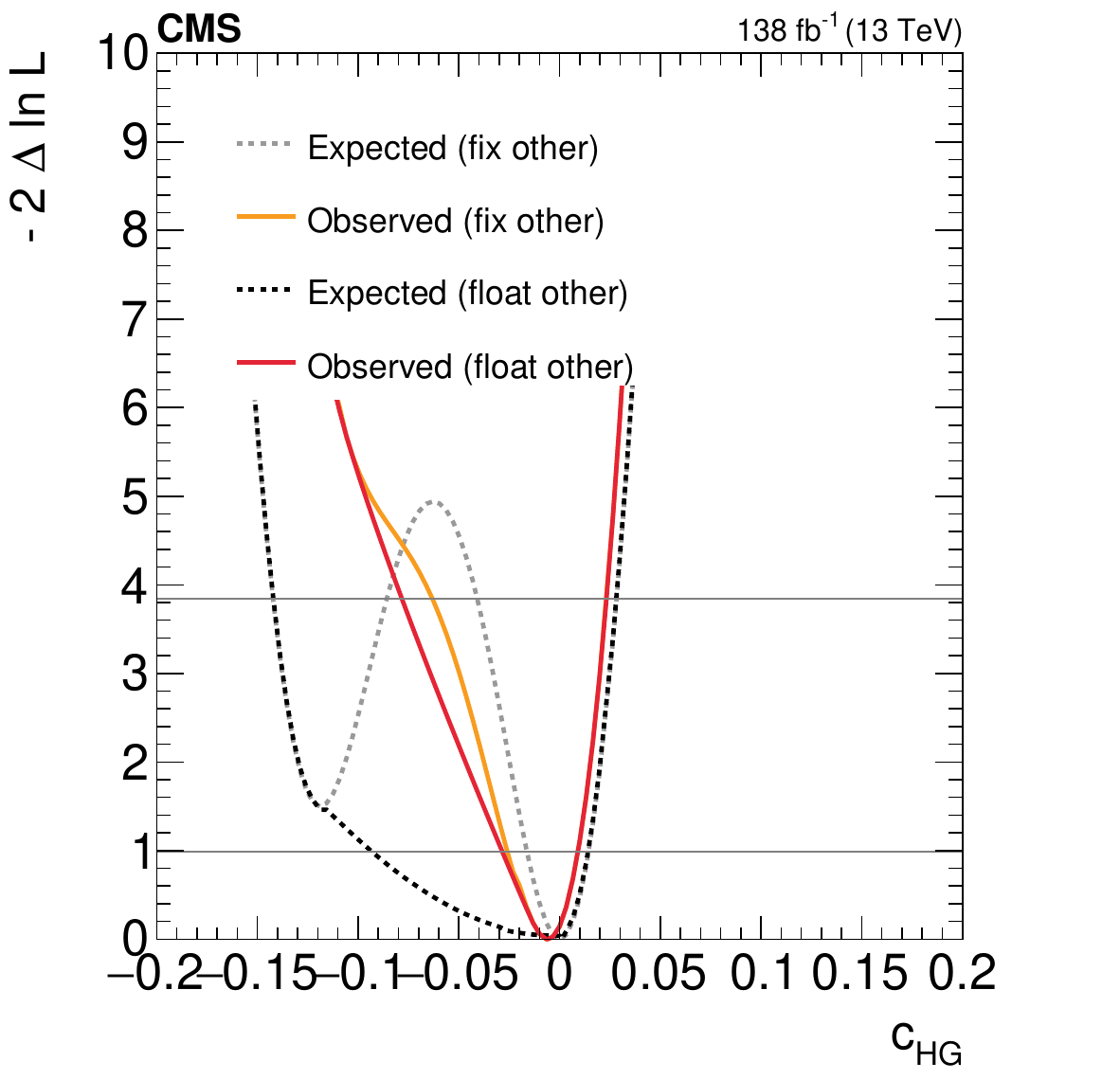}
\includegraphics[width=0.475\textwidth]{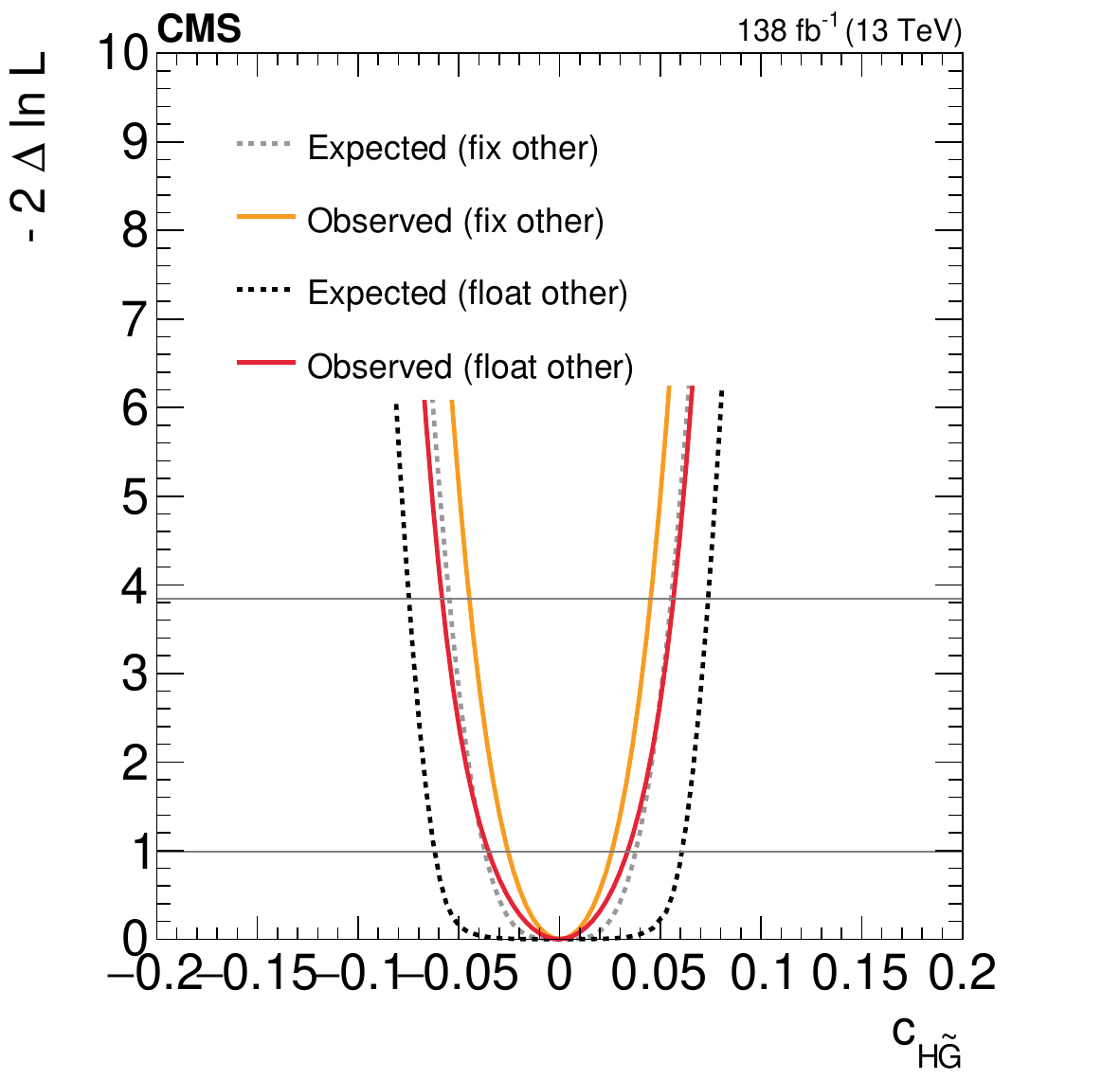} 
	\caption{Expected and observed scans for the Wilson coefficients $c_\text{HG}$ (left) and $c_{\text{H}\tilde{\text{G}}}$ (right). The results are presented for two scenarios: one where all other coefficients are fixed to their SM values (grey) and another where the coefficient with opposite \CP-parity is allowed to float in the fit (black). Horizontal lines indicate the one-dimensional 68\% and 95\% \CL values.}
\label{fig:scans1DggH}
\end{figure}

Additionally, individual profile likelihood fits have been performed for each Wilson coefficient, both with all other coefficients fixed to their SM values (``fix other") and with one additional coefficient allowed to float (``float other"). In the latter case, the choice of the second floating coefficient follows the pairing logic used in the two-dimensional simultaneous fits. For $c_{\text{H}\Box}$ and $c_\text{HD}$, the approach with all other coefficients fixed to zero is used. The corresponding one-dimensional constraints, defined as the ranges allowed by the profile likelihood scan, are shown in Figs.~\ref{fig:scans1DVBF}--~\ref{fig:scans1DVBFHqq} for the fit utilizing $\mathcal{D}_\mathrm{VBF}$, and in Fig.~\ref{fig:scans1DggH} for the fit based on the two-dimensional \Dtwodim distribution. Best fit values and their confidence intervals, summarized in Table~\ref{tab:SMEFTobserved}, were obtained from the individual fits using the scenario where the other Wilson coefficient is allowed to float, except for $c_{\text{H}\Box}$ and $c_\text{HD}$ where the ``fix other" scenario is employed. The observed significance with respect to SM scenario is also quoted. The individual fits reported above are in agreement with the two-dimensional constraints shown in Fig.~\ref{fig:scans2DExpAndObs}. Notably, an apparent excess of 2 s.d. observed in the ``float other" scenario of the profiled likelihood scan for the $c_{\text{H}\tilde{\text{B}}}$ coefficient in Fig.~\ref{fig:scans1DVBF} (lower right) is consistent with the result shown in Fig.~\ref{fig:scans2DExpAndObs} (upper right), where the two-dimensional constraints correspond to the quantiles of a $\chi^2$ distribution with 2 degrees of freedom.  

\begin{figure}[!htb]
\centering
\includegraphics[width=0.5\textwidth]{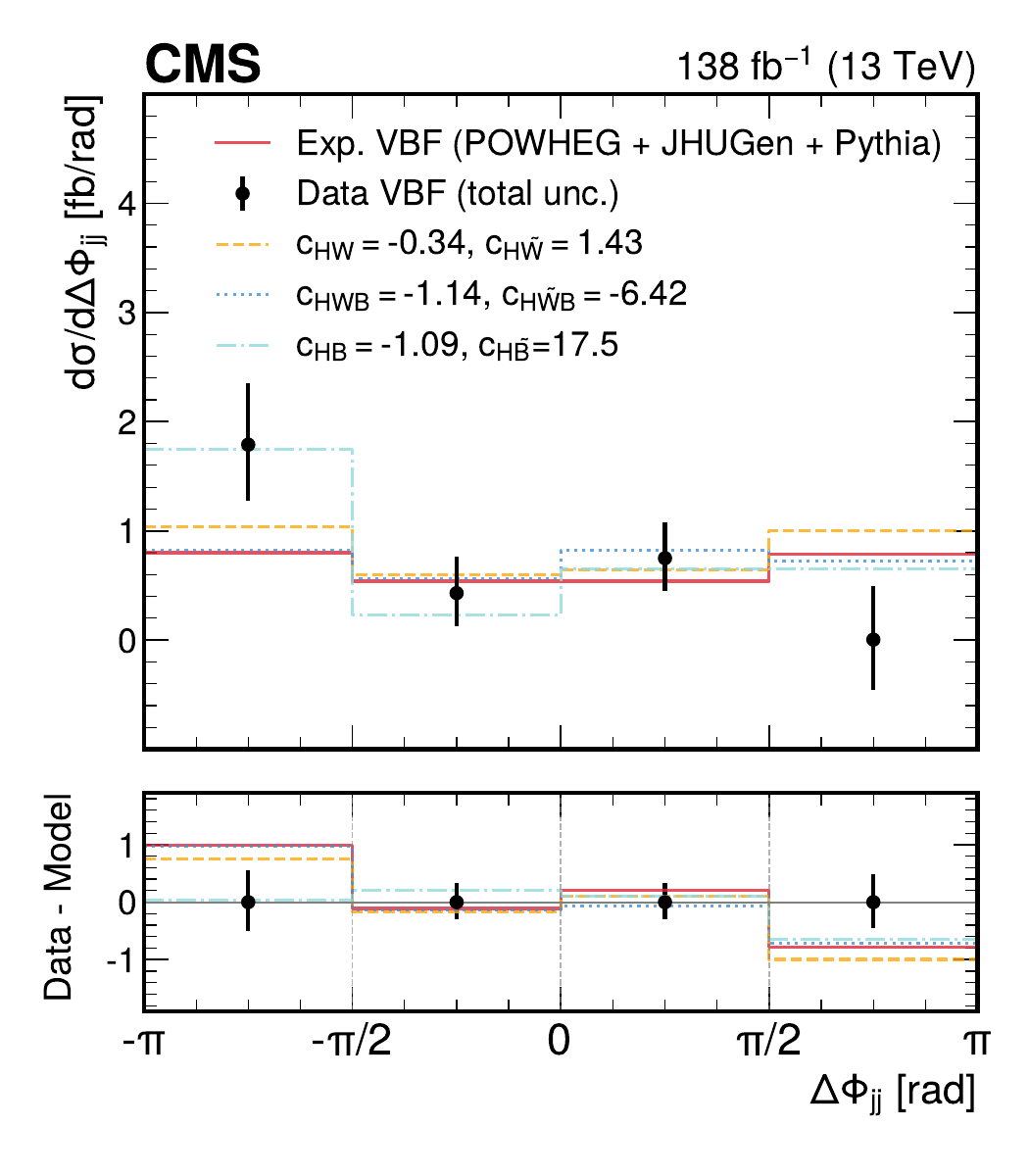}
	\caption{Measured fiducial cross section for VBF production as a function of \dphijj (black) compared to various predictions. The cross section predictions include: the SM (red), the ones obtained from the best fit of Wilson coefficients of $c_\text{HW}$, $c_{\text{H}\tilde{\text{W}}}$ (yellow), $c_\text{HWB}$, $c_{\text{H}\tilde{\text{W}}\text{B}}$ (blue) and $c_\text{HB}$, $c_{\text{H}\tilde{\text{B}}}$ (green). The difference between the data and the predictions are displayed in the lower panel.}
\label{fig:xsection_SMEFT_VBF}
\end{figure}

\begin{figure}[!htb]
\centering
\includegraphics[width=0.5\textwidth]{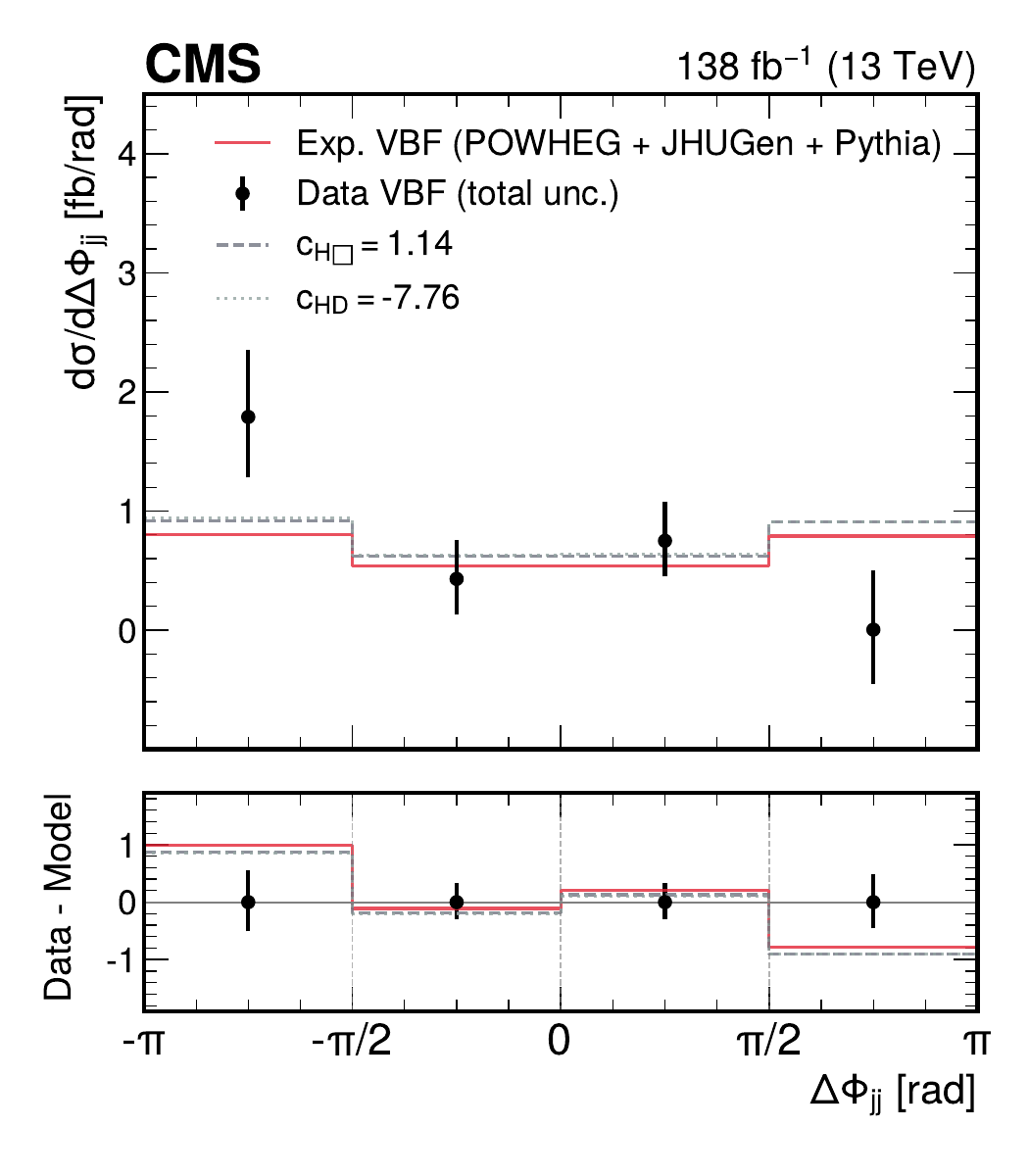}
	\caption{Measured fiducial cross section for VBF production as a function of \dphijj (black) compared to various predictions. The cross section predictions include: the SM (red), the ones obtained from the best fit of Wilson coefficients of $c_{\text{H}\Box}$ (dark grey) and $c_\text{HD}$ (light grey). The difference between the data and the predictions are displayed in the lower panel.}
\label{fig:xsection_SMEFT_VBFkin}
\end{figure}

\begin{figure}[!htb]
\centering
\includegraphics[width=0.5\textwidth]{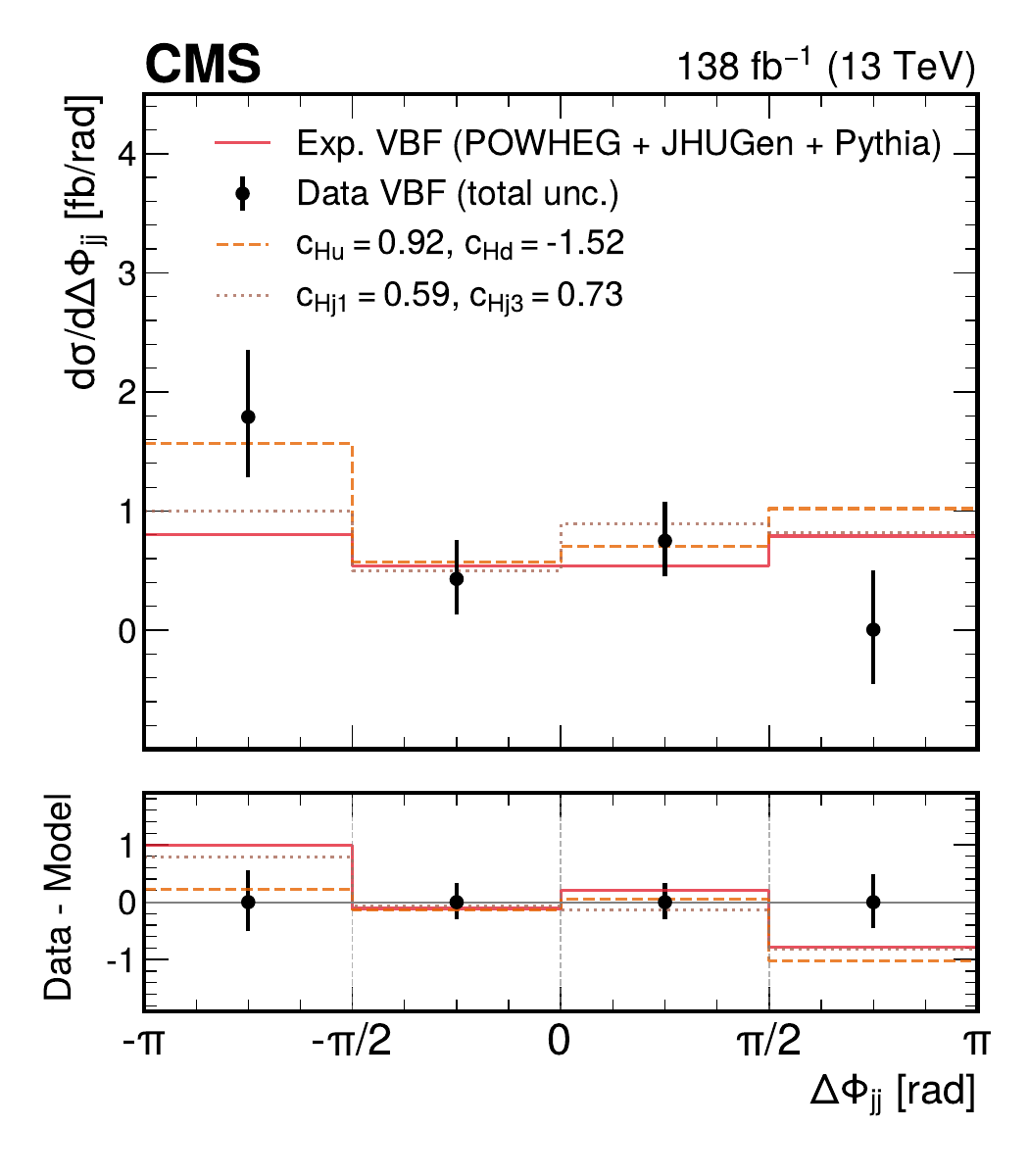}
        \caption{Measured fiducial cross section for VBF production as a function of \dphijj (black) compared to various predictions. The cross section predictions include: the SM (red), the ones obtained from the best fit of Wilson coefficients of $c_{\text{Hu}}$, $c_{\text{Hd}}$ (light orange) and $c_\text{Hj1}$, $c_{\text{Hj3}}$ (dark orange). The difference between the data and the predictions are displayed in the lower panel.}
\label{fig:xsection_SMEFT_VBFHqq}
\end{figure}

\begin{figure}[!htb]
\centering
\includegraphics[width=0.5\textwidth]{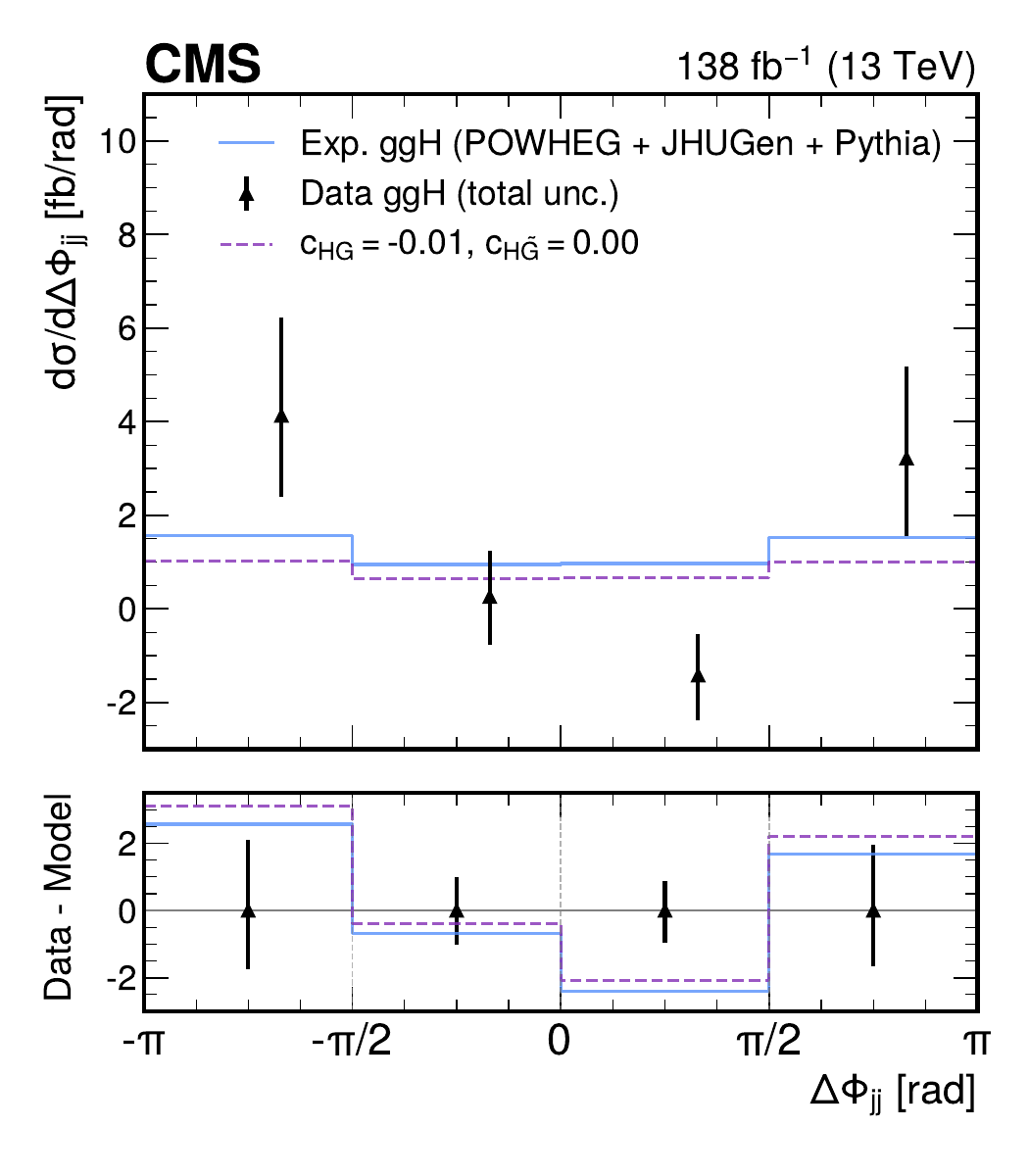}
	\caption{Measured fiducial cross section for ggH production as a function of \dphijj (black) compared to various predictions. The cross section predictions include: the SM (blue) and the ones obtained from the best fit of Wilson coefficients of $c_\text{HG}$, $c_{\text{H}\tilde{\text{G}}}$ (violet). The difference between the data and the predictions are displayed in the lower panel.}
\label{fig:xsection_SMEFT_ggH}
\end{figure}

The differential cross section was recomputed using the observed best fit results from Table~\ref{tab:SMEFTobserved} and compared with the observed results presented in Section~\ref{sec:results}. In Figs.~\ref{fig:xsection_SMEFT_VBF}--\ref{fig:xsection_SMEFT_VBFHqq}, the measured fiducial cross section for VBF production, obtained from the fit to the \text{\Dvbf} distribution and shown as a black circular marker, is compared to various models, including the SM (solid line) and several BSM scenarios (dotted or dashed lines). These BSM interpretations of the cross section correspond to the measured best fit values of the Wilson coefficients within the assumed model, obtained using the same data. Similarly, in Fig.~\ref{fig:xsection_SMEFT_ggH}, the measured fiducial cross section for the ggH process, obtained from the two-dimensional \Dtwodim distribution and depicted as a black triangular marker, is compared to the SM prediction (solid line) and to a scenario where $c_\text{HG}$ and $c_{\text{H}\tilde{\text{G}}}$ are simultaneously fitted to the data. The lower panels of these figures display the difference between the differential cross section reported in Section~\ref{sec:results} and the corresponding predictions from the SM and from the various BSM models as measured in data.

\begin{figure}[!htb]
\centering
\includegraphics[width=0.475\textwidth]{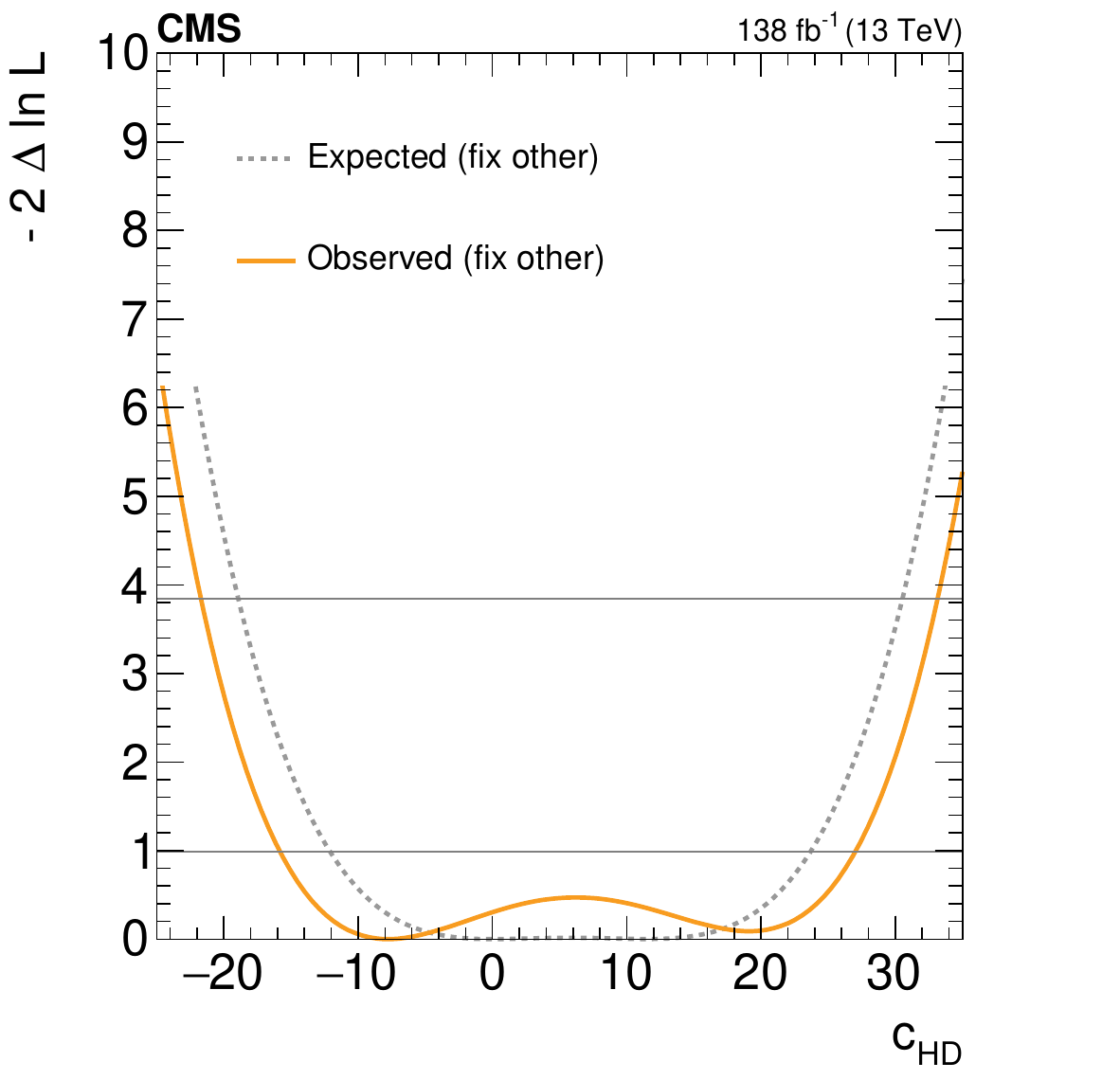} 
\includegraphics[width=0.475\textwidth]{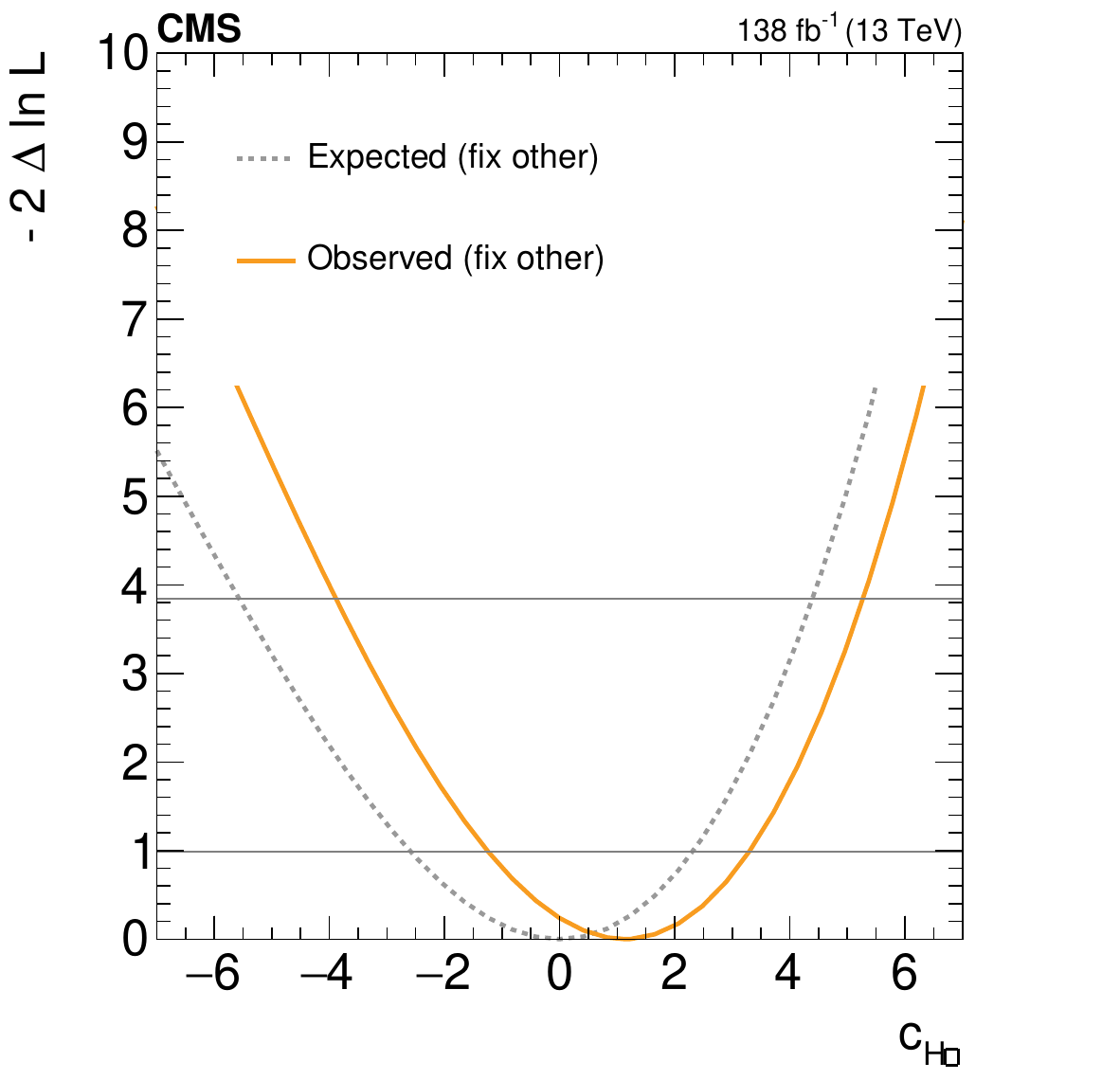}
	\caption{Expected and observed scans for the Wilson coefficients $c_\text{HD}$ (left) and $c_{\text{H}\Box}$ (right). The results are presented for the scenario where all other coefficients are fixed to their SM values. Horizontal lines indicate the one-dimensional 68\% and 95\% \CL values.}
\label{fig:scans1DVBFkin}
\end{figure}

\begin{table}[!htb]
    \centering
	\topcaption{Summary of the constraints on Wilson coefficients, including best fit values, 68\% and 95\% \CL intervals. The observed significance with respect to the SM scenario is shown in the last column. The constraints on $c_{\text{H}\Box}$ and $c_\text{HD}$ were obtained from individual fits with all other coefficients fixed to their SM values. For the remaining coefficients, results were obtained from fits where the corresponding \CP-even or \CP-odd partner was allowed to float, while all other coefficients were fixed to their SM values.}
    \resizebox*{!}{\dimexpr\textheight-3cm\relax}{%
    \begin{tabular}{ccccc}
	    Wilson coefficients & & Observed & Expected & Significance (s.d.) \\
        \hline
	    $c_\text{HW}$ & Best fit & -0.34 & 0.00 & 0.0 \\
	    & 68\% \CL & [-1.46,1.15] & [-1.17,1.76] & \\
	    & 95\% \CL & [-2.35,3.57] & [-2.07,6.50] & \\
        [\cmsTabSkip]
	    $c_\text{HWB}$ & Best fit & -1.14 & 0.00 & 0.5 \\
	    & 68\% \CL & [-1.92,1.82] & [-1.47,1.50] & \\
	    & 95\% \CL & [-2.44,2.34] & [-2.06,2.10] & \\
        [\cmsTabSkip]
	    $c_\text{HB}$ & Best fit & -1.09 & 0.00 & 1.7 \\
	    & 68\% \CL & [-1.35,-0.72] $\cup$ [0.92,1.51] & [-0.80,0.96] & \\
	    & 95\% \CL & [-1.56,1.75] & [-1.13,1.27] & \\
        [\cmsTabSkip]
	    $c_\text{HG}$ & Best fit & -0.01 & 0.00 & 0.0 \\
	    & 68\% \CL & [-0.03,0.01] & [-0.10,0.02] & \\
	    & 95\% \CL & [-0.08,0.02] & [-0.14,0.03] & \\
        [\cmsTabSkip]
	    $c_{\text{H}\tilde{\text{W}}}$ & Best fit & 1.43 & 0.00 & 0.7 \\
	    & 68\% \CL & [-5.95,-5.22] $\cup$ [-2.21,2.32] & [-1.87,1.91] & \\
	    & 95\% \CL & [-7.12,3.00] & [-5.60,4.53] & \\
        [\cmsTabSkip]
	    $c_{\text{H}\tilde{\text{W}}\text{B}}$ & Best fit & -6.42 & 0.00 & 0.9 \\
	    & 68\% \CL & [-9.84,8.90] & [-7.40,7.52] & \\
	    & 95\% \CL & [-12.3,11.4] & [-10.51,10.62] & \\
        [\cmsTabSkip]
	    $c_{\text{H}\tilde{\text{B}}}$ & Best fit & 17.5 & 0.00 & 2.3 \\
	    & 68\% \CL & [-20.6,-14.5] $\cup$ [13.0,21.1] & [-12.2,11.9] & \\
	    & 95\% \CL & [-24.5,-5.9] $\cup$ [6.0,24.1] & [-17.3,16.6] & \\
	[\cmsTabSkip]   
	    $c_{\text{H}\tilde{\text{G}}}$ & Best fit & 0.00 & 0.00 & 0.0 \\
	    & 68\% \CL & [-0.04,0.03] & [-0.06,0.06] & \\
	    & 95\% \CL & [-0.06,0.06] & [-0.08,0.07] & \\
	[\cmsTabSkip]
        $c_{\text{H}\Box}$ & Best fit & 1.14 & 0.00 & 0.5 \\
        & 68\% \CL & [-1.26,3.31] & [-2.60,2.33] & \\
        & 95\% \CL & [-3.88,5.27] & [-5.56,4.39] & \\
	[\cmsTabSkip]    
        $c_\text{HD}$ & Best fit & -7.76 & 0.00 & 0.0 \\
        & 68\% \CL & [-15.8,27.0] & [-12.1,23.7] & \\
        & 95\% \CL & [-21.7,33.1] & [-19.0,30.5] & \\
	[\cmsTabSkip]
	    $c_{\text{Hu}}$ & Best fit & 0.92 & 0.00 & 1.4 \\
        & 68\% \CL & [-1.21,-0.51] $\cup$ [0.49,1.22] & [-0.76,0.76] & \\
        & 95\% \CL & [-1.46,1.46] & [-1.09,1.09] & \\
	[\cmsTabSkip]    
	    $c_{\text{Hd}}$ & Best fit & -1.52 & 0.00 & 0.4 \\
        & 68\% \CL & [-2.65,2.38] & [-2.22,2.08] & \\
        & 95\% \CL & [-3.45,3.24] & [-3.14,3.00] & \\
	[\cmsTabSkip]    
	    $c_{\text{Hj1}}$ & Best fit & 0.59 & 0.00 & 1.0 \\
        & 68\% \CL & [-1.51,1.32] & [-1.28,1.01] & \\
        & 95\% \CL & [-1.95,1.70] & [-1.72,1.45] & \\
	[\cmsTabSkip]    
	    $c_{\text{Hj3}}$ & Best fit & 0.73 & 0.00 & 1.2 \\
        & 68\% \CL & [-0.20,1.24] & [-0.48,1.05] & \\
        & 95\% \CL & [-0.83,1.50] & [-0.77,1.35] & \\
    \end{tabular}
    }
    \label{tab:SMEFTobserved}
\end{table}

\clearpage

\section{Summary}\label{sec:conclusions}

This paper presents a model-independent measurement of the Higgs boson differential production cross section in its decay to a pair of W bosons, with a final state that includes two jets, two different-flavor leptons $(\Pe\PGm)$, and missing transverse momentum. The model independence of the measurement is maximized by employing a discriminating variable, developed through machine learning, that is agnostic to the signal hypothesis. The measurement is based on proton-proton collision data recorded with the CMS detector between 2016 and 2018, corresponding to an integrated luminosity of 138\fbinv at a center-of-mass energy of 13\TeV. The production cross section is measured as a function of the azimuthal angle difference between the two jets. Three different signal extraction configurations are employed to measure the Higgs boson production cross section in association with two jets via vector boson fusion (VBF) and gluon-gluon fusion (ggH). Differential cross section measurements are further utilized to constrain Wilson coefficients within the standard model effective field theory framework. The most stringent constraints are obtained on the Charge conjugation Parity (\CP)-even $c_\text{HW}$ and $c_\text{Hj3}$ coefficients from the VBF cross section measurement, and on the \CP-even $c_\text{HG}$ coefficient from the ggH cross section measurement. All results are found to be consistent with the SM expectations.

\begin{acknowledgments}
We congratulate our colleagues in the CERN accelerator departments for the excellent performance of the LHC and thank the technical and administrative staffs at CERN and at other CMS institutes for their contributions to the success of the CMS effort. In addition, we gratefully acknowledge the computing centers and personnel of the Worldwide LHC Computing Grid and other centers for delivering so effectively the computing infrastructure essential to our analyses. Finally, we acknowledge the enduring support for the construction and operation of the LHC, the CMS detector, and the supporting computing infrastructure provided by the following funding agencies: SC (Armenia), BMBWF and FWF (Austria); FNRS and FWO (Belgium); CNPq, CAPES, FAPERJ, FAPERGS, and FAPESP (Brazil); MES and BNSF (Bulgaria); CERN; CAS, MoST, and NSFC (China); MINCIENCIAS (Colombia); MSES and CSF (Croatia); RIF (Cyprus); SENESCYT (Ecuador); ERC PRG, TARISTU24-TK10 and MoER TK202 (Estonia); Academy of Finland, MEC, and HIP (Finland); CEA and CNRS/IN2P3 (France); SRNSF (Georgia); BMFTR, DFG, and HGF (Germany); GSRI (Greece); NKFIH (Hungary); DAE and DST (India); IPM (Iran); SFI (Ireland); INFN (Italy); MSIT and NRF (Republic of Korea); MES (Latvia); LMTLT (Lithuania); MOE and UM (Malaysia); BUAP, CINVESTAV, CONACYT, LNS, SEP, and UASLP-FAI (Mexico); MOS (Montenegro); MBIE (New Zealand); PAEC (Pakistan); MES, NSC, and NAWA (Poland); FCT (Portugal); MESTD (Serbia); MICIU/AEI and PCTI (Spain); MOSTR (Sri Lanka); Swiss Funding Agencies (Switzerland); MST (Taipei); MHESI and NSTDA (Thailand); TUBITAK and TENMAK (T\"{u}rkiye); NASU (Ukraine); STFC (United Kingdom); DOE and NSF (USA).

\hyphenation{Rachada-pisek} Individuals have received support from the Marie-Curie program and the European Research Council and Horizon 2020 Grant, contract Nos.\ 675440, 724704, 752730, 758316, 765710, 824093, 101115353, 101002207, 101001205, and COST Action CA16108 (European Union); the Leventis Foundation; the Alfred P.\ Sloan Foundation; the Alexander von Humboldt Foundation; the Science Committee, project no. 22rl-037 (Armenia); the Fonds pour la Formation \`a la Recherche dans l'Industrie et dans l'Agriculture (FRIA-Belgium); the Beijing Municipal Science \& Technology Commission, No. Z191100007219010, the Fundamental Research Funds for the Central Universities, the Ministry of Science and Technology of China under Grant No. 2023YFA1605804, and the Natural Science Foundation of China under Grant No. 12061141002 (China); the Ministry of Education, Youth and Sports (MEYS) of the Czech Republic; the Shota Rustaveli National Science Foundation, grant FR-22-985 (Georgia); the Deutsche Forschungsgemeinschaft (DFG), among others, under Germany's Excellence Strategy -- EXC 2121 ``Quantum Universe" -- 390833306, and under project number 400140256 - GRK2497; the Hellenic Foundation for Research and Innovation (HFRI), Project Number 2288 (Greece); the Hungarian Academy of Sciences, the New National Excellence Program - \'UNKP, the NKFIH research grants K 131991, K 133046, K 138136, K 143460, K 143477, K 146913, K 146914, K 147048, 2020-2.2.1-ED-2021-00181, TKP2021-NKTA-64, and 2021-4.1.2-NEMZ\_KI-2024-00036 (Hungary); the Council of Science and Industrial Research, India; ICSC -- National Research Center for High Performance Computing, Big Data and Quantum Computing, FAIR -- Future Artificial Intelligence Research, and CUP I53D23001070006 (Mission 4 Component 1), funded by the NextGenerationEU program (Italy); the Latvian Council of Science; the Ministry of Education and Science, project no. 2022/WK/14, and the National Science Center, contracts Opus 2021/41/B/ST2/01369, 2021/43/B/ST2/01552, 2023/49/B/ST2/03273, and the NAWA contract BPN/PPO/2021/1/00011 (Poland); the Funda\c{c}\~ao para a Ci\^encia e a Tecnologia, grant CEECIND/01334/2018 (Portugal); the National Priorities Research Program by Qatar National Research Fund; MICIU/AEI/10.13039/501100011033, ERDF/EU, "European Union NextGenerationEU/PRTR", and Programa Severo Ochoa del Principado de Asturias (Spain); the Chulalongkorn Academic into Its 2nd Century Project Advancement Project, and the National Science, Research and Innovation Fund via the Program Management Unit for Human Resources \& Institutional Development, Research and Innovation, grant B39G680009 (Thailand); the Kavli Foundation; the Nvidia Corporation; the SuperMicro Corporation; the Welch Foundation, contract C-1845; and the Weston Havens Foundation (USA).
\end{acknowledgments}

\bibliography{auto_generated} 

\cleardoublepage \appendix\section{The CMS Collaboration \label{app:collab}}\begin{sloppypar}\hyphenpenalty=5000\widowpenalty=500\clubpenalty=5000
\cmsinstitute{Yerevan Physics Institute, Yerevan, Armenia}
{\tolerance=6000
A.~Hayrapetyan, V.~Makarenko\cmsorcid{0000-0002-8406-8605}, A.~Tumasyan\cmsAuthorMark{1}\cmsorcid{0009-0000-0684-6742}
\par}
\cmsinstitute{Institut f\"{u}r Hochenergiephysik, Vienna, Austria}
{\tolerance=6000
W.~Adam\cmsorcid{0000-0001-9099-4341}, J.W.~Andrejkovic, L.~Benato\cmsorcid{0000-0001-5135-7489}, T.~Bergauer\cmsorcid{0000-0002-5786-0293}, M.~Dragicevic\cmsorcid{0000-0003-1967-6783}, C.~Giordano, P.S.~Hussain\cmsorcid{0000-0002-4825-5278}, M.~Jeitler\cmsAuthorMark{2}\cmsorcid{0000-0002-5141-9560}, N.~Krammer\cmsorcid{0000-0002-0548-0985}, A.~Li\cmsorcid{0000-0002-4547-116X}, D.~Liko\cmsorcid{0000-0002-3380-473X}, M.~Matthewman, I.~Mikulec\cmsorcid{0000-0003-0385-2746}, J.~Schieck\cmsAuthorMark{2}\cmsorcid{0000-0002-1058-8093}, R.~Sch\"{o}fbeck\cmsAuthorMark{2}\cmsorcid{0000-0002-2332-8784}, D.~Schwarz\cmsorcid{0000-0002-3821-7331}, M.~Shooshtari, M.~Sonawane\cmsorcid{0000-0003-0510-7010}, W.~Waltenberger\cmsorcid{0000-0002-6215-7228}, C.-E.~Wulz\cmsAuthorMark{2}\cmsorcid{0000-0001-9226-5812}
% ...
\par}
\vskip\cmsinstskip
\dag:~Deceased\\
$^{1}$Also at Yerevan State University, Yerevan, Armenia\\
$^{2}$Also at TU Wien, Vienna, Austria\\
% ...
\end{sloppypar}
\end{document}